\documentclass[letterpaper]{emulateapj}
\usepackage{epsfig,graphicx}
\def\lap{\lower.5ex\hbox{$\; \buildrel < \over \sim \;$}}
\def\gap{\lower.5ex\hbox{$\; \buildrel > \over \sim \;$}}

\def\ergcm2s{${\rm erg\ cm^{-2}\ s^{-1}}$}

\def\ergscm2s{${\rm erg\ cm^{-2}\  s^{-1}}$}

\def\cm-2{${\rm cm^{-2}}$}

\begin{document}

\title{The Panchromatic Hubble Andromeda Treasury  X. Ultraviolet to Infrared Photometry of 117 Million Equidistant Stars}

\author{Benjamin F. Williams\altaffilmark{1},
Dustin Lang\altaffilmark{2},
Julianne J. Dalcanton\altaffilmark{1},
Andrew E. Dolphin\altaffilmark{3},
Daniel R. Weisz\altaffilmark{1,4,5},
Eric F. Bell\altaffilmark{6},
Luciana Bianchi\altaffilmark{7},
Eleanor Byler\altaffilmark{1},
Karoline M. Gilbert\altaffilmark{8},
L\'eo Girardi\altaffilmark{9},
Karl Gordon\altaffilmark{8},
Dylan Gregersen\altaffilmark{10},
L.~C. Johnson\altaffilmark{1},
Jason Kalirai\altaffilmark{7,8},
Tod R. Lauer\altaffilmark{11},
Antonela Monachesi\altaffilmark{6},
Philip Rosenfield\altaffilmark{12},
Anil Seth\altaffilmark{10}, and
Evan Skillman\altaffilmark{13}
}
\altaffiltext{1}{Department of Astronomy, Box 351580, University of Washington, Seattle, WA 98195; ben@astro.washington.edu, jd@astro.washington.edu, byler@astro.washington.edu, lcjohnso@astro.washington.edu,}
\altaffiltext{2}{McWilliams Center for Cosmology, Department of Physics, Carnegie Mellon University, 5000 Forbes Ave., Pittsburgh, PA; dstn@cmu.edu}
\altaffiltext{3}{Raytheon, 1151 E. Hermans Road, Tucson, AZ 85706; adolphin@raytheon.com}
\altaffiltext{4}{Department of Astronomy, University of California at Santa Cruz,1156 High Street, Santa Cruz, CA, 95064; drw@ucsc.edu}
\altaffiltext{5}{Hubble Fellow}
\altaffiltext{6}{Department of Astronomy, University of Michigan, 830 Denninson Building, Ann Arbor, MI 48109-1042; antonela@umich.edu,ericbell@umich.edu}
\altaffiltext{7}{Department of Physics and Astronomy, Johns Hopkins University, 3400 North Charles Street, Baltimore, MD 21218; bianchi@pha.jhu.edu}
\altaffiltext{8}{Space Telescope Science Institute, Baltimore, MD 21218; kgilbert@stsci.edu, kgordon@stsci.edu, jkalirai@stsci.edu}
\altaffiltext{9}{Osservatorio Astronomico di Padova -- INAF, 
  Vicolo dell'Osservatorio 5, I-35122 Padova, Italy, \email{leo.girardi@oapd.inaf.it}}
\altaffiltext{10}{Department of Astronomy, University of Utah; dylan.gregersen@utah.edu, aseth@astro.utah.edu}
\altaffiltext{11}{NOAO, 950 North Cherry Avenue, Tucson, AZ 85721; lauer@noao.edu}
\altaffiltext{12}{Dipartimento di Fisica e Astronomia Galileo Galilei,
  Universit\`a di Padova, Vicolo dell'Osservatorio 3, I-35122 Padova, Italy \email{philip.rosenfield@unipd.it}}
\altaffiltext{13}{Department of Astronomy, University of Minnesota, 116 Church SE, Minneapolis, MN 55455; skillman@astro.umn.edu}

\begin{abstract}

We have measured stellar photometry with the {\it Hubble Space
  Telescope} (HST) {\it Wide Field Camera 3} (WFC3) and {\it Advanced
  Camera for Surveys} (ACS) in near ultraviolet (F275W, F336W),
optical (F475W, F814W), and near infrared (F110W, F160W) bands for 117
million resolved stars in M31.  As part of the Panchromatic Hubble
Andromeda Treasury (PHAT) survey, we measured photometry with
simultaneous point spread function fitting across all bands and at all
source positions after precise astrometric image alignment ($<$5-10
milliarcsecond accuracy). In the outer disk, the photometry reaches a
completeness-limited depth of F475W$\sim$28, while in the crowded,
high surface brightness bulge, the photometry reaches
F475W$\sim$25. We find that simultaneous photometry and optimized
measurement parameters significantly increase the detection limit of
the lowest resolution filters (WFC3/IR) providing color-magnitude
diagrams that are up to 2.5 magnitudes deeper when compared with
color-magnitude diagrams from WFC3/IR photometry alone.  We present
extensive analysis of the data quality including comparisons of
luminosity functions and repeat measurements, and we use artificial
star tests to quantify photometric completeness, uncertainties and
biases.  We find that largest sources of systematic error in the
photometry are due to spatial variations in the point spread function
models and charge transfer efficiency corrections.  This stellar
catalog is the largest ever produced for equidistant sources, and is
publicly available for download by the community.

\end{abstract}

\section{Introduction}

The stellar content of galaxies probes fundamental quantities such as
the initial mass function, the cluster mass function, the distance
scale, stellar evolution, galaxy growth, the history of star
formation, and star formation feedback energetics.  Resolved stellar
photometry for a massive spiral galaxy therefore provides a superb
testbed for validating the details of many astrophysical processes.

Interpreting large libraries of Galactic stars is currently
challenging due to the uncertain and wide-ranging distances and
extinctions. In contrast, M31, the nearest large spiral galaxy to our
own, is far enough away that all of the disk stars are at the same
distance to within $\sim$1\%.  Furthermore, M31 is the only other
galaxy in the Local Group that is similar in mass, morphology, and
metallicity to those that host most of the stellar content of the
universe \citep[massive disk-dominated galaxies of roughly solar
  metallicity;][]{driver2007,gallazzi2008}.

\subsection{Ground-based M31 Disk Catalogs}

Although M31 seems an ideal target for producing a valuable library of
stellar photometry, work has historically been inhibited by its large
angular size \citep[190$'$ major axis;][]{devaucouleurs1991}. Most
ground based studies of M31 that have concentrated on the field star
population, on OB associations, or on portions of the halo
\citep{massey1986,mould1986,pritchet1988,haiman1994,davidge1993,durrell1994,durrell2001,ibata2001,mcconnachie2009}.

Due to the poor angular resolution available, ground-based surveys of
the disk are limited to photometry of only the brightest stars.  The
first resolved star study covering the optical disk was that of
\citet{magnier1992}, providing a catalog of 360,000 objects in 4
bands.  This catalog has been superseded by \citet{massey2006}, which
contains a similar number of stars in 5 bands with improved
photometric and astrometric precision and which has led to a census of the
brightest and most massive M31 members.  This photometry also provided
some constraints on the age distribution and star formation rate of
the disk \citep{williams2003}.

Ground-based halo studies have taken advantage of the very large
fields of view available on many telescopes, leading to the discovery
of extended density features from accreted smaller galaxies
\citep{ibata2001,ferguson2002}.  Further ground-based photometry and
spectral work showed the halo metallicity gradient and more structure
\citep{mcconnachie2009,gilbert2012}.

\subsection{HST M31 Catalogs}

Until now, HST imaging has been limited in coverage because of the
large angular size of M31; however, even small studies have had
significant scientific impact.  Studies of the stellar populations in
the bulge in the UV \citep{bertola1995} resolved hot stars, later
found to be post-AGB stars \citep{brown2008}.  Studies of the
luminosity function of the field stars in the bulge have also been
done in the IR \citep{stephens2003}.

HST observations of the resolved stellar populations of halo fields
provided detailed evidence of its metal-rich nature \citep{rich1996},
and very deep observations have provided the detailed metallicity
and age distribution of a handful of halo pointings
\citep{brown2003,brown2006,brown2007,brown2008,brown2009}

There have been several isolated HST fields in the disk analyzed as
well
\citep{sarajedini2001,williams2002,bellazzini2003,ferguson2005,brown2006}.
These studies have provided insight into how portions of the disk
formed, including finding signs for early metal enrichment, a typical
age older than 1 Gyr and extended star formation in many of the
currently star forming regions.  Furthermore, many individual star
forming regions in the disk have been observed in detail in the UV
\citep[e.g.,][]{bianchi2012}, allowing initial measurements of
hierarchical clustering and dispersion timescales of young stars.

Many other HST resolved star studies of M31 have focused on star
clusters, measuring metallicities, structural parameters for old
clusters \citep[e.g.,][]{holland1997,barmby2000,barmby2002} and ages
for young clusters \citep{williams2001a,williams2001b}, finding many
similarities to the Milky Way cluster population, but a larger number
of young, massive clusters in M31.  Improved catalogs and measurements
of star cluster parameters have also been ongoing
\citep{krienke2008,barmby2009,hodge2010,perina2010,tanvir2012,wang2013,agar2013}

All of these studies to date made significant advances in our
knowledge of M31, even though they were limited to scattered fields
across the disk and halo of M31.

\subsection{The Panchromatic Hubble Andromeda Treasury}

Given that M31 offers the best opportunity to study the resolved
stellar populations of a large spiral galaxy, we carried out the
Panchromatic Hubble Andromeda Treasury (PHAT) survey, covering
$\sim$1/3 of the star-forming disk of M31 in six bands from the
near-UV to the near-IR using HST's imaging cameras.  This survey combines the wide field coverage typical of ground-based surveys with the precision of HST observations.  The overall
survey strategy, initial photometry and data quality assessments were
described in detail in \citet{dalcanton2012}. Our wavelength
coverage provides the data necessary to constrain masses,
metallicities, and extinctions of {\it individual stars}, from which
we can infer the spatially-resolved metallicity, age, and extinction
distributions over a very large contiguous area.

The PHAT survey has been instrumental in several M31 discoveries
already.  These include improved constraints on post-AGB and
AGB-manqu$\acute{\rm e}$ evolutionary phases \citep{rosenfield2012},
tracing the stellar mass distribution of the inner M31 halo
\citep{williams2012}, new techniques for measuring robust ages and
masses of star clusters \citep{beerman2012} and for measuring the
initial mass function from resolved stellar photometry
\citep{weisz2013}, a major increase in the number of cataloged star
clusters in M31 \citep{johnson2012,morgan2014}, evidence for a
metallicity ceiling for Carbon stars \citep{boyer2013}, and additional
complexity in the structural components of M31 \citep{dorman2013}.
These results were based on our first generation of photometry, where
measurements were made for each camera separately, and then combined at
the catalog level.

In this paper, we report on our second generation of photometric
measurements of the resolved stars in the PHAT imaging, in which we
take advantage of all available information by carrying out photometry
simultaneously in all 6 filters.  This new approach gives a
significant increase in the depth and accuracy of our photometry over
that presented in \citet{dalcanton2012}.  Section 2 describes our
technique for performing and merging simultaneous 6-filter photometry,
fitting data from 3 HST cameras with different point spread functions
(PSFs), distortion corrections, and pixel scales.  Section 3 details
the resulting color-magnitude diagrams (CMDs).  Section 4 discusses
contamination from non-M31 sources.  Section 5 gives the detailed
analysis of the quality of the photometry, including a full assessment
of random and systematic uncertainties.  Section 6 compares this
generation of PHAT photometry to the previous version.  Section 7
describes individual fields whose photometry may differ from that of
the survey in general.  Section 8 describes the available data
products.  Finally, Section 9 provides a summary of our work.

\section{Data}

\subsection{Survey Overview}

The data for the PHAT survey were obtained from July 12, 2010 to
October 12, 2013 using the Advanced Camera for Surveys (ACS) Wide
Field Channel (WFC), the Wide Field Camera 3 (WFC3) IR (infrared)
channel, and the WFC3 UVIS (Ultraviolet-Optical) channel.  The
observing strategy is described in detail in \citet{dalcanton2012}.
In brief, we performed 414 2-orbit visits.  Each visit was matched by
another visit with the telescope rotated at 180 degrees, so that each
location in the survey footprint was covered by the WFC3/IR,
WFC3/UVIS, and ACS/WFC cameras.  The different orientations were
schedulable 6 months apart from one another, so that the WFC3 data and
ACS observations of each region were separated by 6 months.  A list of
the target names, observing dates, coordinates, orientations,
instruments, exposure times, and filters is given in
Table~\ref{obs_table}. We note that each field includes a short ($<$20
sec) exposure in F475W and F814W to avoid saturation of the brightest
stars.  In total, the area covered by all 6 filters is 0.5 deg$^2$.

\begin{figure}
\centerline{\includegraphics[width=7.5in]{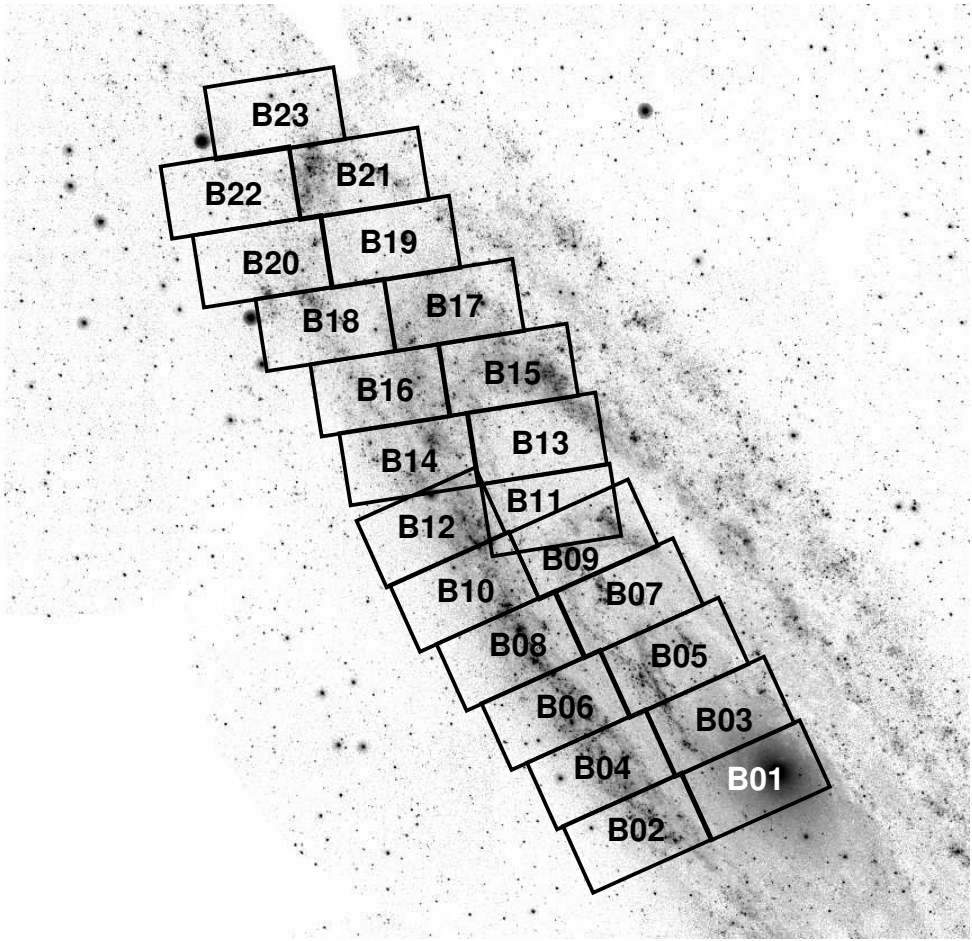}}
\caption{PHAT footprints.  The pattern of the 23 ``bricks'' of the PHAT survey are shown on a GALEX NUV image oriented with north up and east left.}
\label{bricks}
\end{figure}

\begin{figure}
\includegraphics[width=6.5in]{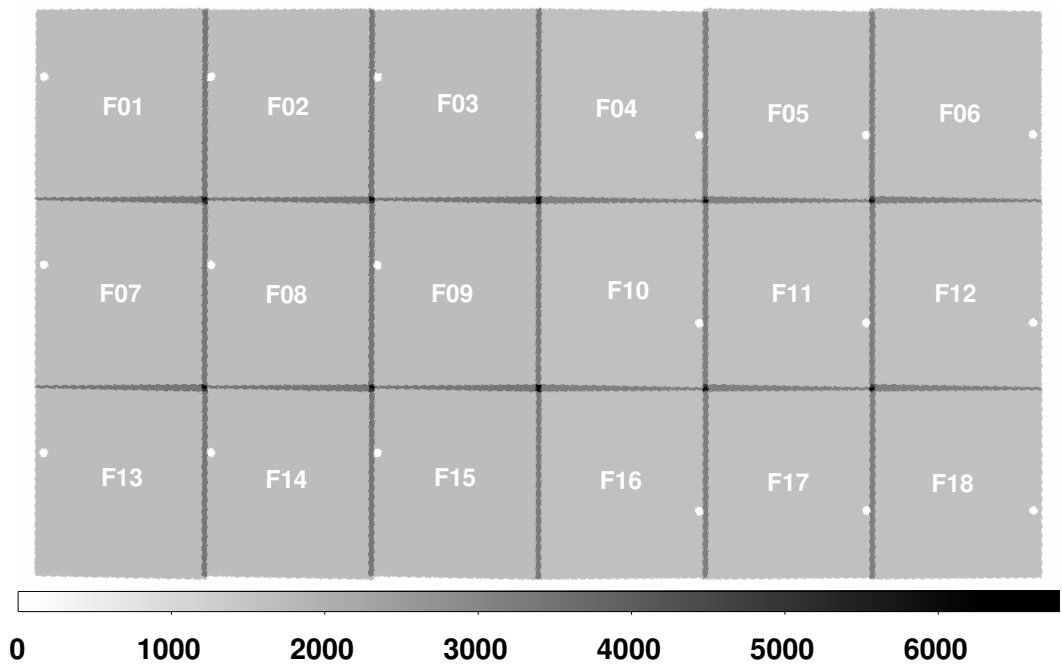}
\caption{IR exposure map with labeled field numbers for a generic PHAT
  brick. The grayscale is in units of seconds of exposure time.}
\label{ir_exposuremap}
\end{figure}

\begin{figure}
\includegraphics[width=6.5in]{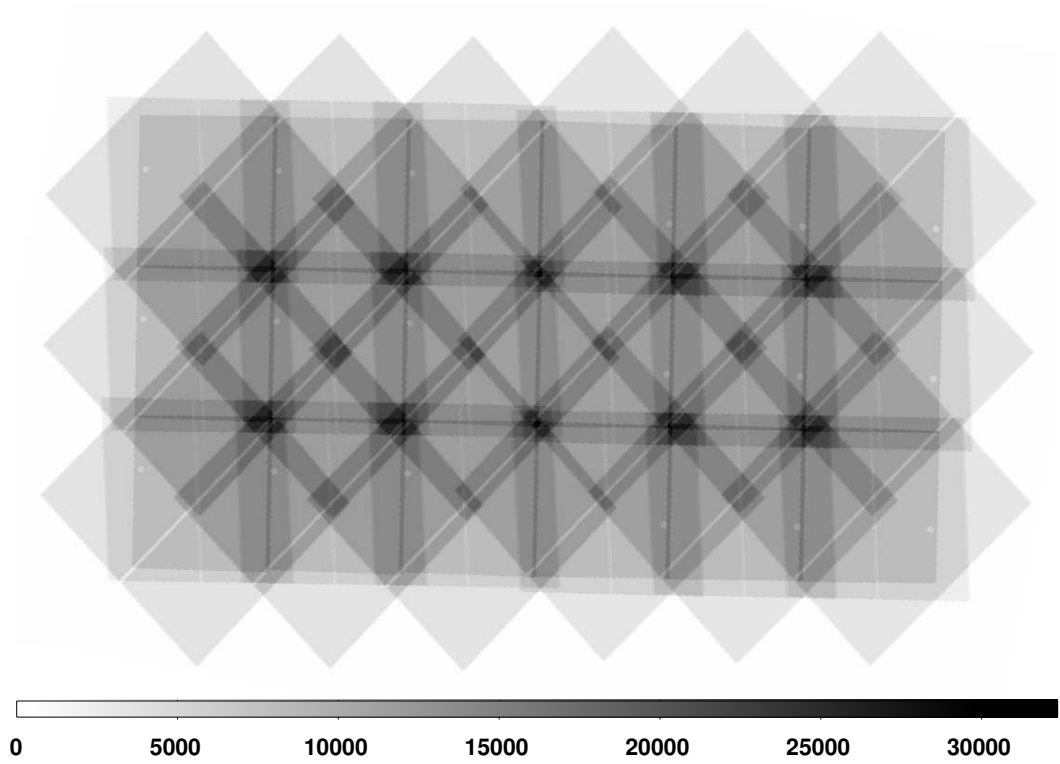}
\caption{Total exposure map for a generic PHAT brick including all
cameras.  The grayscale is in units of seconds of exposure time.}
\label{exposuremap}
\end{figure}

The survey was designed around the camera with the smallest footprint,
the WFC3/IR.  Thus the tiling is most easily navigated by plotting the
WFC3/IR footprints, as shown in the F160W exposure map in
Figure~\ref{ir_exposuremap}. The area plotted in Figure~\ref{ir_exposuremap} corresponds to a single
rectangle in Figure~\ref{bricks}.  The ACS/WFC and WFC3/UVIS exposure
maps are shown in Figure~\ref{exposuremap}.  This tiling was
necessarily much more complex to maximize the contiguous WFC3/IR
coverage and make the most efficient use of the ACS parallels.

Each survey region is described by two identifiers, a ``brick'' number
and a ``field'' number. ``Bricks'' are rectangular areas of
$\sim$6$'{\times}12'$ that correspond to a 3$\times$6 array of WFC3/IR
footprints.  The bricks are numbered, 1-23, starting from the brick at
the nucleus, and counting west to east, south to north.  Odd numbered
bricks move out along M31's major axis, and even numbered bricks
follow the eastern side of the odd-numbered bricks, as shown in
Figure~\ref{bricks}.  Within each of the 23 bricks, there are 18
``fields'' that each correspond to an IR footprint.  These fields are
numbered, 1-18, from the northeast corner, counting east to west,
north to south (see Figure~\ref{ir_exposuremap}).  Thus,
Brick~23-Field~18 is the IR footprint in the southwest corner of the
brick farthest out along the major axis.  Each pointing has its own
unique brick, field number.  Images in other cameras are labeled by
the brick and field of the IR footprint they primarily overlap (e.g.,
M31-B01-F13-ACS overlaps M31-B01-F13-IR even though it was taken in
parallel during the primary observations of M31-B01-F16-IR).

\begin{figure}
\includegraphics[width=6.5in]{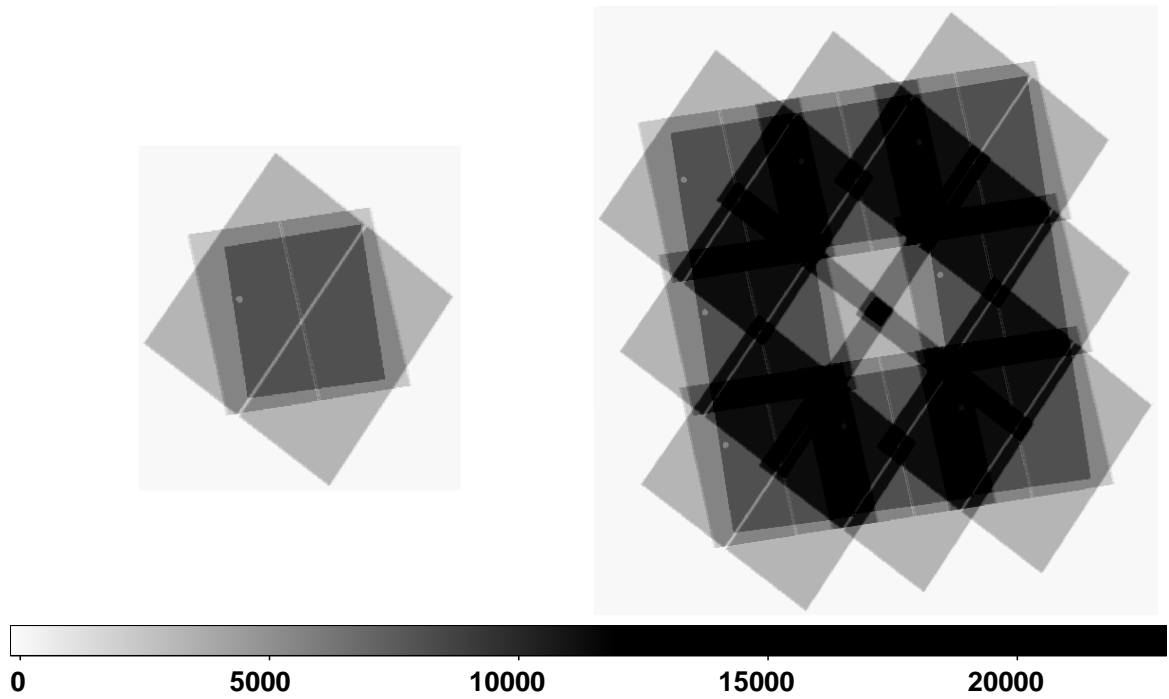}
\caption{{\it Left:} Total exposure map for one PHAT field including
  all cameras. This shows the data stack that is included for one
  group of CCD reads into DOLPHOT. {\it Right:} Exposure map of all of
  the neighboring observations that overlap with IR footprint of the
  field shown in {\it left}.  It was too computationally expensive to
  include all of these fields in the simultaneous stack in this
  generation of photometry, but their addition may improve future
  generations.  The grayscale is in units of seconds of exposure
  time.}
\label{onefield}
\end{figure}

Although the neighboring fields overlap significantly in ACS and UVIS,
for the purposes of this generation of photometry, we were not able to
take advantage of this extra exposure due to computational
limitations.  Including all exposures that overlap a location in our
survey requires putting more than 200 CCD reads into memory and slows
down the measurements, requiring a large amount of processing time on
expensive high-memory machines.  Therefore, when carrying out the
6-filter photometry for each survey brick+field, we included only the
UVIS, ACS, and IR data specifically pointed at that brick+field.  The
effective exposure for each field as measured is shown in
Figure~\ref{onefield}.  Note that the single-field data leaves the chip gap in ACS uncovered; we discuss how these areas were treated in Section~\ref{catalog_merge}. The Figure also shows the other exposures in
the survey that could be included in the photometry of this single
field.  For our next generation of photometry, the computing power
should be available to include all overlapping exposures when
measuring photometry of each field, which will improve the depth and
homogeneity of the photometry.

\subsection{Astrometry}\label{alignment}

To measure stars in all 6 bands, our first challenge was to align all
of our data to a relative precision of $\sim$0.1 ACS pixel (5
milliarcseconds). We aligned the images based on preliminary star
catalogs using initial photometry performed on each individual CCD readout to generate
initial positions of stars for each CCD of each exposure.  These
catalogs were then cross-correlated to solve for the relative
astrometry of each CCD of each exposure in each brick.  We now describe the
process in detail.

We first cleaned the images of  cosmic rays (CRs)  to avoid them being mistakenly matched and
hindering alignment.  The {\tt flt}
images as initially downloaded from MAST at the time did not have
well-determined CRs, partially due to our observing strategy, which
had only 2 exposures per band in UVIS and a very wide range of
exposure times in ACS.  Therefore, we performed out own CR flagging
with the PyRAF task {\it astrodrizzle} \citep{gonzaga2012}, which
identifies CRs by finding pixels that are bright in one exposure and
not in the other exposures in the same band.  The CRs were identified
and flagged in the DQ extensions of the {\tt flt} images.  We note
that the MAST data reduction pipeline is always improving, so that
future downloads may not require this extra step for reliable CR
flagging.
 
We used the point spread function photometry package DOLPHOT --- an
updated version of HSTPHOT \citep{dolphin2000} --- to find and measure the
positions of stars on each CCD of each exposure of the survey.  These
positions provide the basis for astrometric alignment.  The DOLPHOT
task {\it acsmask} or {\it wfc3mask} updates the images into units of
counts per pixel appropriate for DOLPHOT photometry.  This task
applies the data quality (DQ) flags as well as the pixel area map to
the {\tt flt} science (SCI) extensions to mask out bad pixels
and CRs, and calibrates the flux in each pixel by the area coverage on
the sky.  The {\tt flt} images were then split into their individual
CCDs using the DOLPHOT task {\it splitgroups}.  These single-CCD
DOLPHOT catalogs were culled for artifacts using the same criteria as
those used for the single-camera photometry in \citet{dalcanton2012}.

For the UVIS and short ACS exposures, in some cases there were fewer than $\sim$10
stars for determining alignment.  For several of these cases, we found that
alignment solutions were not acceptable when we measured star
positions on the flat-fielded ({\tt flt}) images.  However, we
circumvented this problem in many cases by measuring star positions on
charge transfer efficiency (CTE)-corrected ({\tt flc}) images in cases
where convergence was not reached from the positions measured on the
{\tt flt} images.  Thus, we found that the CTE-corrected images
produce more reliable centroids than the {\tt flt} images, showing the
high-quality of the on-image CTE-correction algorithm
\citep{anderson2010}.  However, there were still some exposures that
were not able to be aligned due to the low density of detected point
sources (see Section\ref{anomalous_fields}).

Using the single-CCD catalogs, astrometric alignment was performed as
follows.  We first aligned the ACS catalogs with exposure times $>30$
seconds with each other and with a deep CFHT i-band image (which was
in turn calibrated to 2MASS).  The alignment proceeds by assuming the
input astrometry is roughly correct and then performing an astrometric
cross-match between each pair of ACS images and each ACS image to the
CFHT reference catalog.  We align the brightest stars in each sample
to reduce the computational cost and to reduce the impact of measurement
error in fainter stars.  Each matching provides a set of astrometric
offsets, some of which are correct and some of which are due to
spurious matches, or matches to neighboring stars.  We fit these
offsets as a mixture of a flat background (for spurious matches) and a
Gaussian (for correct matches), using a coarse histogram to initialize
an Expectation-Maximization method.  The result allows us to weight
each matched star by its probability of being drawn from the
foreground distribution.  We then solve the large least-squares
problem to find astrometric shifts and affine transforms (rotations,
scalings, and shears) of each input catalog to minimize the matching
residuals.  We repeat this process several times, using a decreasing
matching radius and including an increasing number of fainter stars.
After having aligned the long ACS exposures with each other and the
reference frame in this way, we repeat this process, aligning the
short ACS exposures and the WFC3 IR and UVIS exposures to the ACS
reference. 

The result of these alignment processes is a set of scripts that
update the FITS WCS astrometric headers of the input images, including
all affine geometric transformations.  We applied the image header
update scripts to their respective {\tt flt} images as originally
downloaded from STScI and processed by OPUS 2012\_4.  We then combined
images with updated astrometry to produce images of each brick in each
band, using the PyRAF task {\tt astrodrizzle} with the following
altered parameters: {\tt context=True, build=True, skysub=False,
  final\_wcs=True, final\_rot=0.0, clean=True}.  These brickwide
astrodrizzled images are available from the Space Telescope Science
Institute through the high-level science products archive.

\begin{figure}
\centerline{\includegraphics[width=6.5in]{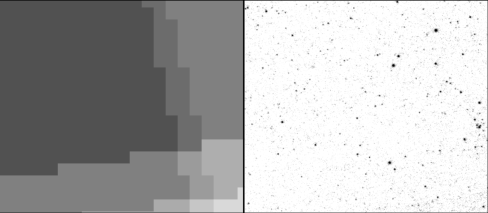}}
\centerline{\includegraphics[width=6.5in]{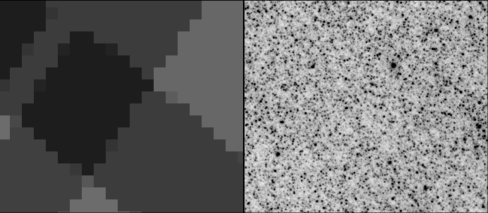}}
\centerline{\includegraphics[width=6.5in]{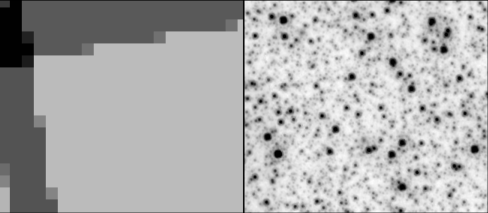}}
\caption{Stacked image quality in each camera.  Left panels show the
  exposure map (1$''$ per pixel; darker gray indicates more exposure)
  of the 17$^{\prime\prime}{\times}17^{\prime\prime}$ location shown
  on the right in UVIS (F336W; top) ACS (F475W; middle), and IR
  (F160W; bottom).  The images (grayscale inverted) show an area of
  complex overlaps including the 180$^{\circ}$ flip at the middle of
  Brick 14.  The precision alignment results in virtually seamless
  images.}
\label{drizzle}
\end{figure}

\begin{figure}
\includegraphics[width=3.1in]{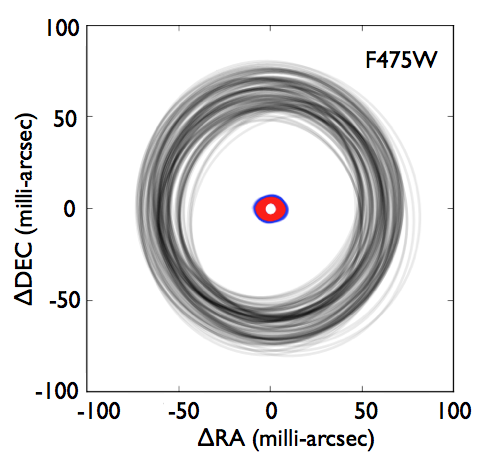}
\centerline{\hspace{-3.5in}\includegraphics[width=3.1in]{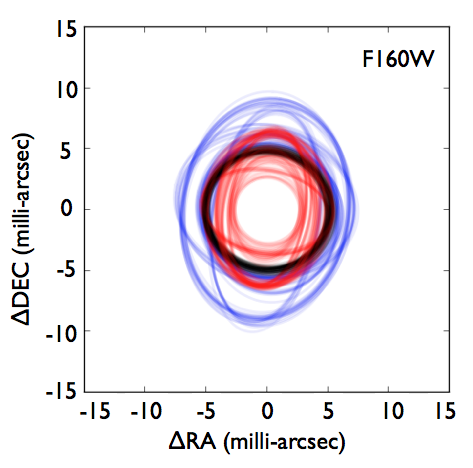}}
\includegraphics[width=3.2in]{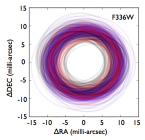}
\includegraphics[width=3.2in]{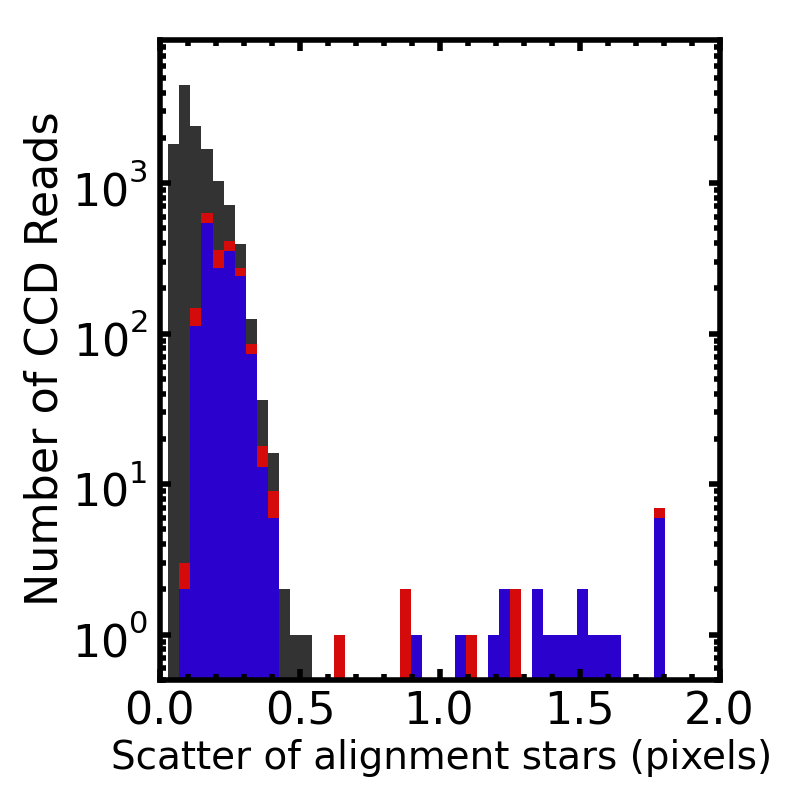}
\caption{Alignment quality of all images in the survey.  {\it Upper
Left:} One-sigma uncertainty ellipses for alignment between pairs of
CCD reads for all F475W data in Brick 15.  The ellipses are fit via
expectation-maximization of a foreground-background model of all
star-to-star matches between the pair of exposures.  Each red ellipse
shows the uncertainty between one pair of F475W CCD reads.  Each blue
ellipse shows the uncertainty between an F475W CCD read and a F814W
CCD read. Each gray ellipse shows the uncertainty between an F475W CCD
read and the ground-based absolute astrometric reference.{\it Upper
Right:} Same as {\it upper left}, but for F160W CCD read, and gray
circles show alignment to the F475W images, red to other F160W reads,
and blue to F110W reads. {\it Lower Left:} Same as {\it upper right},
but for F336W, and gray circles show alignment to the F475W images,
red to other F336W reads, and blue to F275W reads.  {\it Lower Right:}
The scatter of alignment stars is computed as the standard deviation
of the distribution of alignment stars in units of native pixels of
the CCD read being aligned.  Blue indicates short ($<$20 seconds)
exposure times.  Red indicates UVIS images of the North and East edges
of the survey footprint.}
\label{align_hist}
\end{figure}

A zoomed example of a small
(17$^{\prime\prime}{\times}17^{\prime\prime}$) region with many
overlaps in all 3 cameras is shown in Figure~\ref{drizzle}, which
reveals the seamless nature of our stacked, aligned data.  More
quantitative alignment checks are shown in Figure~\ref{align_hist},
where the first 3 panels show error ellipses are plotted for each
individual CCD read compared to the reference frame (gray), to other
CCD reads in the same filter (red), and compared to CCD reads in other
filters (blue).  These one-sigma error ellipses were measured via
expectation-maximization of a model of all star-to-star matches
between each pair of exposures.  The alignment uncertainties were
typically $\lap$0.01$''$ between different HST images in all cameras.
The reference frame is the ground-based global astrometric reference
for F475W, and the F475W images provide the reference frame for the
other bands. The uncertainties on the alignment between the F475W and
the reference frame are the largest (though still $\sim$0.05$''$).

The resulting images were carefully checked by eye to look for any
problems with alignments within each field and across overlaps with
neighboring fields.  These brickwide images are also publicly released
in the Mikulski Archive at Space Telescope (MAST) High Level Science
Products (HLSP).

\subsection{Data Processing}\label{processing}

After updating the original {\tt flt} images with new astrometry, the
images were passed through our multicamera photometry pipeline, which
consists of several pre-processing steps, the run of DOLPHOT, and
several post-processing steps to check quality and produce easy-to-use
data products. We now describe the steps in detail.

\subsubsection{Running DOLPHOT}

The preprocessing divides the data into different groups determined by
their respective brick+field locations.  Data from each brick+field
were processed separately (see left panel of Figure~\ref{onefield}).
This stack contains 31 CCD reads.  There are 9 ACS exposures (2 CCDs
per exposure), 5 IR exposures (1 CCD per exposure), and 4 UVIS
exposures (2 CCDs per exposure).  Thus, there were 18 separate groups
of 31 CCD reads for each brick.  The entire processing pipeline was
run independently for each of these 18 groups.

Within each group, all of the preprocessing of the individual images
was carried out in the same manner as when preparing them for
individual photometry (see previous subsection), but with improved
alignment.  The identification of CRs was also more reliable because
prior to our astrometry updates, misaligned stars were sometimes
identified as CRs.

DOLPHOT uses a reference image as the astrometric standard. It ties
each detection to a location on the reference image.  We used the
stacked, distortion-corrected F475W image as this reference because it
offers the best combination of depth, completeness, and resolution.
To generate the F475W reference image, the F475W astrodrizzle output
was also processed with {\it acsmask}, {\it calcsky}, and {\it
splitgroups} in order to be used as the DOLPHOT reference image.

As described in \citet{dalcanton2012}, the package DOLPHOT was used
for all of the photometry for the PHAT survey.  However, the quality
of the photometry output from DOLPHOT is strongly dependent on the
appropriateness of the parameters set by the user.  In particular, the
precision and completeness of photometry in very crowded fields---like
those in the PHAT survey---can be strongly affected by the technique
DOLPHOT uses to measure the local sky and the size of the aperture
used for determining PSF offsets.

To optimize our photometry, we performed a grid of test runs of
DOLPHOT on several select fields using a variety of DOLPHOT
parameters.  The parameters we varied were {\tt R$_{\tt Aper}$}, {\tt
  R$_{\tt Chi}$}, {\tt FitSky}, and {\tt PSFPhot} which control the
radius inside of which aperture photometry is compared with
PSF-fitting, the radius inside of which the quality of the PSF fit is
assessed, the method for measuring the background level for each star,
and the weighting during PSF fitting, respectively.  Ranges of the
parameters were 2--8, 1.5--8, 1--3, and 1--2 for each of the four
parameters, respectively. One thousand artificial stars were inserted
at each of 2--3 CMD locations roughly corresponding to the main
sequence (UVIS and ACS only), the RGB (ACS and IR only), and $\sim$1
mag above the S/N cutoff in both filters (all 3 cameras). All of our
photometry uses the Anderson (ACS ISR 2006-01) PSF library.

To assess the quality of each set of test parameters, we calculated
the number of detected sources (all 3 cameras), the red clump
tightness (ACS only), and the RGB width (IR only) as measured from the
output photometry for all combinations of processing parameters.  We
also assessed the artificial star completeness, bias, and scatter (all
3 cameras) as measured from the artificial star tests performed on the
test fields for each set of test parameters.

Based on these metrics, the best photometry resulted from small
apertures, standard PSF photometric weighting, and measuring the sky
immediately outside the photometry aperture (but within the PSF).
This requires adjusting the PSF for the fact that sky is being
measured in a region where the star itself contributes.  These
parameters were clearly superior in the most crowded regions, and were
at least as good as other parameter combinations in uncrowded regions,
where the sensitivity to parameter choices is much less.  Our final
DOLPHOT processing parameters are given in Table~\ref{parameters}.  

We note that while CTE correction was turned on ({\tt UseCTE=1}), CTE
corrections were only made for the ACS bands (F475W and F814W) because
when we began our photometry no CTE corrections were available for
WFC3.  In the next generation of photometry, we plan to run our
measurements on CTE-corrected images, and turn the DOLPHOT CTE
corrections off, which may further improve the level of systematic
errors in our photometry.

We were able to improve the results from DOLPHOT by fixing the {\tt
  ref2img} parameters, which control the distortion terms for the
aligning images.  To optimize cross-camera alignment, DOLPHOT uses the
many stars in the image stack to refine the distortion solution for
each camera.  If these parameters were allowed to be freely fitted for
all individual fields, sometimes relatively empty short exposures or
UV exposures could throw off the solution.  Thus, using a few fields
from Brick 21, where the crowding is low, and the density of UV-bright
stars is relatively high, we measured the {\tt ref2img} values that
provided the best image alignment. This DOLPHOT parameter provides
minor changes to the IDCTAB specified distortions, which yield the
best fit for our data, but because we did not include the overlaps at
different cross-camera orientations from the standard PHAT observation
strategy, our precise terms would likely not be relevant for other HST
observing programs.

We fixed the {\tt ref2img} values measured in Brick 21 for the rest of
the survey data, which resulted in excellent alignment across the
survey.  DOLPHOT measures the precision by which the images of the
input stack are aligned with the reference image. Our median
astrometric precision is $\sim$5 milliarcseconds.  In the lower-right
panel of Figure~\ref{align_hist}, we provide the histogram of
alignment values from DOLPHOT, in units of pixels, for all of the CCD
chip readouts for the entire survey, showing 99.8\% of all of the
survey images are aligned to 0.5 pixels or better.  The median value
is 0.11 ACS pixels, or 5 milliarcseconds.  The small tail of CCD reads
with poor alignment are a small fraction of the very short exposure
(10--15~s) ACS images in regions with few bright stars, and a few UVIS
exposures in regions along the NE edge of the survey with little or no
young stars to align the image.  These few images are essentially pure
noise, and therefore contribute very little to any of our photometry
measurements.  Because few, if any, stars are detected in these
exposures, they are not included in the vast majority of combined
photometry measurements.  In cases where a noise spike is measured in
the very short optical exposures (i.e. any star not saturated in the
long exposures), that measurement will be weighted at 0.006 (0.6\%,
the fraction of the exposure time in these frames), making its effects
on the resulting photometry at most 0.6\%.  These fields are described
in more detail, along the other fields that have data issues, in
\S~\ref{anomalous_fields}.

When DOLPHOT is run on a collection of images, all of the individual
CCDs are aligned to the reference image in memory, stacked to search
for any significant peaks, and then each significant peak above the
background level is fitted with the point spread function for the
specific band, and camera of each CCD in the stack.  The measurements
are corrected for charge transfer efficiency (CTE, ACS only) and
calibrated to infinite aperture.  The measurements are then combined
into a final measurement of the photometry of the star.  The final
measurements include a combined value for all data in each band for
the count rate, rate error, VEGA magnitude and error, background,
$\chi$ of the PSF fit, sharpness, roundness, crowding, and
signal-to-noise.  

VEGA magnitudes apply the encircled energy corrections and zero points
from the ACS handbook dated 15-July-2008 and the WFC3 encircled energy
corrections and zero points from 2010 by J. Kalirai.  The rate and
rate error measurements are particularly useful for stars that were
not detected in one or more bands (and therefore have negative rates
or rates of 0, which results in an undefined magnitude), as they
provide upper-limits in these bands that can be employed for
constraints when fitting spectral energy distributions. The sharpness
is zero for a perfectly-fit star, positive for a star that is too
sharp (perhaps a cosmic ray), and negative for a star that is too
broad (perhaps a blend, cluster, or galaxy).  The crowding parameter
is in magnitudes, and tells how much brighter the star would have been
measured had nearby stars not been fit simultaneously. For an isolated
star, the value is zero.  These measurements are also output for each
individual exposure, allowing for the possibility of variability
studies (Wagner-Kaiser et al., in prep.) or searches for artifacts not
masked by our preprocessing techniques.  All measurements are output
to a ``{\tt .phot}'' ascii file.

\subsubsection{Creation of Photometry Catalogs}

\begin{figure}
\includegraphics[width=3.3in]{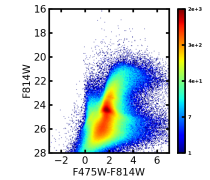}
\includegraphics[width=3.3in]{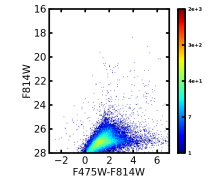}
\caption{Example of raw {\tt phot.fits} output in the optical ({\it
left}) and measurements culled from the raw DOLPHOT output ({\tt
phot.fits}) when producing the ST tables ({\tt st.fits}) because no
single band had a measurement with signal-to-noise$>$4 and a sharpness
value that passed our cut from Table~\ref{gstcuts} ({\it left}). No
CMD feature is seen in the culled measurements, showing the low
quality of these measurements.}
\label{st_cull}
\end{figure}

The star positions from DOLPHOT are then converted to RA and Dec using
the header of the reference image (see above) for astrometry.  RA and
Dec are then added to the phot catalog, and the full photometric
catalog (481 columns) is placed into a tagged FITS table ({\tt
  .phot.fits} file). The data are then culled to include only the
combined output for sources, significantly reducing the size of the
data files for those not interested in the measurements on the
individual CCD chips. At this step, we also kept only sources with
S/N$\geq$4 and reasonable sharpness (see Table~\ref{gstcuts}) in at
least one band (``ST'' catalogs) to limit the number of noise spikes,
CRs, and artifacts surrounding saturated stars in our catalogs.  This
initial cut removed up to 20\% of the objects from the initial DOLPHOT
output, mostly very low signal-to-noise with some scattered brighter
measurements.  An example CMD of the culled objects is shown in the
right side panel of Figure~\ref{st_cull}, and it shows no
overdensities associated with typical CMD features. Thus, users
looking for a specific source may find it in the pre-culled {\tt
  phot.fits} files, if it is not contained in the ST files; however,
any measurement not included in the ST files should only be used with
extreme caution.

The sources in the ST catalog are then flagged so that any band with
measurements of high crowding or high square of the sharpness (see
Table~\ref{gstcuts}) or that have S/N$<$4 can be easily left out of
any analysis, resulting in the ``GST'' sample. The flagged
measurements are likely to be unreliable (strongly affected by
blending, CRs, or instrument artifacts).  The key difference between
all measurements in the ST catalogs and the subset that pass our GST
criteria is that the GST criteria are performed per band.  For
example, if a star is well-measured in F275W, F336W, F475W, F814W, and
F110W it may not be well-measured in F160W.  Measurements in all six
bands for the star will appear in the catalog, but its F160W
measurement will have a GST flag of 0. A further difference between
the full catalog measurements and those with the GST flag is the use
of crowding as a quality metric.  Crowding is one of the quality
metrics considered when determining which bands pass the GST flag
criteria for each star; however, crowding is not considered at all for
inclusion in ST the catalog.

The precise values of sharpness and crowding used for cutting were
different for the UV, optical, and IR, and are provided in
Table~\ref{gstcuts}.  We used both qualitative and quantitative
criteria to determine the photometric quality cuts that maximize the
quality of the stellar CMDs (e.g., number of stars, photometric depth,
tightness of features) and minimize the number of non-stellar objects
(e.g., blends, background galaxies, cosmic rays).  We accomplished
this by a combination of visual inspection and fitting particular
CMDs, which we now describe.

For each permutation of DOLPHOT input parameters (\texttt{raper},
\texttt{rchi}, \texttt{fit sky} \texttt{psfphot}), we constructed 140
per-camera CMDs for various permutations of sharpness and crowding.
For each CMD, we calculated or visually inspected the following
factors: CMD depth for a fixed SNR$=$4, number of stars that passed
the photometric cuts, quality of maximum likelihood fits of a Gaussian
plus line model to the color and luminosity profile of the RC (only
for the ACS and IR CMDs), and the CMD of rejected objects.  In
addition, we also considered the completeness fraction and
color/magnitude biases from sets of artificial star tests (ASTs).
Specifically, for each CMD, we inserted 1000 ASTs 1 magnitude above
the S/N limit, 1000 on the upper MS, and 1000 on the upper RGB.  We
computed the recovered fraction (i.e., completeness) and mean
magnitude and color biases for each set of 1000 ASTs.

Using all of the above criteria, we found that the majority of
photometric quality cut combinations produced CMDs that were clearly
not ideal (e.g., poor completeness, obvious stellar features on the
rejected CMDs, poor fits to the RC).  We readily eliminated these
permutations and focused on the small subset that produced deep CMDs
with clear features.  To distinguish among these, we primarily
considered the completeness fraction and magnitude and color spreads
of the AST sets.  We also visually compared high S/N regions of the
various CMDs for tightness of the luminous CMD features (e.g., upper
MS), and eliminated those that rejected a higher percentage of high
S/N objects that had colors and magnitudes consistent with known
stellar sequences (e.g., MS, HeB sequence).  Based on this iterative
process, we found the parameters listed in Table~\ref{gstcuts}
provided for the best overall CMDs.  We note that these cuts were
chosen to provide an approximate set of values for obtaining
high-quality CMDs from our catalogs over the entire survey.  Those
using our photometry for specific projects will likely want to
determine their own flags optimized for their science needs.

\subsection{Merging Catalogs}\label{catalog_merge}

The final ST photometry catalogs for the individual fields are then
combined into brick-wide catalogs, and then into a survey-wide
catalog.  To avoid duplicate catalog entries, for each brick we
generated a grid in RA and Dec with single-point corners inside the IR
footprint corners of the survey, creating unique regions and trimming
the IR detector edges simultaneously.  Then the catalog of each field
of the grid was cut to include only detections inside of that field
region as defined by the grid.  These cut catalogs were then combined
into a initial brickwide catalogs with no duplicate entries and no
gaps in UVIS or IR coverage.
  
Because ACS exposures from neighboring fields are not included in the
DOLPHOT photometry measurements of an individual field, the ACS chip
gap contains no optical coverage (the UVIS chip gap was covered by
dithers).  Thus, the initial brickwide catalogs had only UV and IR
measurements for stars in the ACS chip gap.  However, the catalogs for
the neighboring fields contain optical measurements at these
locations.  Within the chip gaps, we include ACS measurements from
neighboring fields: for example, the top half of the chip gap in Field
1 is has optical measurements in the catalog for Field 2 (the field to
the right), while the bottom half has optical measurements in the
catalog for Field 7 (the field below).  The resulting final brick-wide
catalogs have full 6-band coverage throughout, although the photometry
in the chip gap has matching that was performed at the catalog level.
Therefore, in the ACS chip gaps, the UVIS and IR photometry will be of
different quality due to the lack of ACS data in that region when the
image stack was being reduced (see Section~\ref{compare}).  To allow
investigators to efficiently leave out the chip gaps if they would
like to avoid these non-uniform measurements, we have included an {\tt
  inside\_chipgap} flag in our catalogs, which is set to 1 for all
stars that were affected by this catalog-level merging.

We defined brick edges in the same way as field edges, with single
corner points.  Then we trimmed the brickwide catalogs to contain only
stars within these single corner points.  Once trimmed in this way,
the catalogs could be merged to produce a single catalog for the
entire survey with no duplicate entries and no gaps in 6-band
coverage.  We note the only gap to be the 0.1 arcmin$^2$ area that we
did not observe due to changes in the observing program that were
necessary to have guide stars for all observations (see
Section~\ref{altered_obs} for details).

The number of stars in each 6-band brick catalog is given in
Table~\ref{stars}, where the total number of ST stars measured in each
brick is given along with the number of reliable (high
signal-to-noise, low crowding) GST measurements in each band. The
entire generation 2 catalog has 116,861,772 stars.  Each one of these
was measured in at least 16 exposures (ignoring overlaps between
fields and the very short ACS exposures), making 1,869,788,352
photometric measurements.  In cases where the star was not detected in
a given band, the {\tt rate} and {\tt rate\_error}, measured based on
the sky level, can be used as a constraint for a non-detection, which
is recorded as a 99.999 in the magnitude column.  The full catalog is
available in machine readable format on Vizier.  The individual brick
catalogs are available in FITS format from
MAST.\footnote{http://archive.stsci.edu/missions/hlsp/phat/} A small
example of the catalog is provided in Table~\ref{catalog}.


It is clear from Table~\ref{stars} that no band has good measurements
of all 117 million stars, but each of the 117 million stars has a good
measurement in at least one band.  The most extreme cases of this are
in the UV, where only 0.3\% of the total 2$^{\rm nd}$ generation
catalog has reliable measurements (other than upper-limits) for both
the F275W and F336W bands. A very high percentage of these UV
measurements are actually full 6-band measurements.  

\begin{figure}
\includegraphics[width=2.3in]{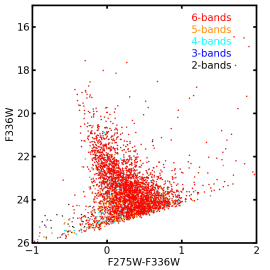}
\includegraphics[width=2.3in]{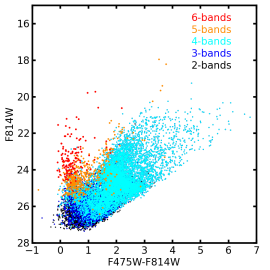}
\includegraphics[width=2.3in]{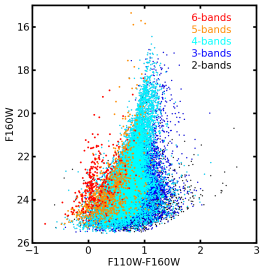}
\caption{UV, optical, and IR CMDs of a small random sample from the catalog, color-coded to show the number of bands with good measurements. In all panels, black, blue, cyan, orange, and red points show stars with good measurements in 2, 3, 4, 5, and 6 bands, respectively.
{\it Left:} UV CMD produced from all
measurements where both bands passed our GST criteria.  {\it
Center:} Optical CMD produced from all
measurements where both bands passed our GST criteria. 
{\it Right:} IR CMD produced from all
measurements where both bands passed our GST criteria.
}
\label{nbands}
\end{figure}

In Figure~\ref{nbands} we plot UV, Optical, and IR CMDs of a small
random sample from our catalog, with the points color-coded by the
total number of bands with reliable measurements.  The UV CMD shows
that nearly every data point corresponds to a star with reliable
measurements in all 6 bands.  The optical CMD shows that a high
fraction of RGB stars are detected in 4 bands (the optical and IR
bands), and only the brightest main-sequence stars tend to be detected
in 6 bands.  The IR CMD shows again that most RGB stars are measured
in 4 bands (optical and IR) while only the brightest main sequence
stars are measured in the UV.  Taken together, these results show that
only faint blue stars tend to have good measurements in fewer than 4
bands.  UV-bright stars tend to have good IR measurements, but
IR-bright stars do not tend to have good UV measurements.  Both tend
to have good optical measurements, given that the ACS data have the
greatest depth in terms of CMD features; however, the reddest RGB
stars are only detected in the IR.

\section{CMDs}

\begin{figure}
\includegraphics[width=3.3in]{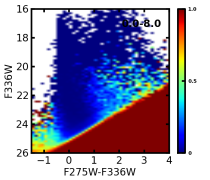}
\includegraphics[width=3.3in]{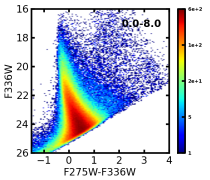}
\caption{UV color magnitude diagram for all F275W and F336W
measurements in the survey.  Numbers in the upper-right of the panels refer to relative stellar densities shown on the map in Figure~\ref{density_map}.  In this case, the full survey, covering all stellar densities, is shown. {\it Left:} Fraction of measurements
flagged as not passing our quality cuts in either of the two bands.
{\it Right:} CMD produced showing only measurements that pass our
quality cuts in both bands.}
\label{uv_cmds}
\end{figure}

We show the UV photometry for the entire survey in the CMDs in
Figure~\ref{uv_cmds}. The UV photometry had roughly equal depth
throughout the survey because it had constant, low crowding everywhere
due to its not being deep enough to include the very numerous RGB
stars, which are very faint in the UV. Thus, only for the UV do we
show the entire survey on a single CMD.  The figure shows the CMD
before our quality cuts, the fraction of sources that pass our quality
cuts as a function of color and magnitude, and the CMD after our
quality cuts.  Stars that survive the cuts but fall in regions where
large fractions of the stars were rejected have a higher probability
of being poor measurements.

\begin{figure}
\centerline{\includegraphics[width=5.5in]{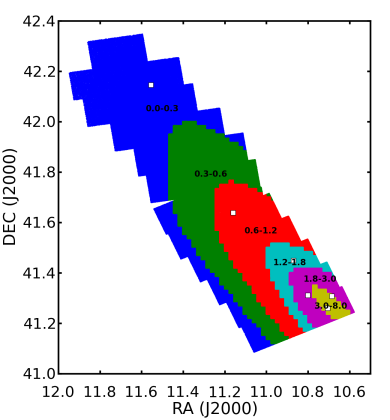}}
\caption{Map of the six levels of model stellar densities used to
  define the six regions that were used to make the different CMDs
  shown in Figures~\ref{uvoptical_bad_cmds}--\ref{ir_gst_cmds}.  Numbers refer to the log of the stellar density of red giant branch stars.  The numbers are relative and determined from a model disk, making the normalization irrelevant.  The areas covered by the defined regions are 770, 481  349, 113, 106, and 31 arcmin$^2$ from lowest to highest stellar density, respectively.  White
  squares mark the centers of the fields where artificial star tests
  were performed to probe a range of stellar densities.  We divided
  our results in this way to account for the changing completeness
  limits with stellar density.}
\label{density_map}
\end{figure}

\begin{figure}
\includegraphics[width=6.5in]{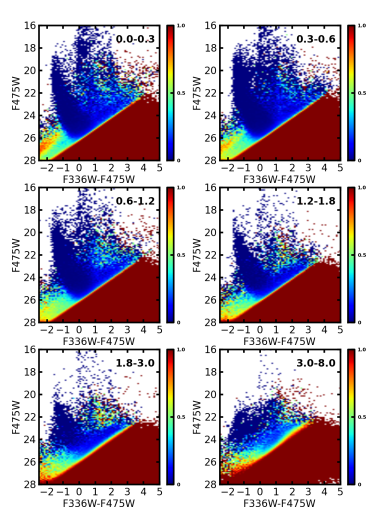}
\caption{Fraction of F336W or F475W measurements flagged as not
passing our quality cuts as a function of color and magnitude.}
\label{uvoptical_bad_cmds}
\end{figure}

\begin{figure}
\includegraphics[width=6.5in]{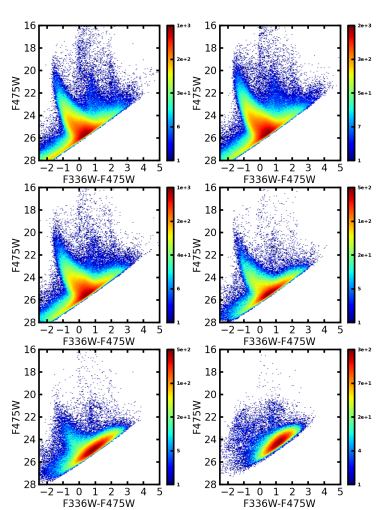}
\caption{F336W-F475W CMDs of the 6 different regions defined by
Figure~\ref{density_map}, showing only measurements that pass our
quality cuts in both bands.}
\label{uvoptical_gst_cmds}
\end{figure}

\begin{figure}
\includegraphics[width=6.5in]{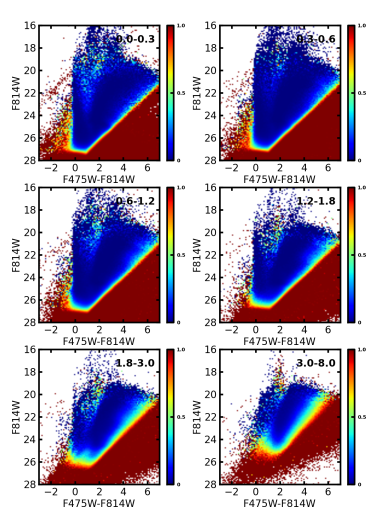}
\caption{F475W-F814W at different stellar densities of the fraction of stars passing our cuts, as in Figure~\ref{uvoptical_bad_cmds}.}
\label{optical_bad_cmds}
\end{figure}

\begin{figure}
\includegraphics[width=6.5in]{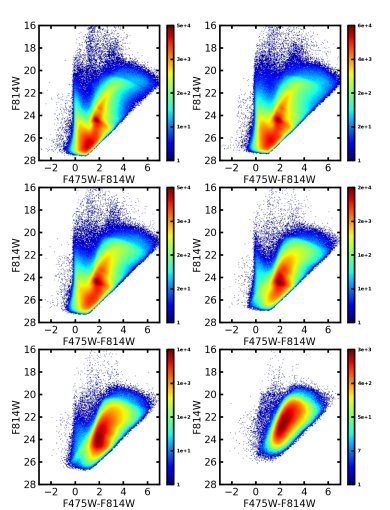}
\caption{F475W-F814W CMDs at different stellar densities after quality cuts, as in Figure~\ref{uvoptical_gst_cmds}.}
\label{optical_gst_cmds}
\end{figure}

\begin{figure}
\includegraphics[width=6.5in]{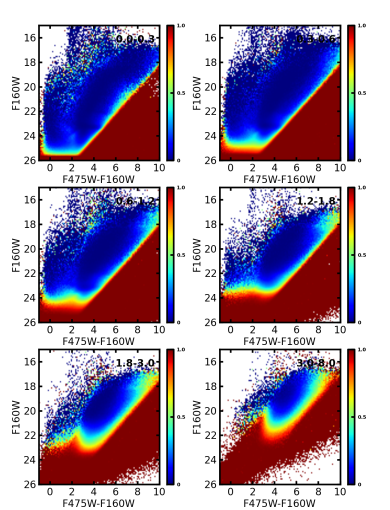}
\caption{F475W-F160W CMDs at different stellar densities of the fraction of stars passing our cuts, as in Figure~\ref{uvoptical_bad_cmds}.}
\label{optir_bad_cmds}
\end{figure}

\begin{figure}
\includegraphics[width=6.5in]{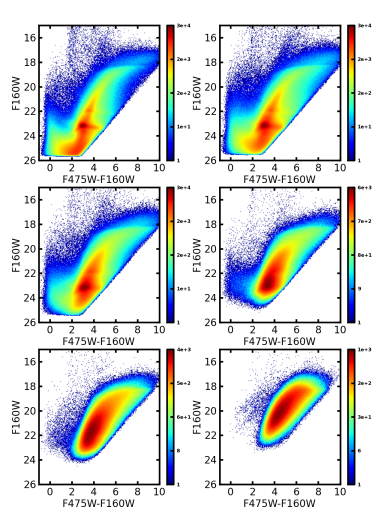}
\caption{F475W-F160W CMDs at different stellar densities after quality cuts, as in Figure~\ref{uvoptical_gst_cmds}.}
\label{optir_gst_cmds}
\end{figure}

\begin{figure}
\includegraphics[width=6.5in]{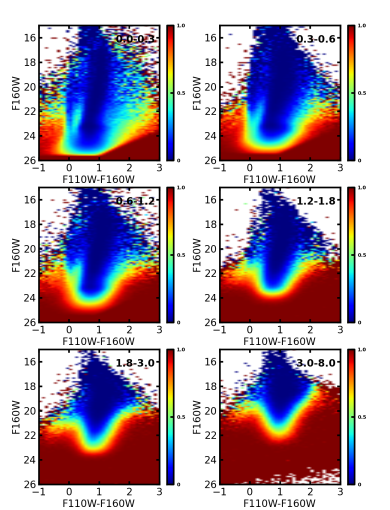}
\caption{F110W-F160W CMDs at different stellar densities of the fraction of stars passing our cuts, as in Figure~\ref{uvoptical_bad_cmds}.}
\label{ir_bad_cmds}
\end{figure}

\begin{figure}
\includegraphics[width=6.5in]{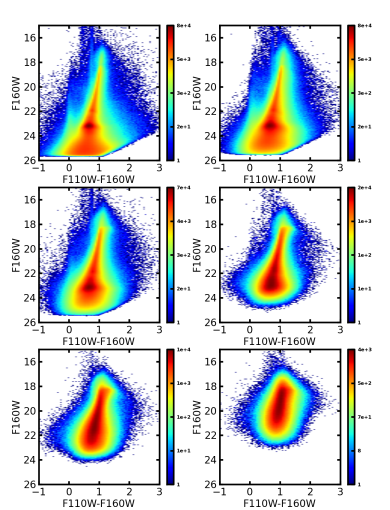}
\caption{F110W-F160W CMDs at different stellar densities after quality cuts, as in Figure~\ref{uvoptical_gst_cmds}.}
\label{ir_gst_cmds}
\end{figure}

Because of the high density of RGB stars in M31, the depth of our
optical and IR photometry varied greatly with position in the galaxy.
We therefore have produced 6 color-magnitude diagrams in each of 4
filter combinations, following the spatial distribution shown in
Figure~\ref{density_map}.  These densities are relative logarithmic
star number densities taken from a model distribution.  They were used
only to provide a reasonable division of the catalog into regions with
similar photometric depths.  The areal coverage of each stellar
density range is different, with 770, 481, 349, 113, 106, and 31
arcmin$^2$ for the 0.0--0.3, 0.3--0.6, 0.6--1.2, 1.2--1.8, 1.8--3.0
and 3.0--8.0 regions, respectively. In
figures~\ref{uvoptical_bad_cmds}-\ref{ir_gst_cmds}, we show the
fraction of measurements that pass our quality cuts, and the CMDs
after quality cuts in each of these filter combinations at each of
these densities.

Apart from the familiar features like the RGB and main sequence, these
CMDs reveal several detailed features that have been already commented
on in \citet[][Figure 20]{dalcanton2012}, but which appear somewhat
more clearly in the present photometry, such as the sharp drop in star
counts at the tip of the RGB at F814W$\sim$20 or F160W$\sim$18, and
spanning a broad range of colors, the red clump at F814W$\sim$24 or
F160W$\sim$23.5.  Other features appear that are associated with the
red clump, such as the extension to red colors caused by differential
extinction, the half-magnitude extension to fainter magnitudes caused
by the secondary red clump \citep{girardi1999}, and a more subtle
extension to bluer colors and fainter F814W magnitudes caused by
horizontal branch stars \citep{williams2012}.  In addition, there is a
clear concentration of stars at the early-AGB about 1.5 mag above the
red clump. Younger helium burning stars draw more feeble and
less-defined sequences towards brighter magnitudes, but their upper
part is clearly seen at F814W$<$20, F475W-F814W$\sim$3. Finally, the
UV CMDs of the bulge region (e.g., Figure~\ref{uvoptical_gst_cmds},
bottom-right panel) reveal the unusual sequences of hot horizontal
branch and post asymptotic giant branch stars and their progeny
\citep{rosenfield2012}, while the NIR CMDs reveal a plume of TP-AGB
stars departing from the TRGB towards brighter magnitudes.

\section{Milky Way and Background Galaxy Contamination}

\begin{figure}
\includegraphics[width=2.3in]{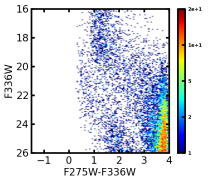}
\includegraphics[width=2.3in]{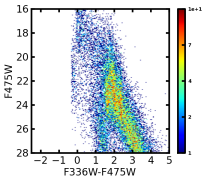}
\includegraphics[width=2.3in]{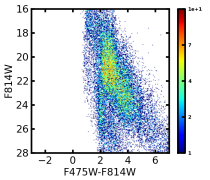}
\includegraphics[width=2.3in]{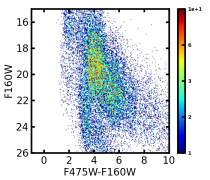}
\includegraphics[width=2.3in]{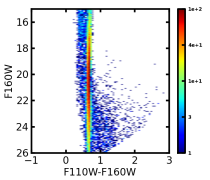}
\caption{Model predictions for the CMDs of the total foreground
  population of the 0.5~deg$^2$ PHAT survey.  The foreground stars
  draw nearly-vertical sequences at colors 1$<$F275W-F336W$<$3,
  0$<$F336W-F475W$<$2, 1$<$F475W-F814W$<$4, 1.5$<$F475W-F160W$<$5, and
  0.4$<$F110W-F160W$<$0.8.  Although the foreground stars make up
  $<$0.02\% of the catalog, these sequences are still evident in the
  observed CMDs show in Figures~\ref{uv_cmds}-\ref{ir_gst_cmds}.}
\label{fg}
\end{figure}

Figures~\ref{uv_cmds}-\ref{ir_gst_cmds} also contain contributions
from foreground MW stars. We plot the CMDs for the model foreground in
Figure~\ref{fg}.  The foreground stars draw nearly-vertical sequences
at colors 1$<$F275W-F336W$<$3, 0$<$F336W-F475W$<$2,
1$<$F475W-F814W$<$4, 1.5$<$F475W-F160W$<$5, and
0.4$<$F110W-F160W$<$0.8. These features are expected given that the
PHAT survey covers 0.5 deg$^2$. For example, the vertical feature at
0.4$<$F110W-F160W$<$0.8 corresponds to the locus of nearby dwarfs, of
masses 0.3--0.5 M$_{\odot}$.  Due to their low effective temperatures,
these stars develop marked water vapor features in their near-infrared
spectra and hence accumulate at those NIR colors
\citep[e.g.,][]{allard2000}.

Because the higher density regions cover less area and the foreground
contamination has constant density on the sky, the foreground features
are less pronounced in the higher density region CMDs.  We ran a
simulation of the \citet{girardi2005,girardi2012} model of the stellar
population of the Milky Way at the location of M31.  The model returns
$\sim$23000 stars down in the range 15$<$I$<$27 in a 0.5 deg$^2$
region.  This total is likely an upper-limit as the models are not
well constrained at the faint end, and other models \citep[such
  as][]{robin2003} predict far fewer.  Thus, the foreground
contamination in our catalog is expected to be an inconsequential
percentage ($<$0.02\%).

Although the percentage is low, the foreground stars occupy regions of
color-magnitude space that could overlap with interesting and
sparsely-populated features for M31 stars.  In general, the bright
portion of our CMDs, where the young He-burning stars and bright AGB
stars in M31 lie, also contain MW foreground.  

In addition to foreground stars, we would expect some contamination
from background galaxies; however, most of these are flagged by our
sharpness cut.  Our previous work has found a background contamination
density in the optical bands of $\sim$60 arcmin$^{-2}$
\citep{dalcanton2009}.  As our deepest bands are the same bands and
are of comparable depth to \citet{dalcanton2009}, we expect our
background galaxy contamination density to be similar.  This density
suggests $\sim$10$^5$ background galaxies in the survey after our
sharpness and crowding cuts, still $\ll$1\% of the total number of
stars.

\section{Data Quality}

We consider a number of tests to characterize photometric quality of
the data.  These include: (1) recovery of artificial stars, (2)
comparing repeated measurements of stars in overlapping data, and (3)
stability of CMD features at fixed magnitude across the survey.  We
now discuss each of these tests in turn.

\subsection{Random Uncertainties, Biases, and Completeness}

We first measured the precision and completeness of our photometry in
each band with a series of artificial star tests in each field.  To
cover the most relevant portions of the 6-band space, we produced
50,000 spectral energy distributions (SEDs) in our 6 bands covering a
grid of \citet{kurucz1979} model spectra, applying our assumed M31
distance, and applying a random extinction to each model of 0${<}{\rm
  A_V}{<}$2.  We only included model stars that had magnitudes $<$1
mag below our limiting magnitude in at least one band.  These 50,000
model SEDs were then assigned random positions within each of 6 fields
that covered the full range of red giant branch stellar densities in
our data (i.e., corresponding to the range of densities sampled in the
Hess diagrams in Figures~\ref{uvoptical_gst_cmds}-\ref{ir_gst_cmds}),
but with higher sampling where the crowding increases most
dramatically.  In each of the six chosen fields (each covering
one-eighteenth of a brick), the artificial stars were added to the
entire stack of data (all 18 ACS and WFC3 exposures).  Thus, while our
tests cover the full range of crowding levels in our data, they only
cover 1.5\% of the survey area.  These tests are computationally
expensive (2000 CPU hours to run 50,000 for one field).  Thus we
performed this efficient set of tests to provide an accurate quality
check at a range of crowding levels.

We applied photon noise to each model SED, and added a star with the
corresponding magnitudes to the images of each field using the PSF of
each image.  We then reran the photometry to determine if the
artificial star was recovered, and if it was recovered, to compare the
input and output magnitudes.  These tests were performed one star at a
time to avoid the artificial stars from affecting one another.  A star
was considered to be recovered if it was detected within two pixels of
the same position, passed our quality criteria, and had a measured
magnitude within 2 magnitudes of the input value.  A catalog of our
artificial star tests is provided on Vizier, as well as in
Table~\ref{fake-catalog}.  The RMS uncertainties and median magnitude
bias from the artificial stars, along with the corresponding DOLPHOT
reported error, and ratio of the RMS to the DOLPHOT reported error as
a function of stellar density, filter, and magnitude are provided in
Table~\ref{ast_results}.  We now describe the results in detail.

\subsubsection{Magnitude Errors and Bias}

\begin{figure}
\includegraphics[width=6.5in]{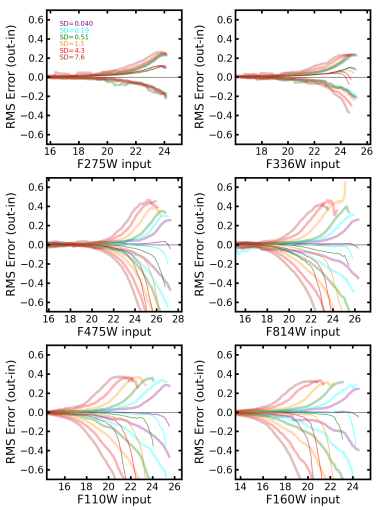}
\caption{Root mean square scatter in the positive and negative
  directions (thick) and median (thin) of artificial star tests as a
  function of (and relative to) input magnitude in all 6 bands at 5
  different stellar densities (number of stars with
  18.5$<$F160W$<$19.5 per arcsec$^2$).  Both the bias and the RMS
  scatter are very small at the bright end.  In the UV bands, the
  errors at the faint end only reach $\sim$0.2 mag, and the bias is
  faintward.  In redder bands with low crowding, the bias is near 0,
  and the errors are $\sim$0.2 mag at the faint end.  In crowded
  regions, the bias and the errors can reach $>$0.5 mag, and the bias
  is brightward.}
\label{rms}
\end{figure}

Figure~\ref{rms} shows the root mean square (RMS) scatter in positive
and negative directions (calculated separately) and the median bias of
the artificial star photometry in each band, as a function of
magnitude for six characteristic fields, color-coded by stellar
density of the region.  The magnitude biases and uncertainties are
clearly a strong function of band and stellar density.  

In the UV, the RGB stellar density has little effect on the photometry
because our UV images do not probe the RGB.  Therefore none of our UV
images suffer from significant crowding.  In all areas of the survey,
the UV photometry goes from a bias of $\sim$0 at the bright end to
$\sim$0.1 mag at the faint end, and the uncertainties go from
$\sim$0.01 mag at the bright end to $\sim$0.2 mag at the faint end.
The bias in the UV is in the direction of stars at the faint end being
recovered at {\it fainter} magnitudes than the input.  This result
clearly shows that crowding is not the cause of magnitude bias in our
UV photometry.  We attribute this bias to charge transfer efficiency
(CTE).  The CTE trails in the image cause the sky level to be slightly
over-estimated, making the brightness of the star systematically low.

All of the other bands show clear crowding effects.  In the lower
density regions that represent the bulk of our survey, the magnitude
bias is $<$0.05, and is not correlated with brightness.  As the RGB
stellar density increases, the bias becomes negative (the measurement
is brighter than the input) at brighter magnitudes and redder bands,
reaching ${\sim}-$0.5 mag at the faint end.  This crowding bias is of
most concern in the most dense portion of the M31 bulge, where it
begins to affect photometry near the TRGB in F814W.  Thus, in the most
crowded regions of the survey, the bias can cause RGB stars to be
measured brightward of the TRGB.  Such biases will be important to
consider for any study seeking to disentangle the stellar populations
of the central M31 bulge.

\begin{figure}
\includegraphics[width=6.5in]{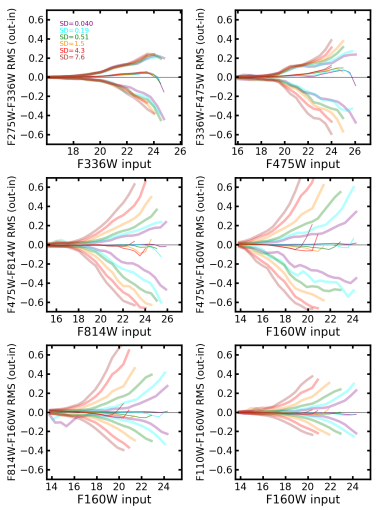}
\caption{Root mean square scatter in the positive and negative
directions (thick) and median (thin) in color as a function of
magnitude and stellar density (number of stars with
18.5$<$F160W$<$19.5 per arcsec$^2$).  Color scatter increases
substantially with stellar density, but the bias remains low, except
right at the magnitude limit, where the bias can be large and
relatively unpredictable.}
\label{color_rms}
\end{figure}

Figure~\ref{color_rms} shows the same quantities as Figure~\ref{rms}
for 6 color combinations from the UV to the IR.  These plots show that
while the scatter is larger in color than magnitude, the color bias in
our photometry is generally $<$0.1 mag outside of our faintest
detections.  This result is not surprising, since our bias is
dominated by contamination from blended stars, which will push all of
the bands to be biased brightward, resulting in less color bias than
magnitude bias.  Again, the fact that crowding is not the source of
magnitude bias in the UV is noticeable, as the color bias in the UV is
worse than in the other bands.  Thus, our artificial star tests appear
to provide a reasonable estimate of the effects of crowding on our
photometric measurements.

\begin{figure}
\includegraphics[width=6.5in]{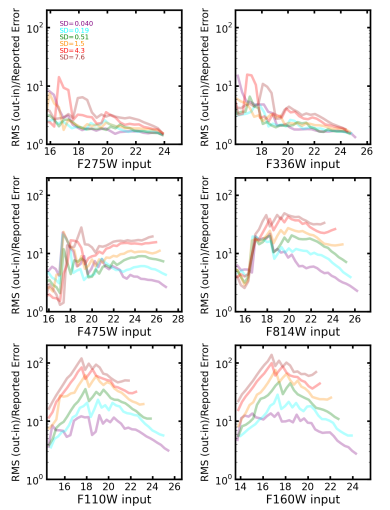}
\caption{Ratio of the RMS from artificial star tests to the
DOLPHOT-reported uncertainty as a function of magnitude for a range of
stellar densities.  Each panel gives the results for a different
filter, as labeled, and each line color represents a different stellar
density (number of stars with 18.5$<$F160W$<$19.5 per arcsec$^2$), as
labeled.}
\label{rms_err_ratio}
\end{figure}

Figure~\ref{rms_err_ratio} shows the ratio of the RMS of the
artificial star tests to the error reported by DOLPHOT (photon
statistics error) as a function of stellar density, filter, and
magnitude.  In the UV, the DOLPHOT reported errors tend to be
underestimated by factors of 2 to 10, with the worst underestimates
coming at the bright end, where photon statistics typically give very
small uncertainties that are much smaller than the dominant systematic
errors.  In the optical, the brightest stars are only measurable in
the very short exposure (they are saturated in the long exposures),
making their photon statistics poor.  At $\sim$18th magnitude, the
stars are measurable in the longer ACS exposures, and the photon
statistics yield very small errors that are underestimated by factors
of $\sim$10--20, again due to the dominance of systematic errors.  At
the faint end in the optical the factor by which the DOLPHOT reported
errors are underestimated is a strong function of crowding, with
errors in low-density regions underestimated by factors of a few and
errors in high-density regions underestimated by factors of 20--30.
In the IR, crowding causes DOLPHOT reported errors to be severely
underestimated at all stellar densities, ranging from factors of
$\sim$4--20 at the bright end, factors of $\sim$10--100 at $\sim$18th
magnitude, and factors of $\sim$3--80 at the faint end, depending on
the degree of crowding. Thus, over most of our survey, the errors are
dominated by sources of uncertainty other than photon counting, and
which are better characterized by using artificial star tests, or an
interpolation of the artificial star test results presented here.

\subsubsection{Completeness}

\begin{figure}
\centerline{\includegraphics[width=3.5in]{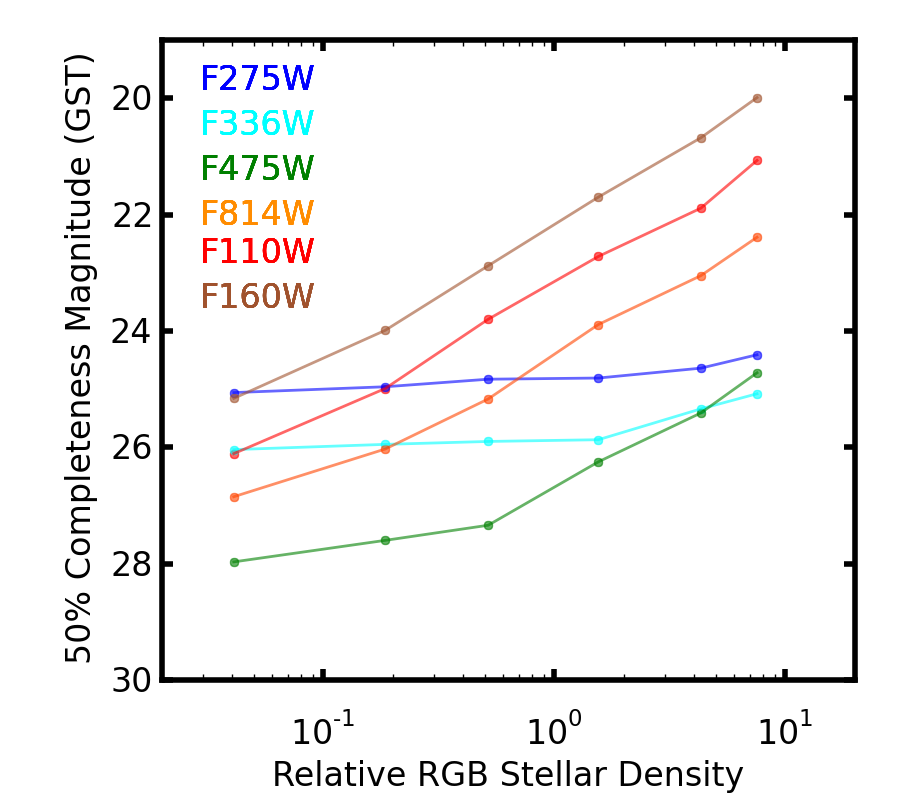}}
\caption{Completeness as a function of stellar density (number of
stars with 18.5$<$F160W$<$19.5 per arcsec$^2$) in the survey.  There
is a clear trend that the red bands are crowding-limited over much of
the survey, while the UV bands are not.}
\label{comp}
\end{figure}

Figure~\ref{comp} and Table~\ref{comp_table} provide the 50\%
completeness limit for each band as a function of the surface density
of stars with 18.5$<$F160W$<$19.5 in stars per arcsec$^2$. These
values are about 1 mag deeper in the IR bands and 1 mag deeper in
F336W compared to those performed on the individual camera data
presented in \citet{dalcanton2012}. This improvement is due to both
our DOLPHOT parameter optimization and in inclusion of data from all
cameras in the DOLPHOT runs (see Section~\ref{compare}).  Essentially,
the deep, high-resolution ACS data increases the fidelity and
precision of source positions, improving deblending and accuracy of
photometry for faint sources in the shallower data from the other
cameras.

The completeness trends for each band reveal the effects of crowding
as a function of spectral window.  In redder bands (F814W--F160W),
which are sensitive to the very numerous RGB stars in the M31 disk,
the completeness is crowding-limited (we detect stars to the point
where they are hopelessly blended with neighbors of similar
brightness), as shown by the monotonic decrease in depth as RGB
stellar density increases.  In the UV, which only contains the much
less numerous massive young stars, the completeness shows no trend
with RGB stellar density, showing that the completeness is
photon-limited (we detect stars to the signal-to-noise limit).
Finally, in the intermediate F475W band, the completeness becomes a
strong function of RGB stellar density only when the stellar density
reaches $\sim$0.5 stars with 18.5$<$F160W$<$19.5 in stars per
arcsec$^2$.

\subsection{Systematic Uncertainties}

\begin{turnpage}
\begin{figure}
\includegraphics[width=3.6in]{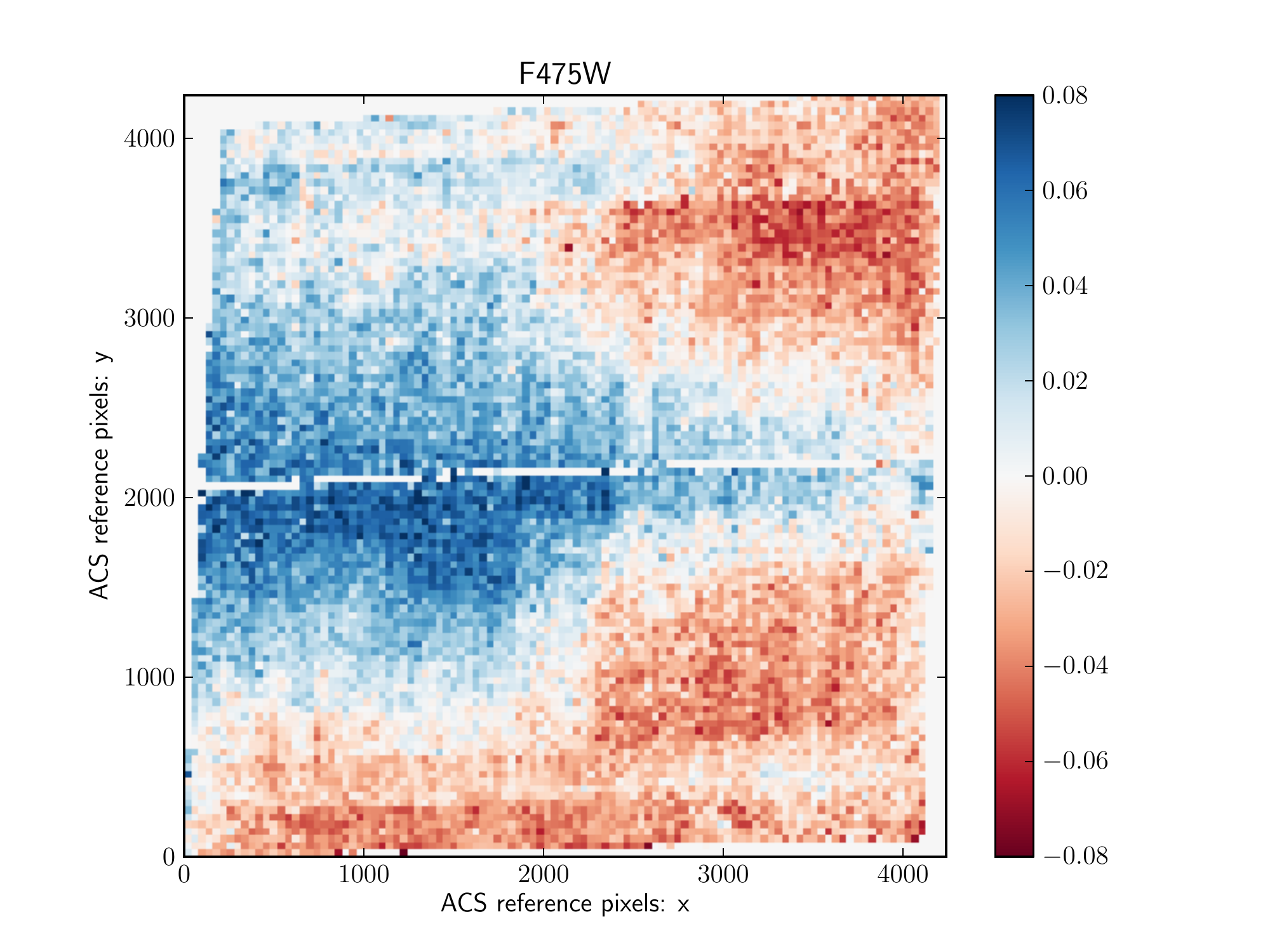}
\hspace{-0.4in}\includegraphics[width=3.6in]{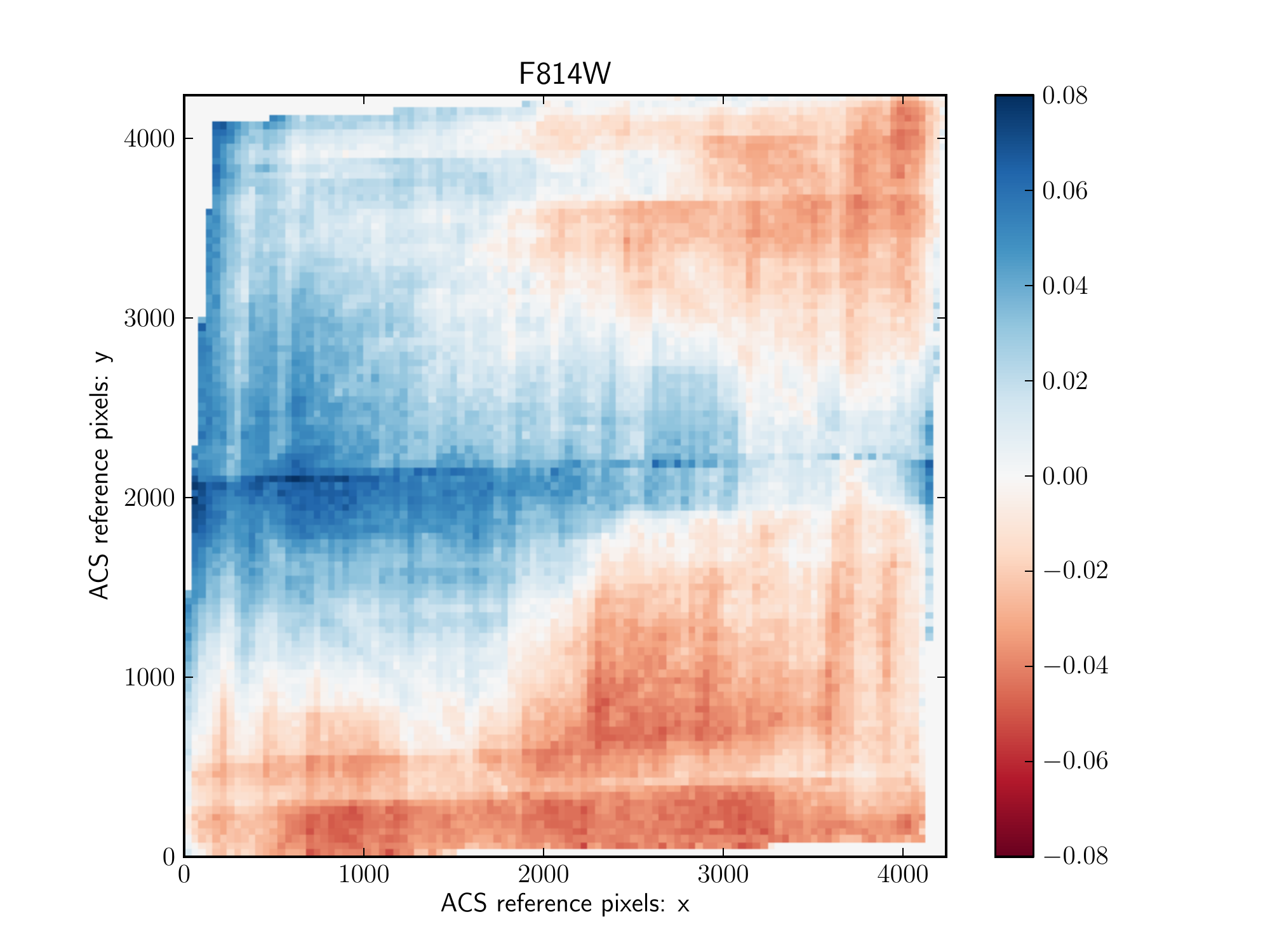}
\hspace{-0.4in}\includegraphics[width=3.6in]{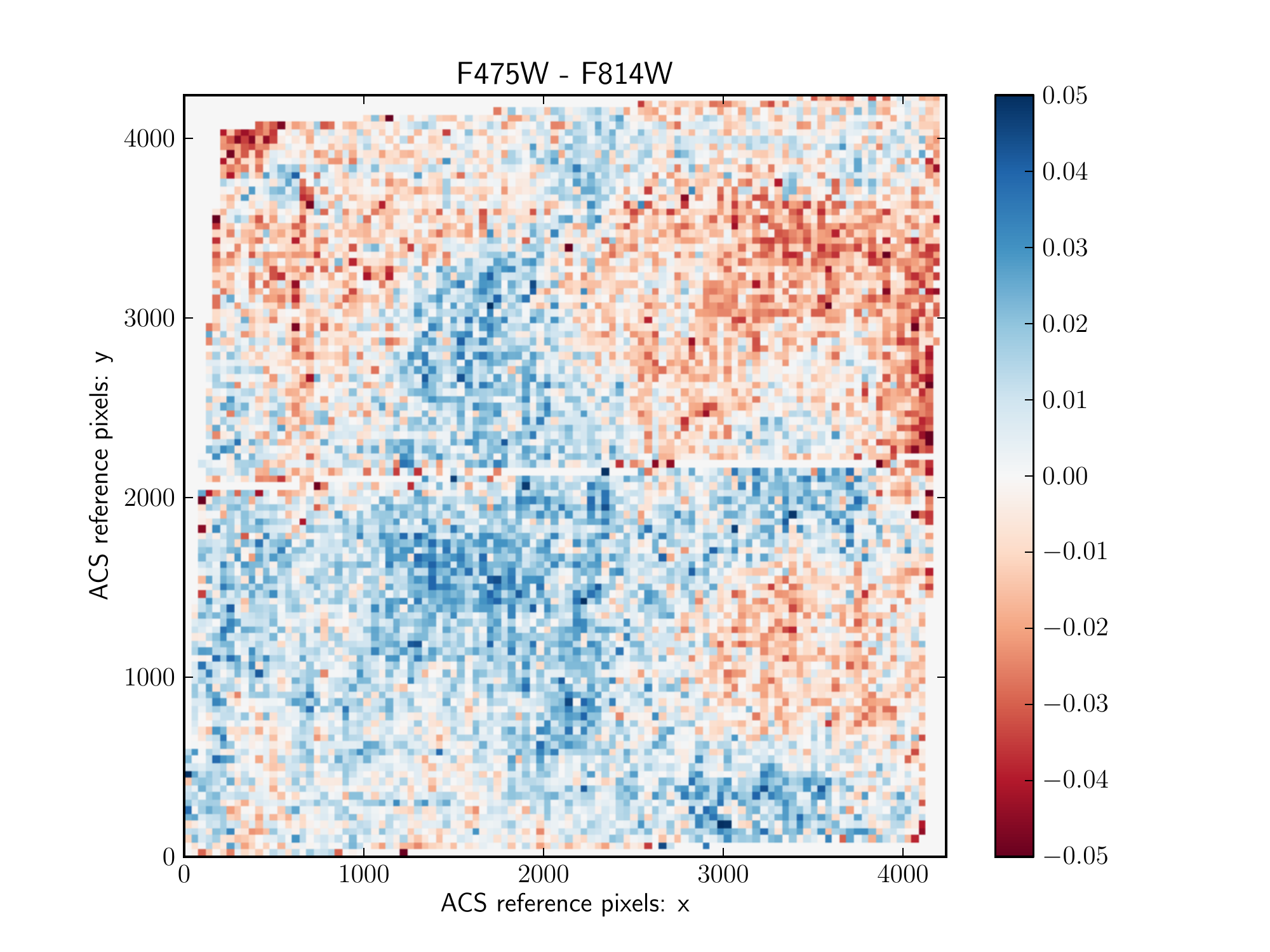}
\includegraphics[width=3.6in]{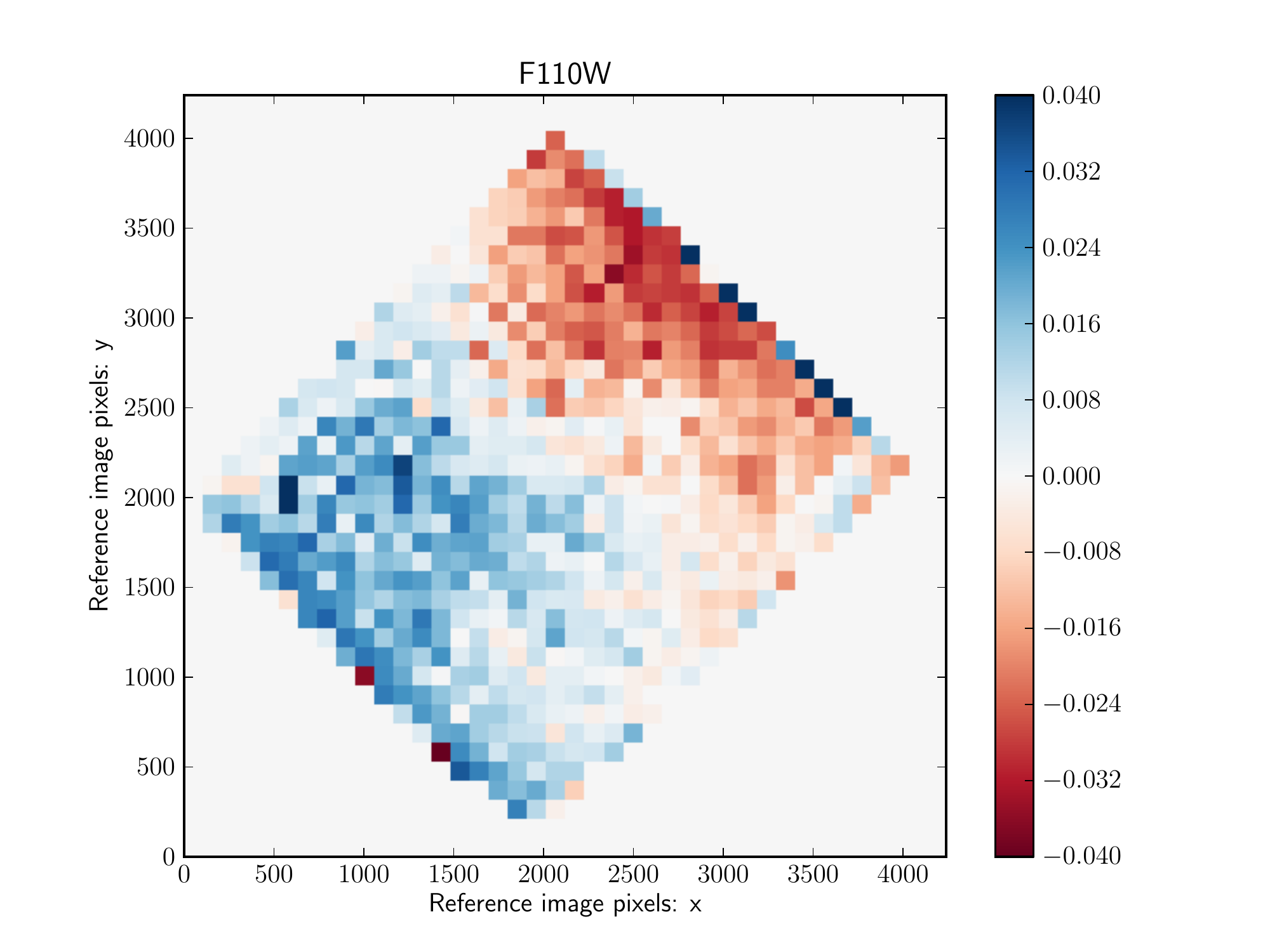}
\hspace{-0.4in}\includegraphics[width=3.6in]{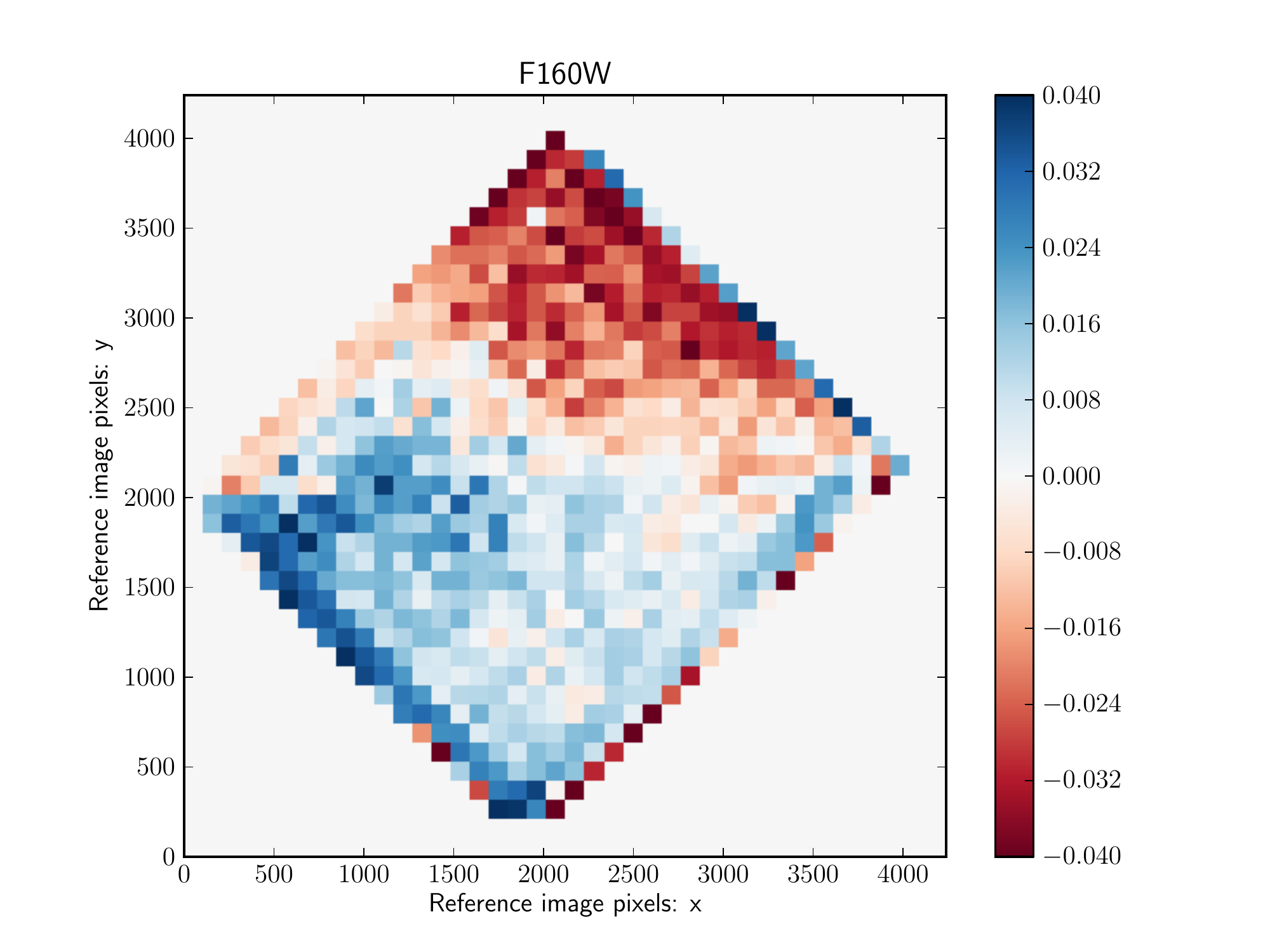}
\hspace{-0.4in}\includegraphics[width=3.6in]{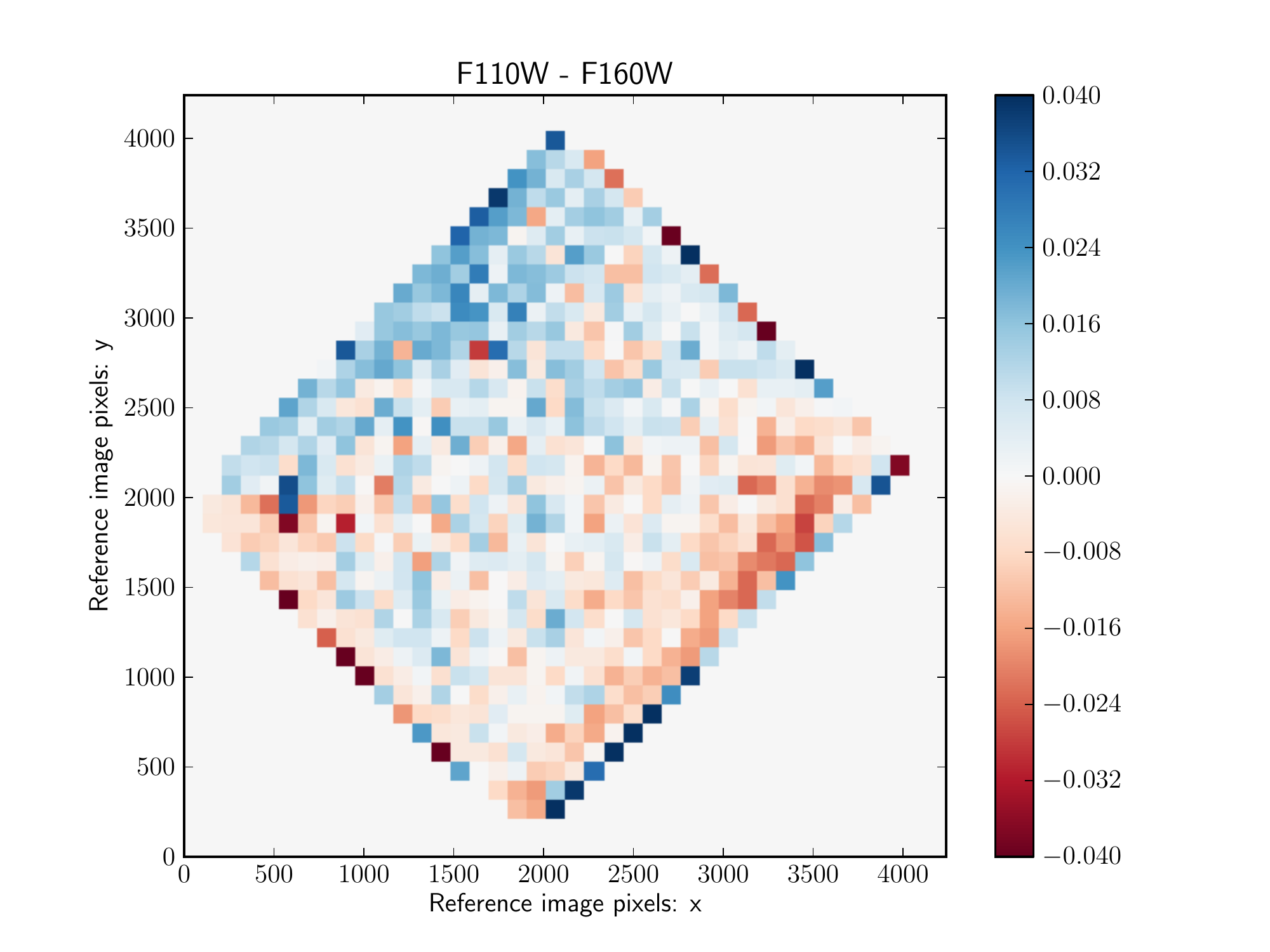}
\caption{Systematic magnitude and color variations for the same stars
  measured on different parts of the ACS/WFC camera in F475W ({\it top
    left}), F814W ({\it top middle}), and F475W-F814W ({\it top right})
  from outside of the bulge, and on different parts of WFC3/IR in
  F110W ({\it bottom left}), F160W ({\it bottom middle}), and
  F110W-F160W ({\it bottom right}) from the overlap between bricks 9 and
  11.  The systematic errors as a function of brightness are
  ${\sim}{\pm}$0.02-0.05 (see Table~\ref{systematics}).}
\label{flats}
\end{figure}

\end{turnpage}

Because of the high stellar density and our overlapping fields in ACS
(entire survey) and WFC3/IR (Brick 9 and 11 boundary), we have
multiple measurements of many stars taken at different locations on
the detectors\footnote{Although UVIS does contain overlapping
  measurements, the stellar density was not high enough to provide a
  useful check for systematic uncertainties in this way}.  To test the
consistency of our photometry, we have matched our overlapping
measurements across locations on the detectors to look for systematic
trends.  We match a large number of stars with magnitude between 22
and 24 in the relevant filter, using a matching radius of 50
milli-arcseconds.  We drop matches with large differences in magnitude
between the two measurements, via sigma clipping.  In
Figure~\ref{flats}, we show the median magnitude residual in spatial
bins in pixel coordinates for ACS and WFC3-IR.

As shown in Figure~\ref{flats}, we find patterns in photometry at RMS
levels of ${\sim}{\pm}$0.02--0.05 mag in F475W, F814W, F110W, and
F160W, depending on stellar brightness (see Table~\ref{systematics}).
These systematics are most likely due to a combination of
flat-fielding, point spread function, and charge transfer efficiency.
As the patterns are similar in F475W and F814W, the colors of stars
are only affected at the $\pm$0.02 mag level (the difference in
amplitude of the two filters).  We also note that the footprint of the
IR camera is visible in these consistency tests at a low level because
bins that contain the detector edge often contain fewer stars.  In any
case, these edges are trimmed from the photometry catalogs (see
Section~\ref{catalog_merge}) and are therefore our final photometry is
not affected by any IR edge effects.

The magnitude of the true systematic errors due to detector position
is likely exaggerated in these maps because our survey strategy
results in a constant overlap pattern.  This pattern results in the
same locations on the detector being paired many times.  Thus each
pixel is not typically being compared with many other pixels around
the detector, but with the same area on the opposite side of the
detector.  Therefore, if a star is too bright by 0.02 mag on one side,
and too faint by 0.02 mag on the other, these Figures will show an
offset of 0.04 on one side and -0.04 on the other, exaggerating the
effect by a factor of 2.  The overlaps aren't exactly the same across
the survey, but they are similar enough that this effect will be
included in the comparison.  As a result, the values for systematic
errors we report should be considered upper limits.

In addition, our systematic error map for F475W and F814W shows a grid
pattern on a scale of 256$\times$256 ACS pixels, at a level smaller
than the large scale $\pm$0.05 and $\pm$0.03 level of the systematic
errors in both bands.  This grid pattern provides a clue as to the
dominant source of our systematic errors because DOLPHOT uses a grid
of this size to define the PSF used for photometry.  In short, the
Anderson (ACS ISR 2006-01) PSFs are binned into a 16$\times$16 grid
when fitted to the data, making small errors in the the PSF models a
likely cause of the grid pattern.  

\begin{figure}
\includegraphics[width=3.4in]{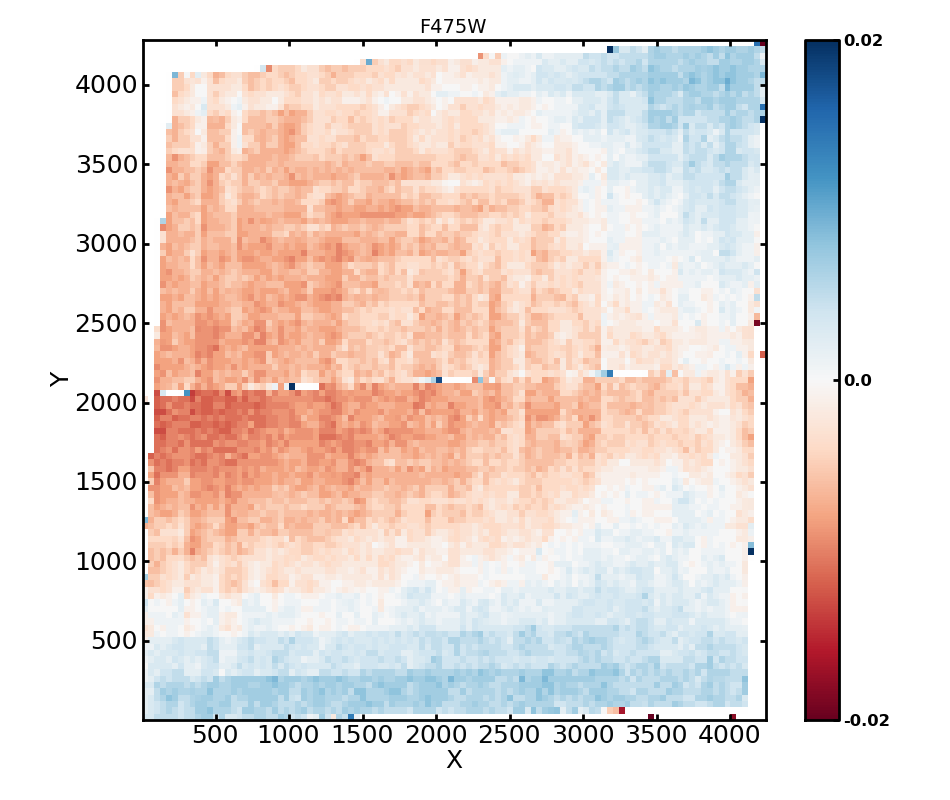}
\centerline{\hspace{-3.5in}\includegraphics[width=3.5in]{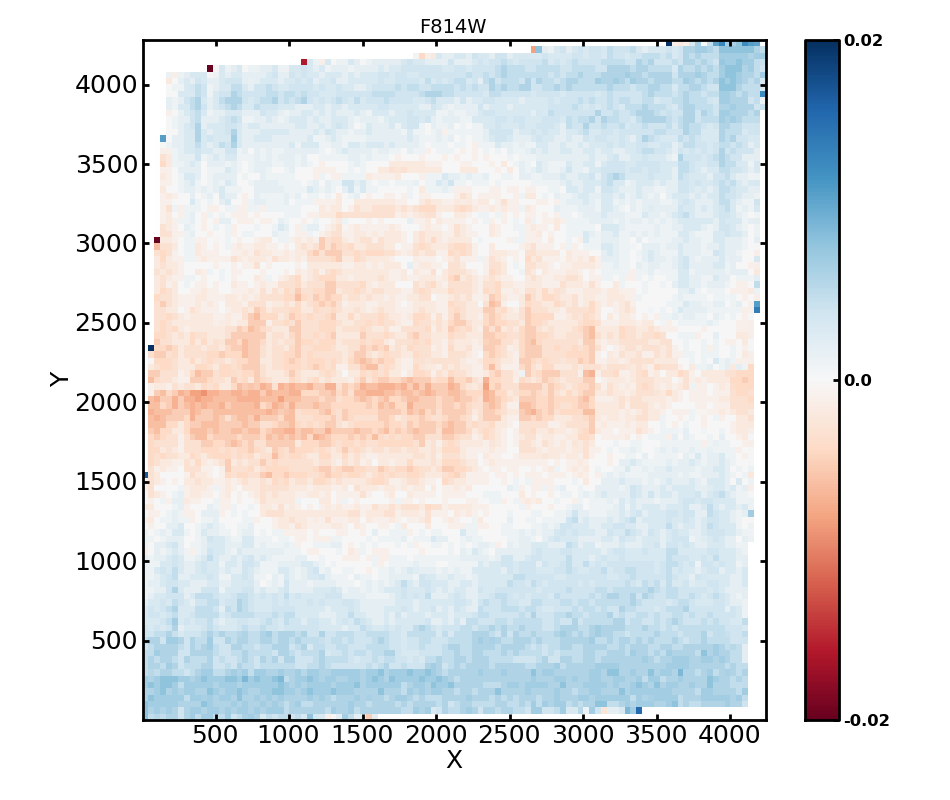}}
\includegraphics[width=3.4in]{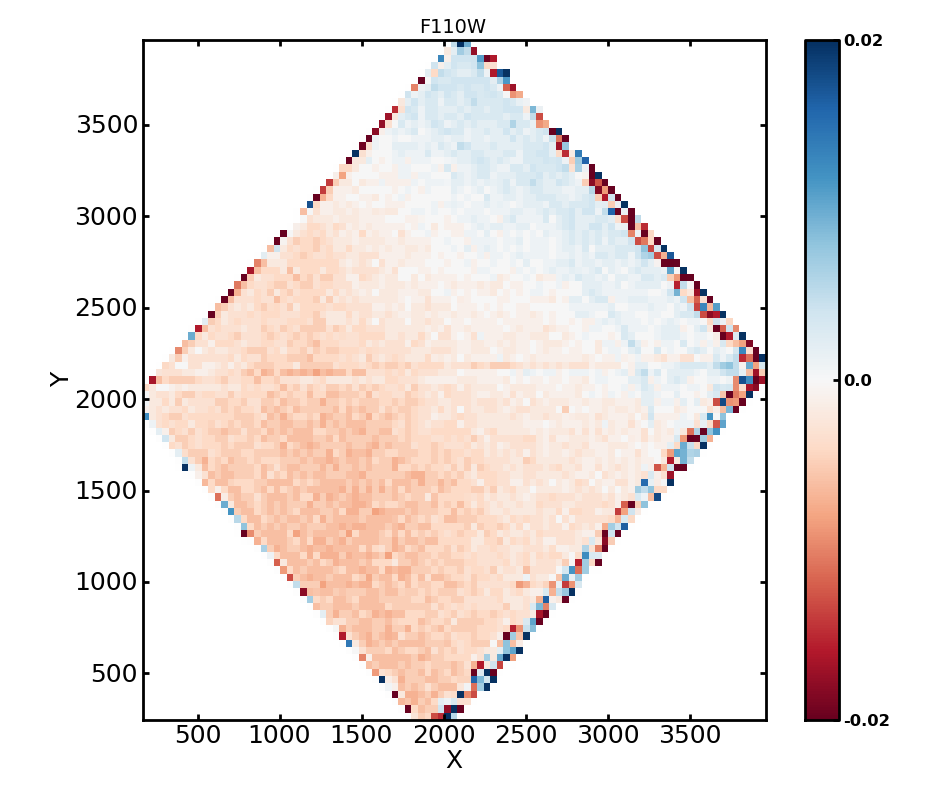}
\includegraphics[width=3.4in]{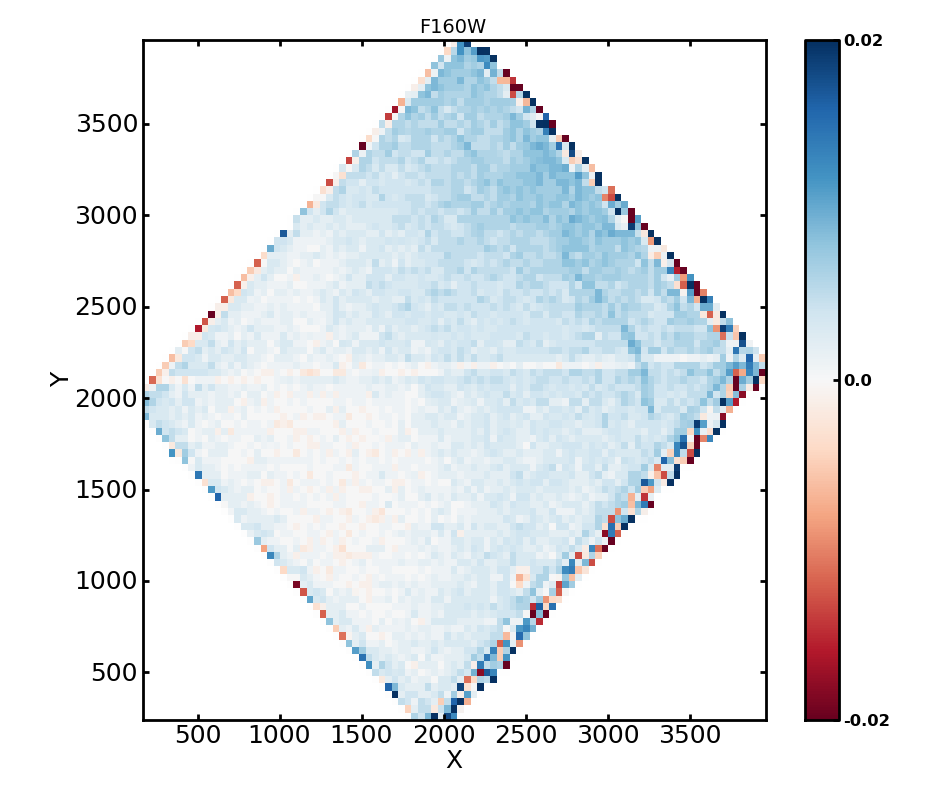}
\caption{Median sharpness values for the all stars outside of the
  bulge as a function of position on the ACS/WFC camera in F475W ({\it
    top left}) and F814W ({\it top right}), and as a function of
  postion on the WFC3/IR camera in F110W ({\it bottom left}) and F160W
  ({\it bottom right}).  The systematic patterns are the same as those
  seen in magnitude, suggesting the PSF library as the dominant source
  of systematic errors.}
\label{sharp}
\end{figure}

To further investigate the PSFs as the dominant source of systematic
error, we produced maps of our median sharpness values as a function
of instrumental position. Negative sharpness means that the PSF
underestimates the peak of the light distribution, causing an
underestimate of the flux, which corresponds to a systematically high
magnitude. The maps for F475W and F814W are shown in
Figure~\ref{sharp}, and clearly show not only the same grid pattern
seen in the systematic magnitude errors, but also the large scale
pattern seen in the systematic magnitude errors.  Furthermore, the
regions with systematically negative sharpness values correspond to
areas with systematically positive magnitude errors, in agreement with
expectation if the PSF is responsible for the errors.  The amplitude
of the effect is larger in F475W than in F814W, also in agreement with
the magnitude errors.  Future DOLPHOT versions will not bin the PSF
models as coarsely, which should eliminate the grid pattern.  However,
only improved PSF models will reduce the larger scale variations.

\begin{figure}
\includegraphics[width=3.4in]{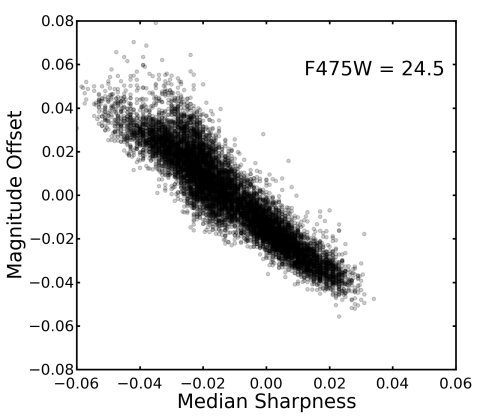}
\centerline{\hspace{-3.5in}\includegraphics[width=3.5in]{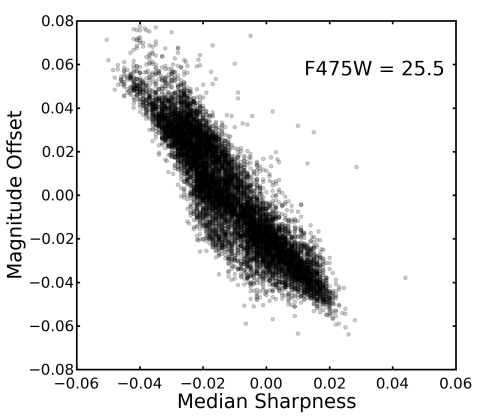}}
\includegraphics[width=3.4in]{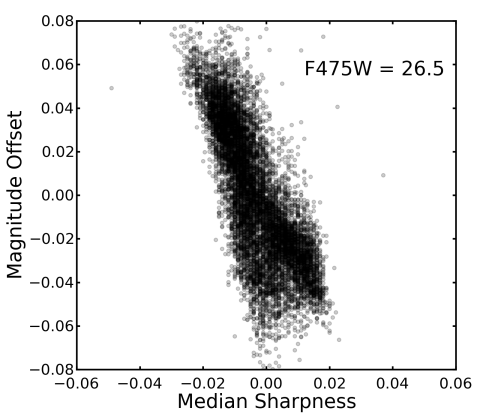}
\includegraphics[width=3.4in]{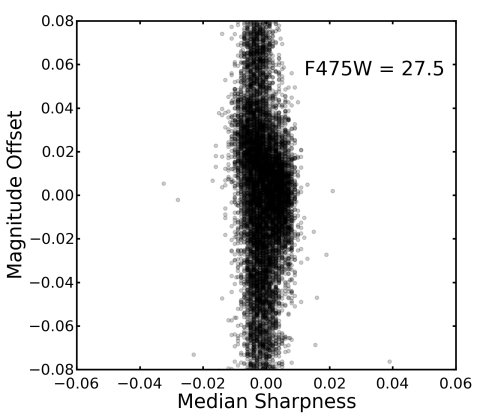}
\caption{Median F475W sharpness values vs. F475W magnitude offset at
  each location on the ACS detector in several magnitude bins. {\it
    Upper-left:} The relation at F475W=24.5, showing a strong
  anti-correlation. {\it Upper-right:} The relation at F475W=25.5,
  showing the correlation begin to steepen as the range of median
  sharpness drops with increasing magnitude.  {\it Lower-left:} The
  relation at F475W=26.5, showing a continued steepening as the
  sharpness range continues to drop.  {\it Lower-right:} The relation
  at F475W=27.5, where the sharpness measurements no longer have the
  necessary precision to detect the correlation.}
\label{sharp_v_dmag}
\end{figure}

Systematic uncertainties related to the model PSF are expected to be
magnitude-dependent.  Unlike aperture photometry, PSF-fitting
photometry applies optimal position-based weighting to the data
contained in each pixel.  Weights are higher at a star's center, so
the star will appear fainter if the model PSF is sharper than the
actual PSF (and vice versa).  We investigated the magnitude dependence
by producing $\Delta$-mag detector maps like those in
Figure~\ref{flats} for a series of magnitude bins.  The standard
deviation of these maps in each filter are provided in
Table~\ref{systematics}, and provide a quantitative measure of the
level of systematic errors in our catalogs.  We also made detector
maps of sharpness (like Figure~\ref{sharp}) in the same magnitude
bins, and we report the standard deviation of those maps in
Table~\ref{systematics} as well.  In Figure~\ref{sharp_v_dmag}, we
plot the median sharpness value vs. the measured magnitude offset in
each location on the detector in four bins of magnitude.  At bright
magnitudes (top row), the magnitude difference correlates strongly
with sharpness, as expected for magnitude errors driven by the PSF
model.  At faint magnitudes, the magnitude difference and sharpness
are uncorrelated suggesting that something other than the PSF is
causing the magnitude offsets at the faint end.

\begin{figure}
\includegraphics[width=3.5in]{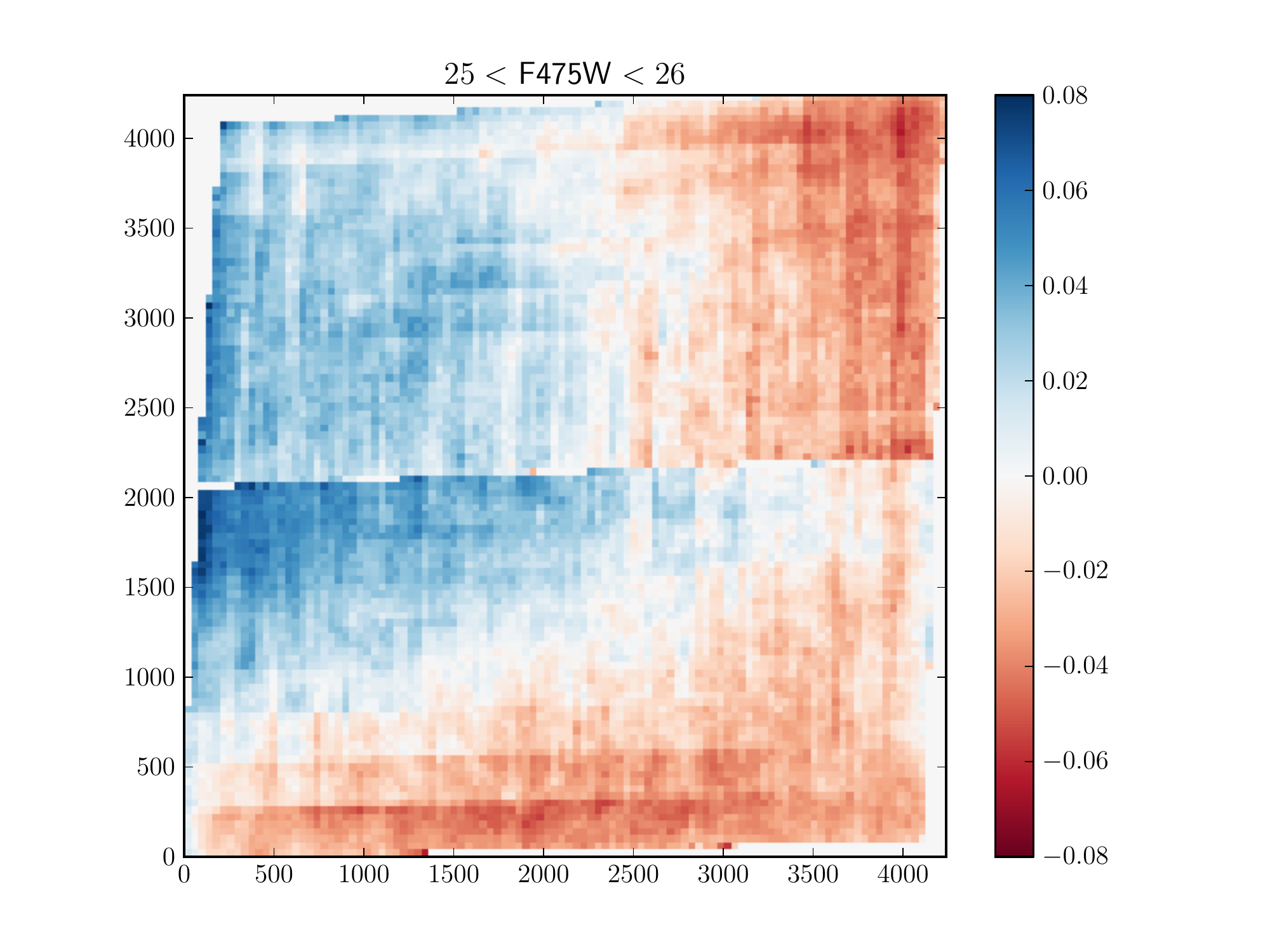}
\centerline{\hspace{-3.5in}\includegraphics[width=3.5in]{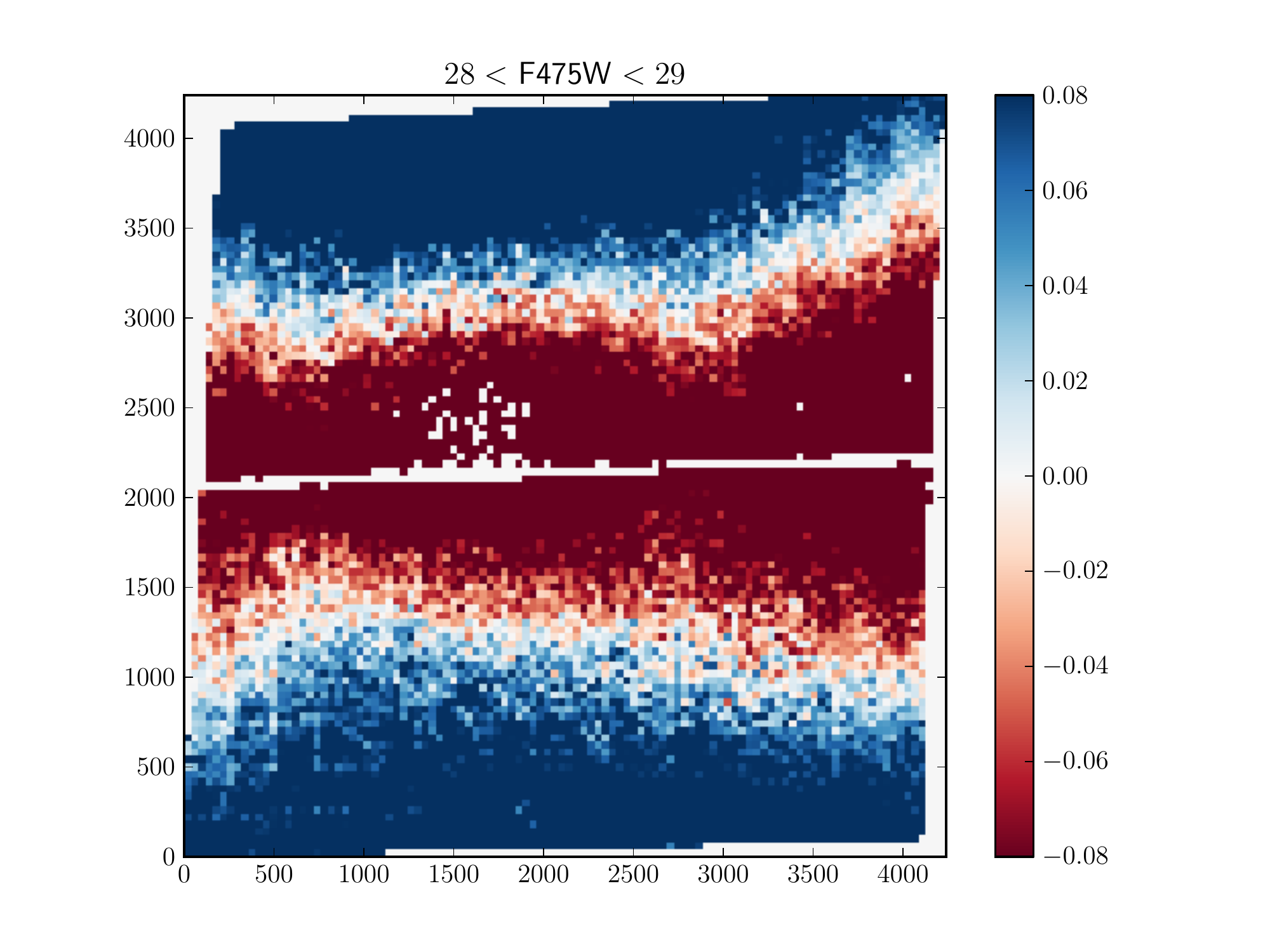}}
\includegraphics[width=3.5in]{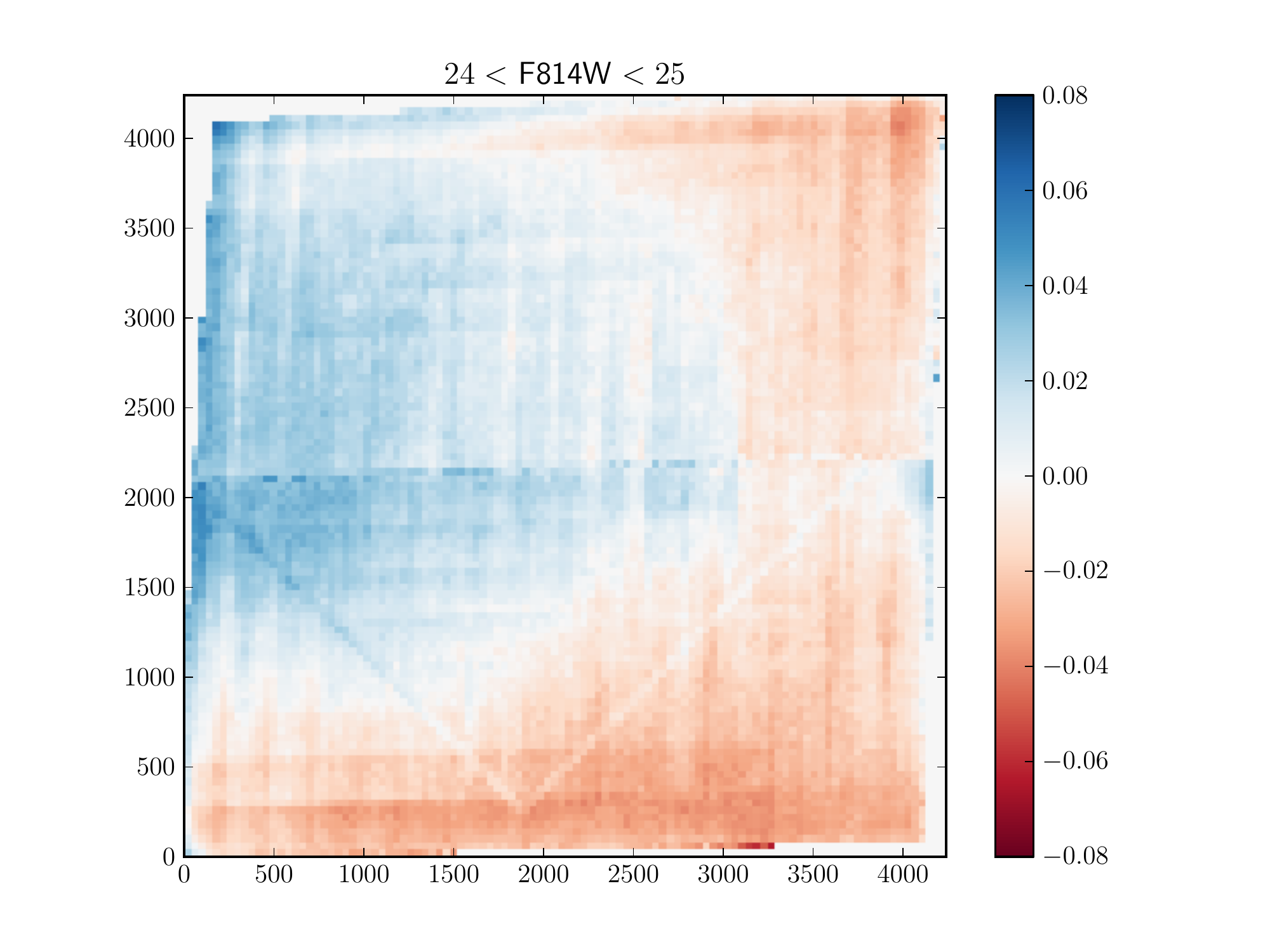}
\includegraphics[width=3.5in]{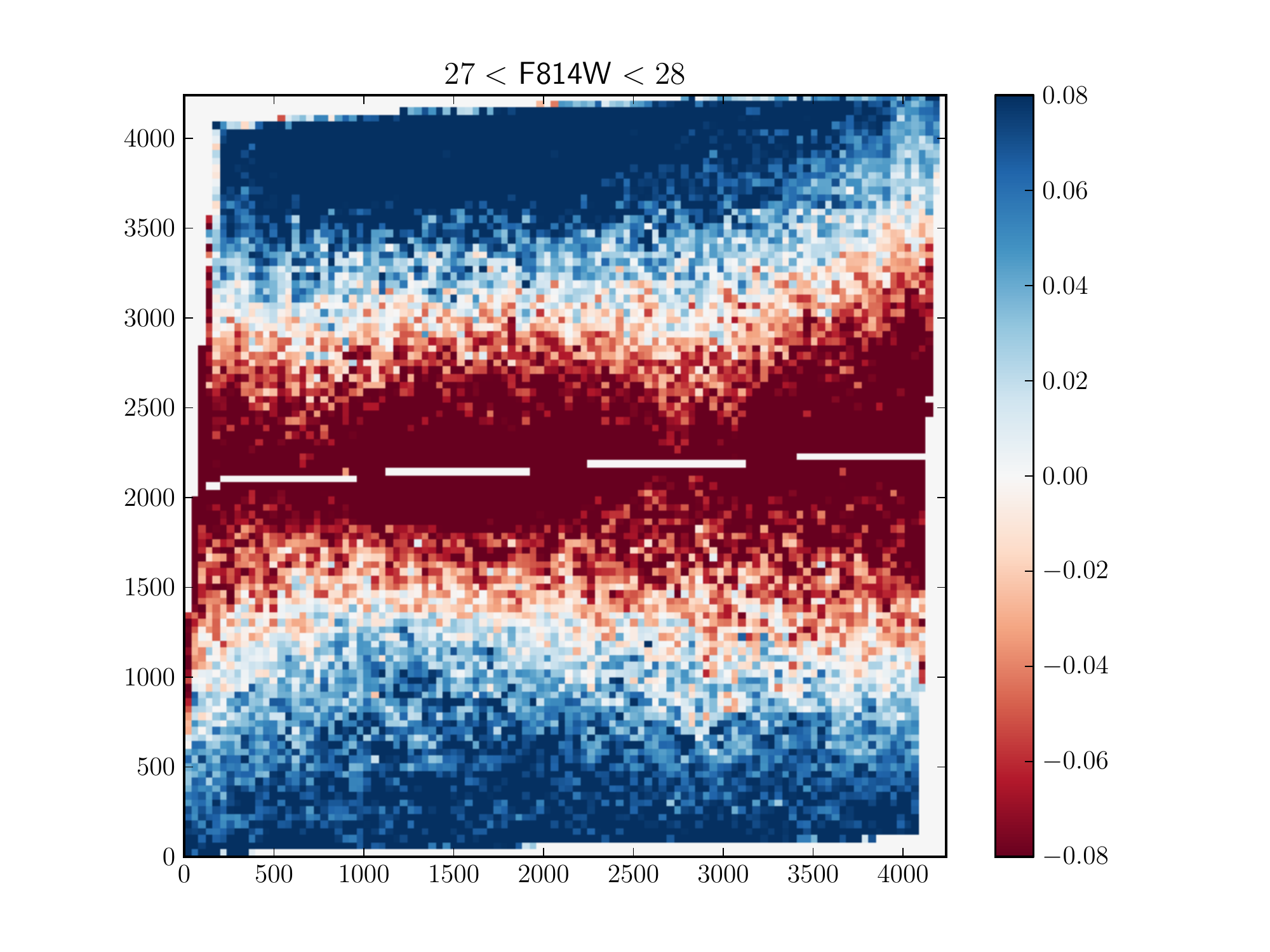}
\caption{Systematic magnitude variations for the same stars measured
  on different parts of the ACS/WFC camera in F475W ({\it top row}) and
  F814W ({\it bottom row}). The pattern at the for brighter stars
  ({\it left side}) shows a grid pattern and correlation with
  sharpness indicative of PSF model origin,
  while the pattern for faint stars ({\it right side}) is symmetric
  about the chip gap, indicative of a CTE-correction origin.}
\label{faint_dmag}
\end{figure}

We explore the faint end systematic differences in magnitude in
Figure~\ref{faint_dmag}, generating the same map of magnitude offset
vs position as in Figure~\ref{flats}, but for faint stars ($m$>28)
alone.  These maps show that at that the systematic magnitude
differences at the faint end are symmetric about the chip gap, and
completely different from those at the bright end. Furthermore, the
amplitude of the magnitude offset variations increases as stellar
brightness decreases, while the amplitude of the sharpness variations
decreases as stellar brightness decreases.  These patterns suggest
that the magnitude offsets at the faint end are dominated not by PSF
model problems but instead by CTE effects.  If so, the systematic
errors at the faint end may improve significantly in the next
generation of photometry, when CTE-corrected images are employed
instead of {\it post~facto} CTE corrections.

In Figures~\ref{flats}--\ref{sharp_v_dmag}, it is clear that the PSF
library alone is not to blame for all off the systematic errors at the
bright end.  The 16$\times$16 grid in the ACS maps is clearly due to
the PSF binning used by DOLPHOT; however, some of the large scale
patterns are slightly different, causing the scatter in the
anti-correlation between sharpness and magnitude offset.  For example,
the upper-right corner of the F475W sharpness map shows a somewhat
different pattern than that of the corresponding $\Delta-$magnitude
map, and the upper-left corner of the F814W sharpness map is different
than the corresponding $\Delta-$magnitude map.  

Finally, there is no grid pattern and little, if any, correlation
between sharpness and magnitude offset in the F110W and F160W,
suggesting some other source for the systematic errors in the
IR. However, the PSF model used by DOLPHOT in the IR contains no
variations with position on the detector, and perhaps a
spatially-varying PSF could improve the systematics there.  In
addition, some of these less severe magnitude offsets may be due to
flat-fielding \citep{dalcanton2012}.

\subsection{CMD Feature Consistency}

\begin{figure}
\includegraphics[width=3.3in]{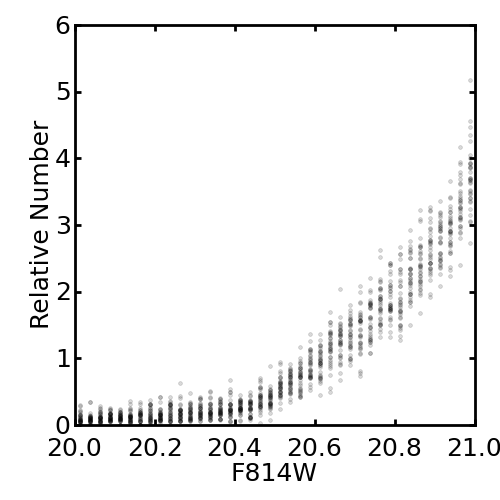}
\includegraphics[width=3.3in]{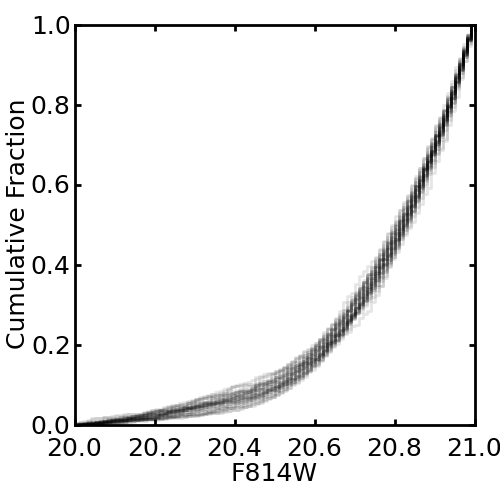}
\caption{The F814W magnitude (left) and cumulative magnitude (right)
distributions around the tip of the red giant branch for 34 $6' \times 6'$
regions of the survey.  Note the tight consistency across the entire
PHAT survey.}
\label{trgb}
\end{figure}

Beyond cross-checking multiple measurements of individual stars, we
can also check the consistency of features in the color-magnitude
diagrams.  In Figure~\ref{trgb}, we overplot the F814W luminosity
function (left) and cumulative luminosity function (right) in a color
slice of 3$<$F475W-F814W$<$3.5 for 34 $6'{\times}6'$ regions of the
survey.  The histograms have all been normalized to have the same
integral over the plotted range.  The luminosity functions of all of
these regions agree, showing that the photometry is exceptionally
consistent across the survey.  From these histograms, there is a clear
change in the slope of the luminosity function just brightward of
F814W$\sim$20.5.  

\begin{figure}
\includegraphics[width=3.3in]{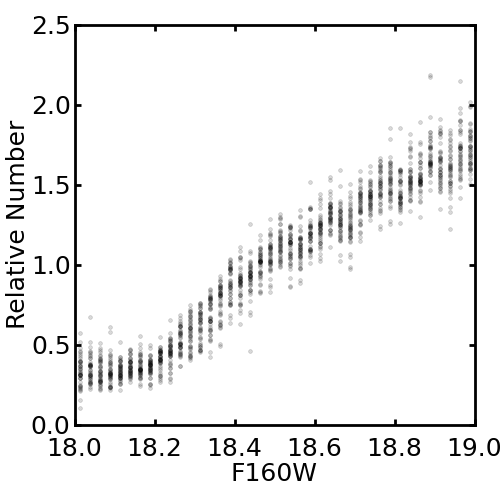}
\includegraphics[width=3.3in]{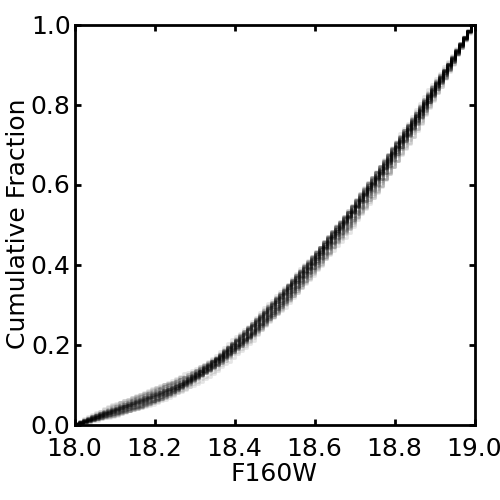}
\caption{The F160W magnitude (left) and cumulative magnitude (right)
distributions around the tip of the red giant branch for 40 $6' \times 6'$
regions of the survey.  Note the tight consistency across the entire
PHAT survey.}
\label{trgb_ir}
\end{figure}

To check the consistency of the IR photometry across the survey, in
Figure~\ref{trgb_ir} we show histograms in the 0.9$<$F110W-F160W$<$1.2
color interval, looking at the F160W luminosity function for 40 $6'
\times 6'$ regions of the survey.  We find the same exquisite
agreement as in the optical, but with a slope change just fainter than
F160W$\sim$18.2.  Determining the precise TRGB magnitude requires
sophisticated modeling of the reddening across the survey, so that
looking at these histograms is likely accurate to only $\pm$0.1 mag.
However, we note that this magnitude for the TRGB, which appears
consistently across our survey, is consistent with the distance
modulus we assume for M31 \citep[24.47;][]{mcconnachie2005} and the
typical M31 foreground extinction of $A_V=0.17$ \citep{schlafly2011}.

\begin{figure}
\includegraphics[width=3.3in]{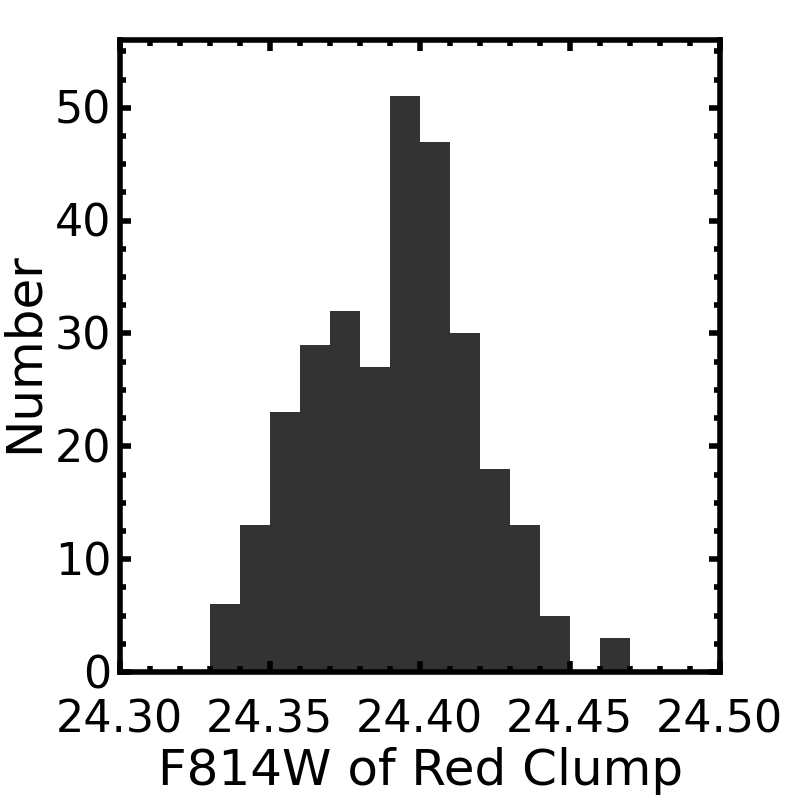}
\caption{Histogram of the F814W magnitude of the center of the red
clump feature for 298 $3' \times 3'$ regions of the survey.  They are all
consistent with one another to $\pm$0.05 mag, and the distribution is
quite symmetric, suggesting that any biases in our photometry are
below this level of precision at this magnitude.}
\label{rc}
\end{figure}

In addition to looking at the TRGB feature, we measured the F814W
magnitude of the red clump across the survey.  This feature is
sufficiently well-populated to obtain a precise measure of its peak.
In Figure~\ref{rc}, we show a histogram of the F814W values of the
peak magnitudes in the color range 1.5${<}F475W{<}$2.0 for 298 $3'
\times 3'$ regions in our survey, which corresponds to the blue end,
or least reddened, portion of the red clump.  The values agree to
within $\pm$0.05 mag, even at this much fainter level in the
photometry, slightly larger than the systematics measured by comparing
multiple F814W measurements of individual stars ($\pm$0.03, see
Figure~\ref{flats}).  Some of the additional scatter is likely
attributable to real changes in the stellar populations, since the red
clump does change somewhat in brightness with metallicity, age, and He
content \citep[e.g.,][]{cole1998,girardi2001}.  A detailed study of
the red clump and AGB bump will be presented in a future paper (Byler
et al., in prep.)  However, even if this entire scatter is due to
photometric errors, our catalog is remarkably homogeneous and clearly
equidistant.

\section{Comparison with First Generation Photometry}\label{compare}

We found that optimizing the DOLPHOT parameters and including data
from all 6 bands significantly affected the resulting photometry
catalogs. The 6-band catalogs have improved matching between bands
over any matched photometry from the individual cameras.  In addition,
the process provides a measurement of every source in every band, so
that upper limits can be used for bands where the source was not
detected at sufficient signal to noise.  Furthermore, completeness of
any combination of filters can be measured, whereas previously there
was no way to assess the completeness for colors that included data
from 2 cameras (such as F336W-F475W or F814W-F160W).

\begin{figure}
\includegraphics[width=2.5in]{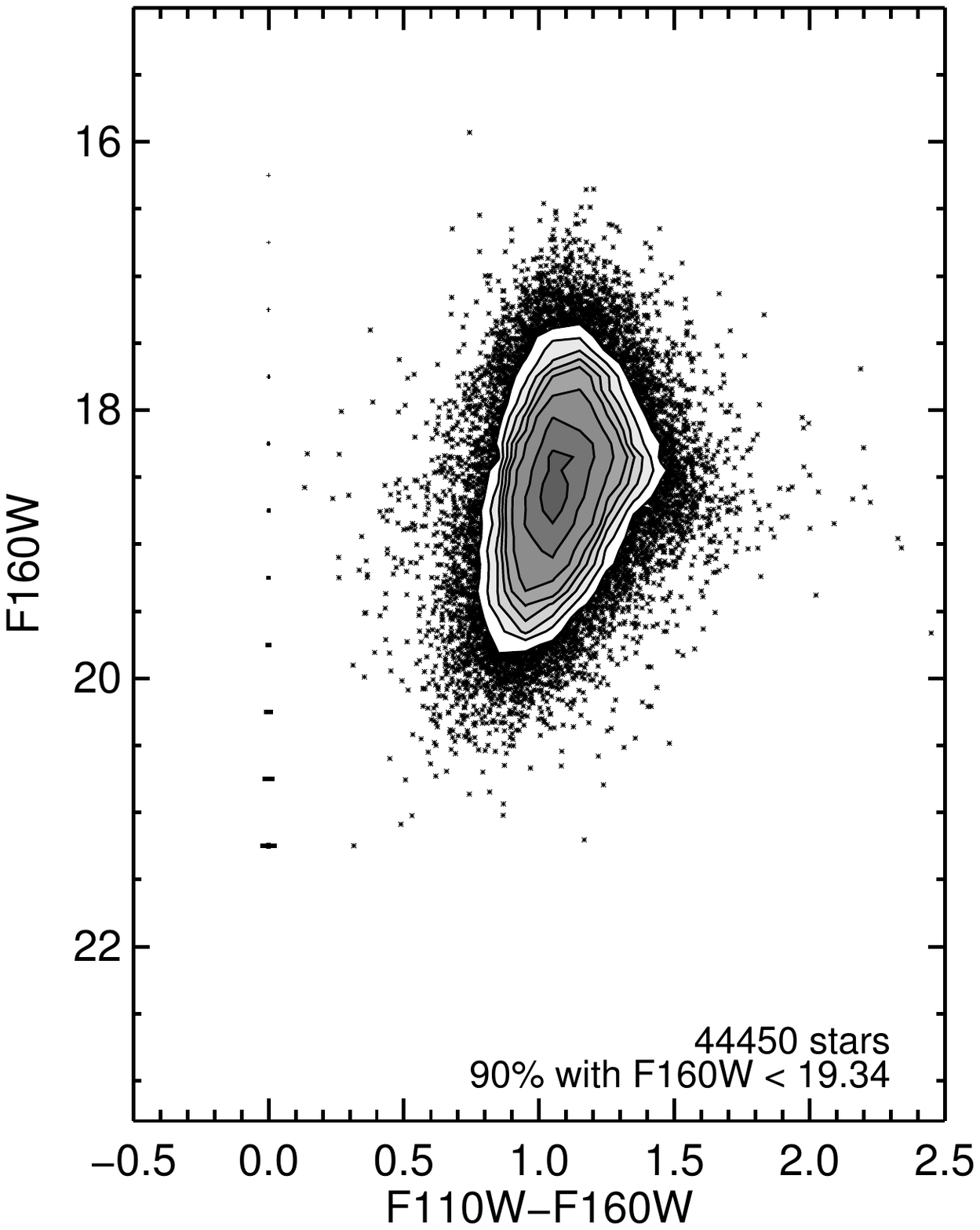}
\hspace{-0.8in}\includegraphics[width=2.5in]{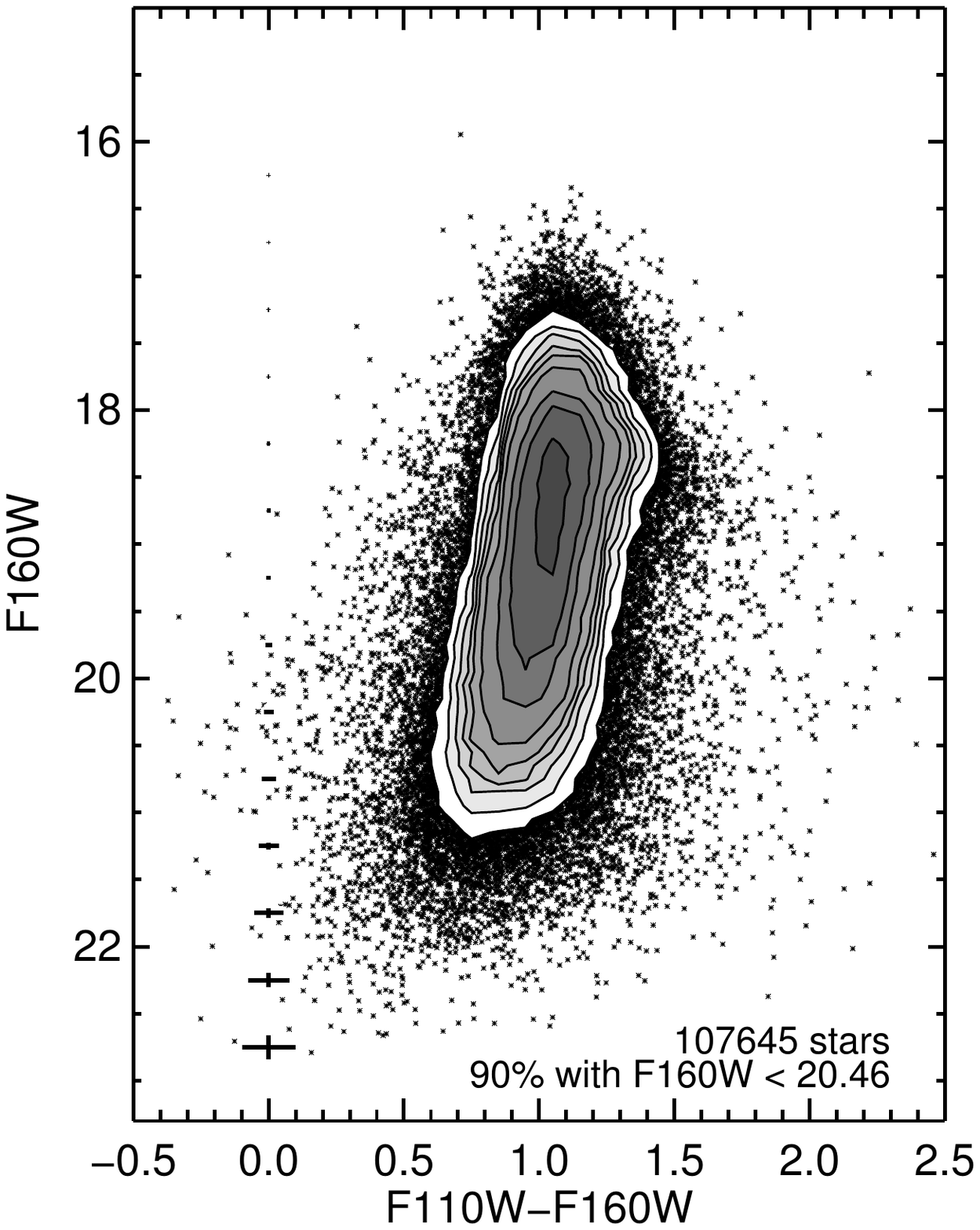}
\hspace{-0.8in}\includegraphics[width=2.5in]{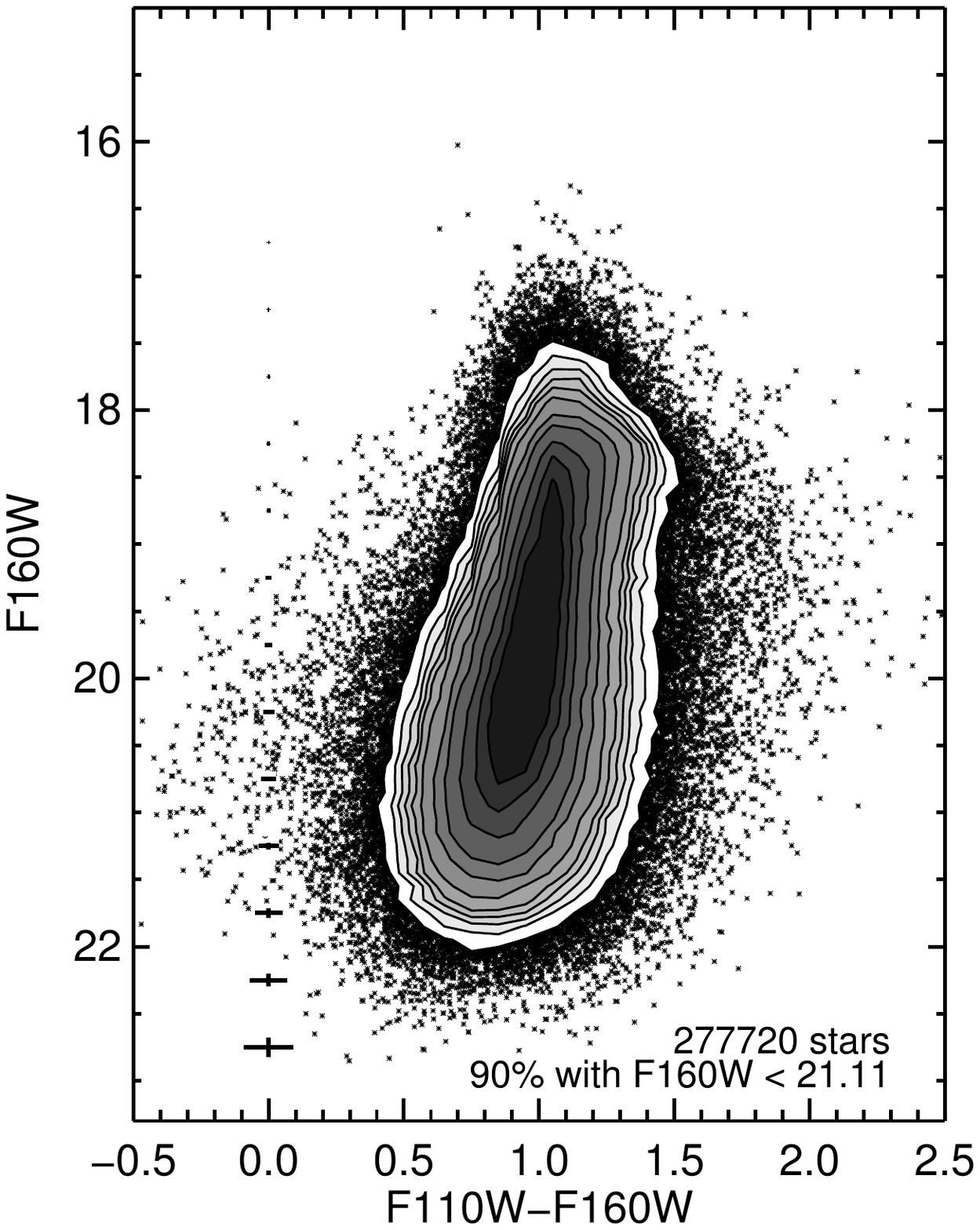}
\caption{{\it Left:} CMD from IR photometry of Brick 1, Field 5 when
  only the IR data are included in the DOLPHOT process. {\it Middle}
  Same as {\it right} with updated DOLPHOT parameters.  {\it Right}
  CMD from IR photometry with updated DOLPHOT parameters when all UV,
  optical, and IR data centered on the Brick 1, Field 5 are included
  in the DOLPHOT process.  All panels provide the total number of
  stars and the magnitude which 90\% of the stars are brighter than.}
\label{better_ir}
\end{figure}

In addition to the cross-matching advantages, our optimization of
DOLPHOT parameters for our data set and including data from multiple
cameras in the DOLPHOT stack dramatically improved the depth and
quality of the IR photometry. While we expected that the crowding
limit of the IR data may improve with the addition of the optical data
to the stack, we did not expect the improvement to be as remarkable as
it was.  Figure~\ref{better_ir} shows side-by-side the old photometry
of Brick 1, Field 5 (with old parameters and including only the IR
data), a re-reduction of this field using improved photometry
parameters alone, and our new photometry, which includes both the new
optimized parameters and the UV and optical data in the stack.  With
the new parameters and the addition of higher resolution optical data
to the stack to separate discrete objects, the photometry extends more
than 2 magnitudes farther down the RGB.

\begin{figure}
\includegraphics[width=3.4in,clip=true,trim=0.5cm 0cm 0.5cm 0cm]{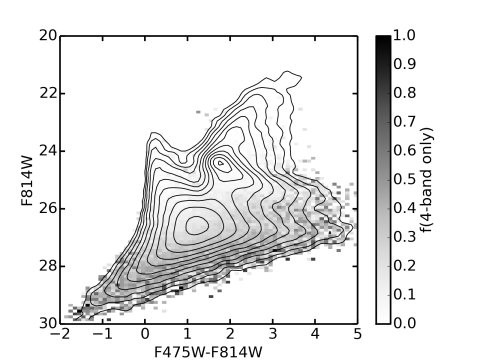}
\includegraphics[width=3.4in,clip=true,trim=0.5cm 0cm 0.5cm 0cm]{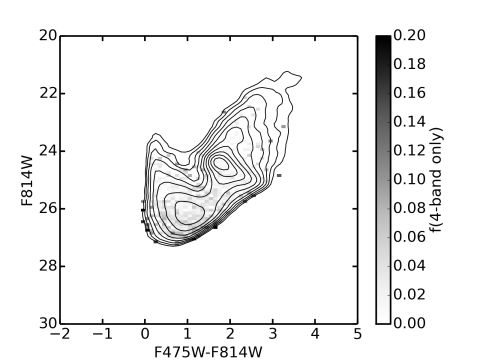}
\caption{{\it Left:} Fraction of raw optical detection lost as a
function of color and magnitude.  Contour levels show the locations of
CMD features in the full sample of the field. Very few ($<$10\%) are
lost inside of CMD features.  {\it Right:} Same as {\it left}, but for
culled optical detections.  The fractions is very low $<$5\% inside of
CMD features.}
\label{exclude_ir_compare}
\end{figure}

In addition to the vastly improved IR photometry, fewer faint sources
were detected in the optical when the IR data were included. To
understand the reason for the lower number, we looked at the CMD of
discrete sources that were measured in stacks with and without the IR
data. The comparison is shown in Figure~\ref{exclude_ir_compare}.  The
plots show the fraction of optical measurements lost when including
the IR data as a function of optical color and magnitude.  We found
$<$5\% of detections were lost brightward of F814W$\sim$25, but a
significantly higher percentage (up to $\sim$30\%) were lost in the
highly-uncertain and relatively amorphous faint points at the bottom
of the CMD.  Furthermore, including the IR data into the DOLPHOT stack
decreases the number of low-quality measurements in the most crowded
regions at the faintest flux levels (clusters and the inner disk).

Overall, the addition of the IR data appears to improve the fidelity
of the catalogs, working as another assessment of the quality of a
measurement. Essentially, the IR data helped remove many low-quality
measurements in a similar way that our quality cuts do.  Thus while
including the IR data may result in lower numbers of optical
detections at the faint end, it also results in higher reliability of
measurements that remain.

\begin{figure}
\includegraphics[width=3.4in,clip=true,trim=1.0cm 0cm 0cm 0cm]{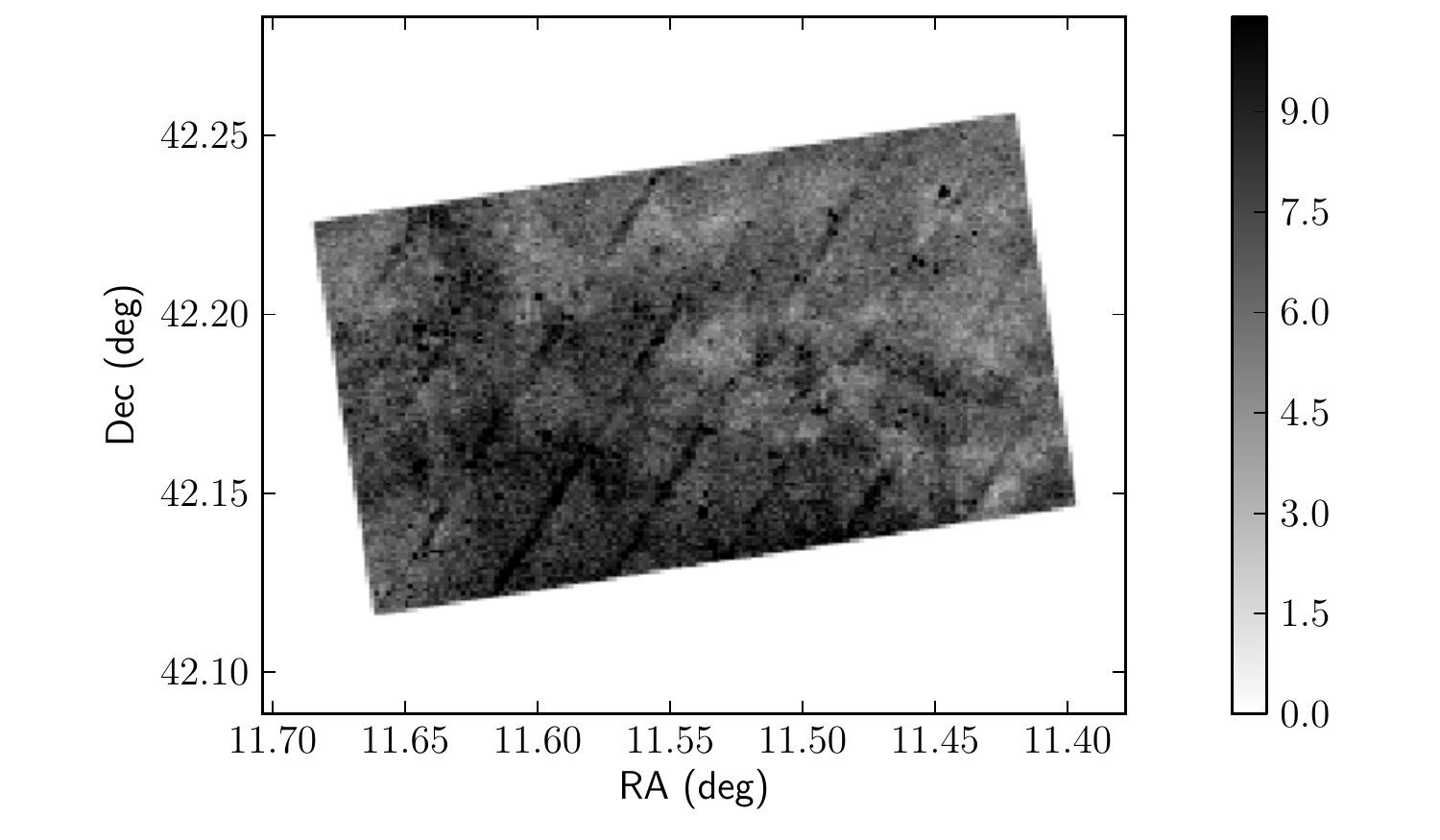}
\includegraphics[width=3.4in,clip=true,trim=1.0cm 0cm 0cm 0cm]{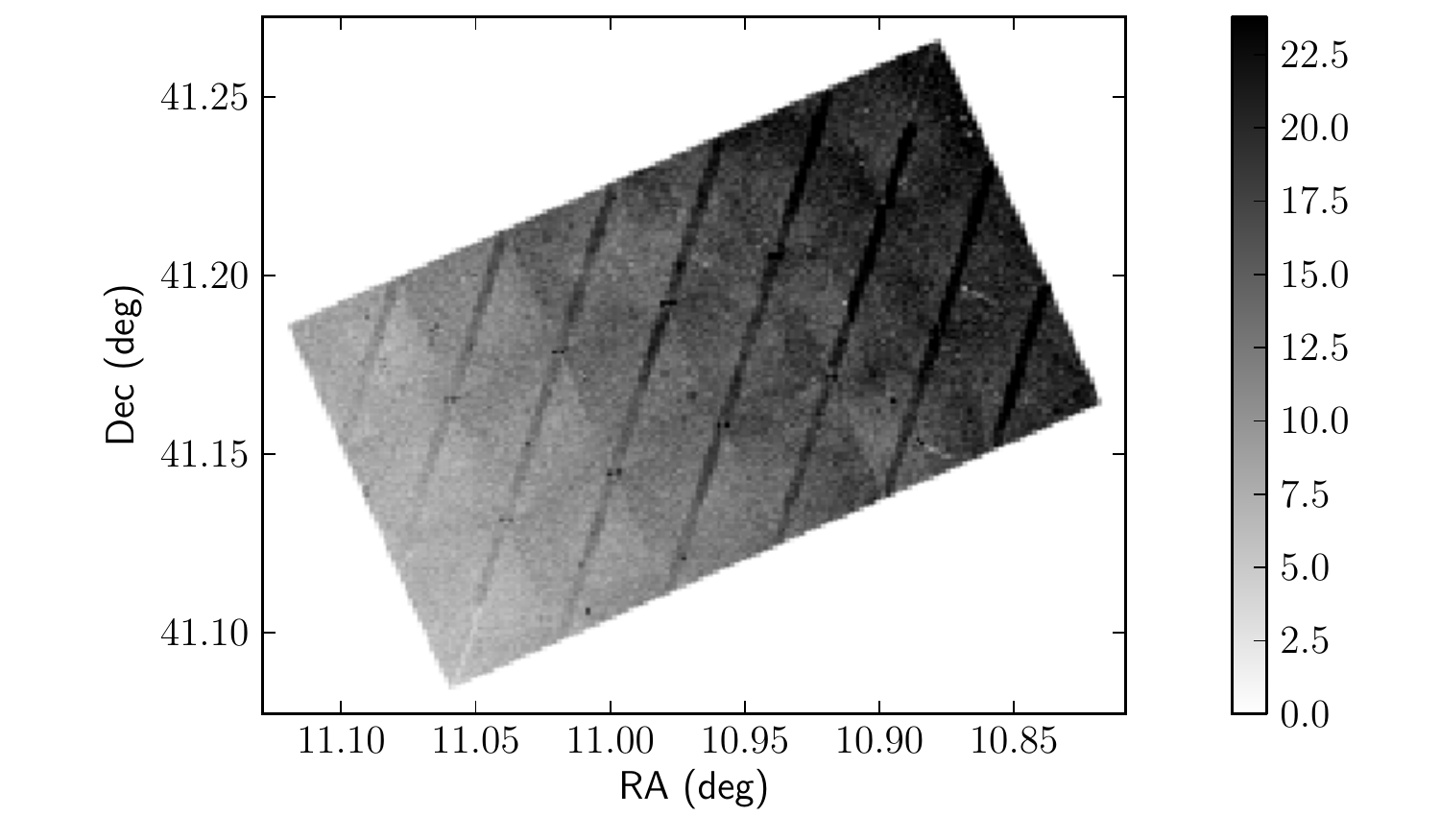}
\caption{{\it Left:} Stellar density map in Brick 17, where crowding
  is less severe. {\it Right:} Same as {\it Left}, but for Brick 2,
  where crowding is more severe.  Grayscale units are in stars per
  arcsec$^2$.  The chip gap features become much more apparent as
  crowding increases.}
\label{6bandgaps}
\end{figure}

\begin{figure}
\includegraphics[width=3.4in]{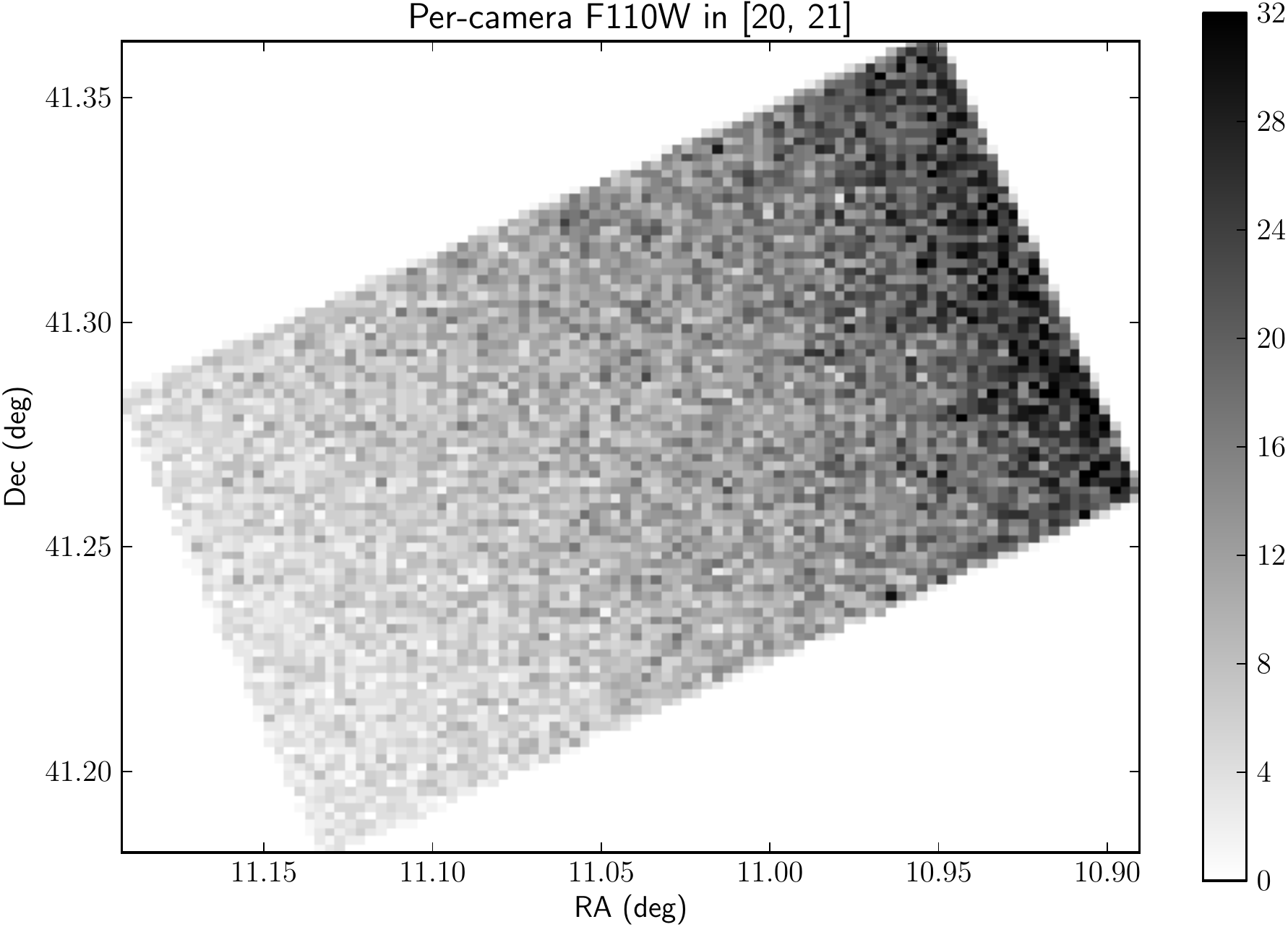}
\includegraphics[width=3.4in]{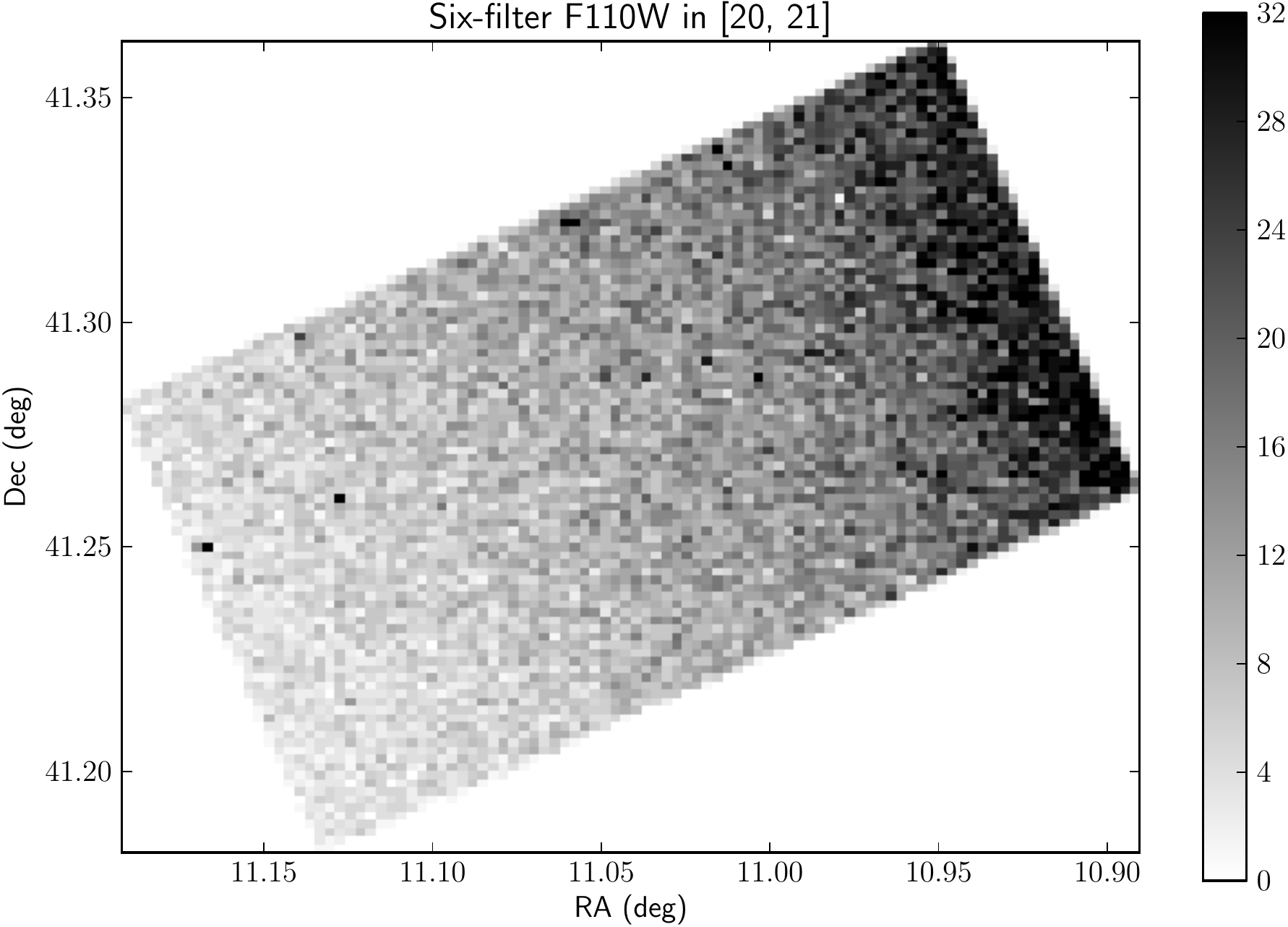}
\caption{Stellar density maps of Brick 02 showing bright stars
  (20$<$m$<$21) with good measurements in F110W. {\it Left:} Map
  produced using results from single-camera photometry. {\it Right:}
  Map produced using results from multi-camera photometry. Grayscale
  units are stars per 5$''{\times}5''$ bin.}
\label{bright_ir_density}
\end{figure}

\begin{figure}
\includegraphics[width=3.4in]{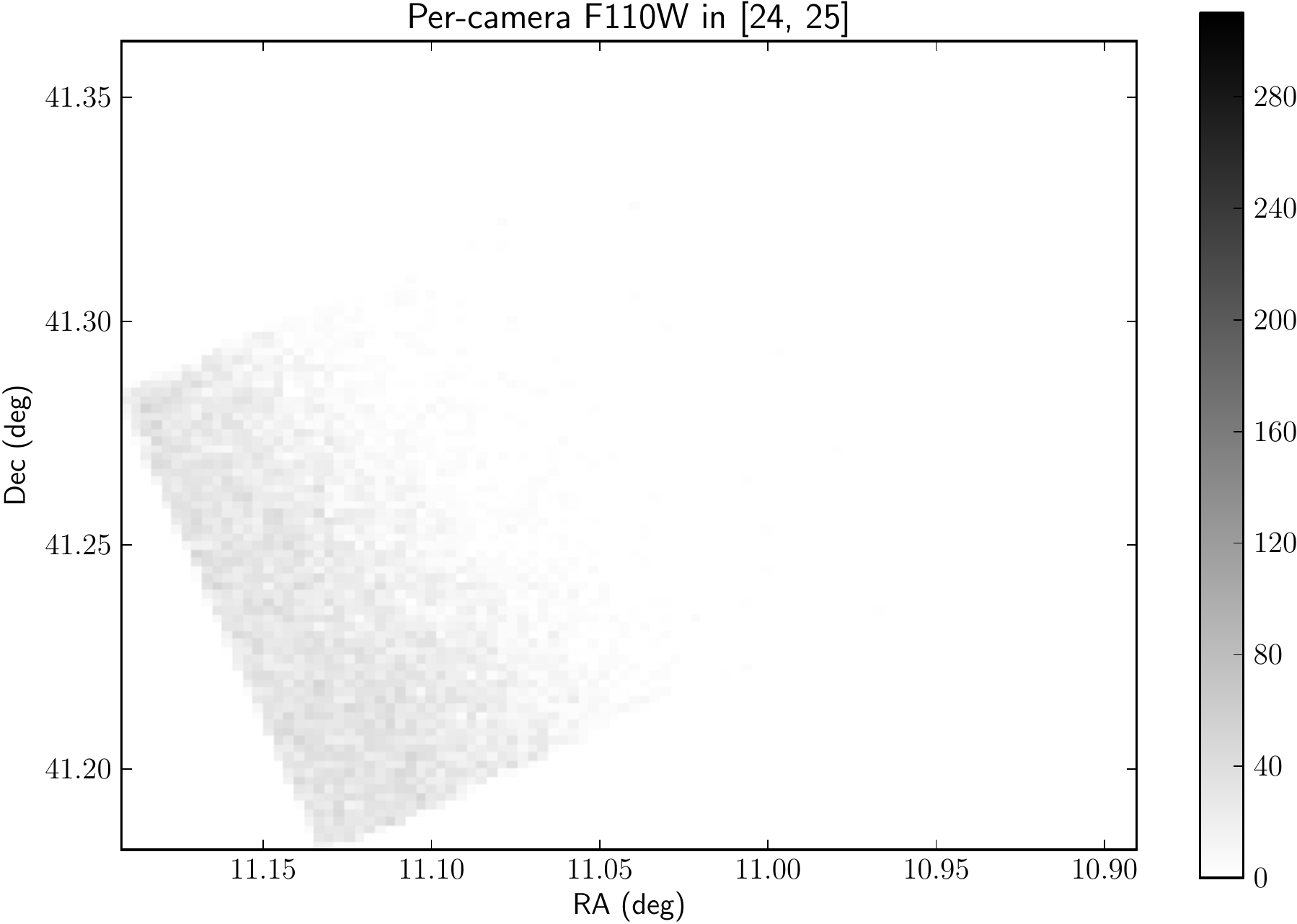}
\includegraphics[width=3.4in]{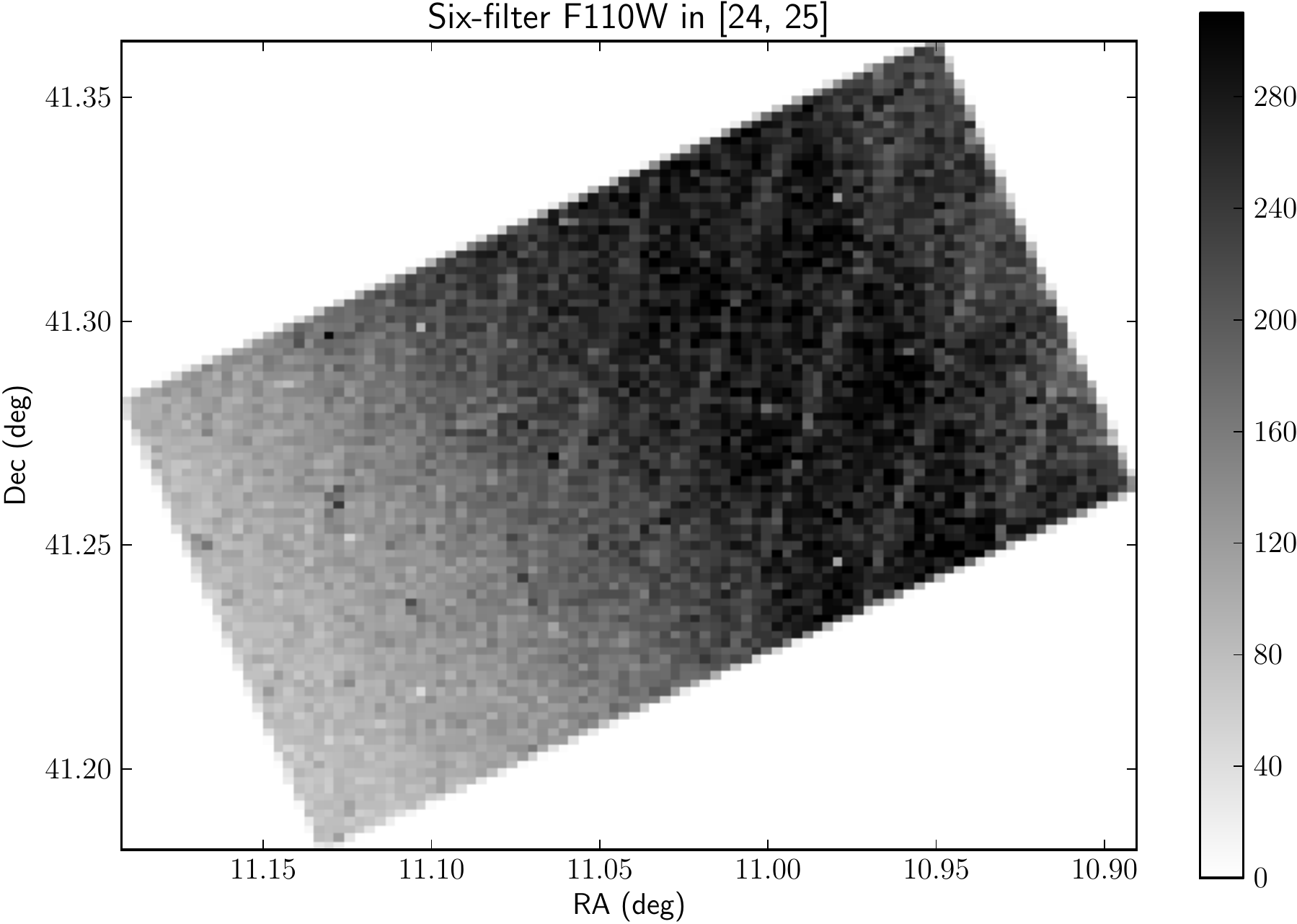}
\caption{Stellar density maps of Brick 02 showing faint stars
  (24$<$m$<$25) with good measurements in F110W. {\it Left:} Map
  produced using results from single-camera photometry. {\it Right:}
  Map produced using results from multi-camera photometry. Grayscale
  units are stars per 5$''{\times}5''$ bin.}
\label{faint_ir_density}
\end{figure}

As shown in Figure~\ref{onefield}, the portions of a Field's catalog
outside of the IR footprint do not contain any IR data.  Therefore, as
described above, these areas will have higher densities of optical
detections. Thus, one feature of merging neighboring ACS and WFC3
catalogs is higher detection densities in the ACS chip gap. We show
this effect in Figure~\ref{6bandgaps}.  This feature only occurs at
the faint magnitudes where crowding has the largest impact on the
quality of the photometry as shown in Figures~\ref{bright_ir_density}
and \ref{faint_ir_density}.  In the outer bricks, the feature is
hardly noticeable.  Future reductions of the data on machines with
significantly more memory will allow the full collection of
overlapping fields to be measured in the same DOLPHOT run, eliminating
the need for this merging and greatly reducing the severity of this
feature.

%
\section{Fields with Caveats}\label{anomalous_fields}

\subsection{Poor UV alignment}

As described in Section~\ref{alignment}, some the some UVIS exposures
were not able to be reliably aligned due to a lack of bright stars.
The UV photometry for these fields may be strongly affected, as no UV
exposures have low weighting.  Thus, UV photometry of the six fields
with fewest UV detections is not reliable.  These include the
following fields: Brick 2, Field 1; Brick 6, Field 13; Brick 12, Field
7; Brick 20, Field 13; Brick 22, Field 13; Brick 23, Field 2.

\subsection{High IR Background}

Some of our fields had strong effects from Earthlimb glow in an IR
exposure \citep{dalcanton2012ir}, which can cause problems with PSF
fitting and background subtraction.  These fields with elevated IR
background levels are listed in Table~\ref{highir}.  In all but one
case, the resulting photometry was of similar quality to the rest of
the survey.  That is, DOLPHOT was able to measure the local background
to the stars and provide CMDs and detection densities similar to
surrounding fields.  In one case, this background issue caused the IR
photometry to be significantly worse than the neighboring fields.
This field was Brick 22, Field 8, where the IR photometry should be
used with extra caution.

\subsection{Fields with Altered Observing Setup}\label{altered_obs}

There were 3 fields that could not be observed at our standard
orientation angles due to a lack of guide stars.  For these fields,
the ACS and UVIS data were taken with different parts of the detector
overlapping than the rest of the survey.  In order to obtain 6-band
coverage for these fields, we needed to have 6 additional orbits.
These orbits were taken at the expense of the southwestern edge of
Brick 11 (Fields 15, 17, and 18 in WFC3, and Fields 13, 14, and 18 in
ACS), because this location was covered almost completely by the
northern edge of Brick 9.

The ACS data for 3 fields were obtained at different orientations than
our standard, because the necessary orientation to have the proper ACS
parallel location did not have available guide stars.  In these cases,
the orientation of the original (WFC3 primary) observation was changed
to provide guide stars, and the ACS data for the optical coverage was
obtained in a separate observation with ACS as the primary instrument,
also with a non-standard orientation.

In detail, the ACS data of Field 12, Brick 3 was observed with in ACS
as primary instrument, and with an orient of 159. In all other bricks
south of the survey bend, the ACS data for Field 12 were taken as the
parallel for Field 9 at an orient of 54, but in this case the WFC3
data for Field 9 were taken at an orient of 249 (instead of the usual
orient of 69) to make guide stars available.  

Brick 10 Field 17 was taken at an orient of 249 with ACS as the
primary instrument. In all other bricks, the ACS data for Field 17 was
taken as the parallel for Field 14 at an orient of 69 degrees, but in
this case, the WFC3 data for Field 14 was observed at an orient of 249
to make guide stars available.

The ACS data for Field 14 of Brick 16 was observed at orient 54 with
ACS as the primary instrument.  In all other bricks north of the
survey bend, the ACS data for Field 14 was taken as the parallel for
Field 17 at an orient of 234, but in this case, the WFC3 data for
Field 17 was observed at an orient of 54 degrees to make guide stars
available.

These 6 observations with non-standard orientations resulted in
additional parallel observations which were not included in this
release of the survey data as they were not needed to produce 6-band
coverage of the PHAT footprint.  They did result in a different
structure of the overlap between the different cameras, but this
difference appeared to have no effect on the resulting photometry.


After the changes for guide star acquisition were finished, we had one
0.1 arcmin$^2$ triangle at the boundary between bricks 9 and 11
without 6-filter coverage (where Brick 11, Field 13 meets Brick 9,
Field 1).

\section{Data Products}

In addition to the machine readable photometry and fake star tables
available through this publication, we are releasing several data
products through the MAST high level science products (HLSP).  These
include, for each field, a fits format table of the original 6-filter
photometry output as returned from DOLPHOT, as well as ST and GST
filtered versions of this table.  We also include drizzled images in
all filters for each field with the correct astrometric solution to
our survey precision of 5 milliarcseconds.

We also provide brickwide catalogs trimmed so that they contain only
unique measurements at each location on the sky.  Each brick has its
own catalog.  These catalogs contain all of the unique measurements
from each ST single-field catalog as described in
Section~\ref{catalog_merge}, along with a GST column for each filter
that has a value of 1 if the measurement passes the GST criteria in that
bandpass.

\section{Conclusions}

We have produced the largest, highest-fidelity, and most homogeneous UV
to IR photometric catalog of equidistant stars ever assembled.  Using
the ACS and WFC3 cameras aboard HST, we have photometered 414
contiguous WFC3/IR footprints covering 0.5 square degrees of the M31
star-forming disk.  The resulting catalog contains 6-band photometry
of over 100 million stars, with very little ($\ll$1\%) contamination
from the Milky Way foreground or background galaxies.  Our photometric
quality as a function of stellar density is homogeneous throughout the
survey; however, the stellar density covers 2.5 orders of magnitude in
the I-band, causing our limiting magnitude in the optical and IR to
vary by 4-5 magnitudes over the full extent of the survey, and causing the
photometric bias to brighter magnitudes to increase with decreasing
radius.

Photometry of artificial stars shows that our UV photometry tends to
measure stars fainter than their true brightness, while all other
bands tend to measure stars brighter than their true brightness due to
crowding effects.  These tests also show that the DOLPHOT
uncertainties, which account only for photon noise, are dominated by
other effects except in the UV at the faint end.  In some cases, the
photon noise error accounts for only $\sim$1\% of the total
photometric error, such as for IR measurements of stars in areas of
high stellar density.

Analysis of the systematic magnitude offsets as a function of detector
position suggests that our systematics are largest in F475W
($\sim$0.04 mag) and less than $\sim$2\% in redder bands.  The spatial
pattern of our magnitude offsets, as well as its magnitude dependence,
suggests that most of our systematic error from ACS photometry is due
to the model PSFs for all but the faintest stars in our catalog.  At
the faint end, however, the systematic errors with position are
symmetric about the chip gap, suggesting that they are dominated by
CTE effects.  These errors at the faint end attributable to CTE are as
large as 0.1 mag.

Support for this work was provided by NASA through grant GO-12055 from
the Space Telescope Science Institute, which is operated by the
Association of Universities for Research in Astronomy, Incorporated,
under NASA contract NAS5-26555. Support for DRW is provided by NASA
through Hubble Fellowship grants HST-HF-51331.01 awarded by the Space
Telescope Science Institute.  We thank D. Pirone and K. Rosema for
their help in engineering the data reduction pipeline.  We thank
Amazon cloud services, for donating some of the computing time
necessary to make these measurements. We thank Alison Vick, for all of
her help in carrying out these complex observations, S. Casertano for
supplying the code to make exposure maps from our Astronomer's
Proposal Tool (APT) files, and J. Anderson for the model PSFs.  IDL
PyRAF, STSDAS, STSCI\_PYTHON, and PyFITS are products of the Space
Telescope Science Institute, which is operated by AURA for NASA.  If
using the PHAT photometry for future science projects, please
reference this work and \citet{dalcanton2012} and acknowledge grant
GO-12055 in their publications.

\clearpage

\begin{deluxetable}{lcccccccc}\tablewidth{18cm}
\tablecaption{Observations, full 2710 row table available electronic only}
\tablehead{
\colhead{Target Name} &
\colhead{RA (J2000)} &
\colhead{Dec (J2000)} &
\colhead{Start Time} &
\colhead{Exp. (s)} &
\colhead{Inst.} &
\colhead{Aper.} &
\colhead{Filter} &
\colhead{PA-V3}
}
\startdata
M31-B01-F01-IR & 00 43 23.674 & +41 15 17.96 & 2010-12-14 06:05:28 & 1197.694 & WFC3   & IR-FIX & F160W & 249.000\\
M31-B01-F01-IR & 00 43 23.687 & +41 15 18.10 & 2010-12-14 06:13:09 & 799.232 & WFC3   & IR-FIX & F110W & 249.000\\
M31-B01-F01-IR & 00 43 23.681 & +41 15 18.15 & 2010-12-14 06:44:46 & 499.232 & WFC3   & IR-FIX & F160W & 249.000\\
M31-B01-F01-UVIS & 00 43 23.973 & +41 15 18.28 & 2010-12-14 03:24:58 & 1350.000 & WFC3   & UVIS-CENTER & F336W & 249.000\\
M31-B01-F01-UVIS & 00 43 23.973 & +41 15 18.28 & 2010-12-14 04:29:38 & 1010.000 & WFC3   & UVIS-CENTER & F275W & 249.000\\
M31-B01-F01-WFC & 00 43 22.717 & +41 15 04.89 & 2010-07-22 21:05:37 & 1520.000 & ACS    & WFC & F814W & 69.000\\
M31-B01-F01-WFC & 00 43 22.717 & +41 15 04.89 & 2010-07-22 22:34:36 & 1720.000 & ACS    & WFC & F475W & 69.000\\
M31-B01-F02-IR & 00 43 14.050 & +41 16 06.28 & 2010-12-16 17:12:31 & 1197.694 & WFC3   & IR-FIX & F160W & 249.000\\
M31-B01-F02-IR & 00 43 14.063 & +41 16 06.42 & 2010-12-16 17:43:36 & 799.232 & WFC3   & IR-FIX & F110W & 249.000\\
M31-B01-F02-IR & 00 43 14.057 & +41 16 06.47 & 2010-12-16 19:27:29 & 499.232 & WFC3   & IR-FIX & F160W & 249.000\\
\nodata & \nodata & \nodata & \nodata & \nodata & \nodata  & \nodata & \nodata & \nodata\\
\enddata
\label{obs_table}
\end{deluxetable}

\clearpage

\begin{deluxetable}{lccc}\tablewidth{8cm}
\tablecaption{DOLPHOT Parameters\tablenotemark{a}}
\tablehead{
\colhead{Detector} &
\colhead{Chip} &
\colhead{Parameter} &
\colhead{Value}
}
\startdata
IR &  1  & xform & ``1 0 0''\\
IR &  1 &  raper & 2\\
IR &  1 &  rchi & 1.5\\
IR &  1 &  rsky0 & 8\\
IR &  1 &  rsky1 & 20\\
IR &  1  &  rpsf & 10\\
IR &  1 &  ref2img & 20 Value Array\\ 
UVIS &  1,2 &  raper & 3\\
UVIS &  1,2  & rchi & 2.0\\
UVIS &  1,2 &  rsky0 & 15\\
UVIS &  1,2 &  rsky1 & 35\\
UVIS &  1,2 &  rpsf & 10\\
UVIS &  1 &  ref2img & 20 Value Array\\ 
UVIS &  2 &  ref2img & 20 Value Array\\ 
WFC &  1,2 &  raper & 3\\
WFC &  1,2 &  rchi & 2.0\\
WFC &  1,2 &  rsky0 & 15\\
WFC &  1,2 &  rsky1 & 35\\
WFC &  1,2 &  rpsf & 10\\
WFC &  1 &  ref2img & 20 Value Array\\ 
WFC &  2 &  ref2img & 20 Value Array\\ 
All &  \nodata &  apsky & ``15 25'' \\
All &  \nodata &  UseWCS & 1 \\  
All &  \nodata &  UseCTE & 1 \\     
All &  \nodata &  PSFPhot & 1 \\       
All &  \nodata &  FitSky & 2  \\         
All &  \nodata &  SkipSky & 2   \\       
All &  \nodata &  SkySig & 2.25  \\      
All &  \nodata &  SecondPass & 5 \\     
All &  \nodata &  SearchMode & 1  \\    
All &  \nodata &  SigFind & 3.0 \\      
All  & \nodata &  SigFindMult & 0.85 \\ 
All &  \nodata &  SigFinal & 3.5\\      
All &  \nodata &  MaxIT & 25  \\       
All &  \nodata &  NoiseMult & 0.10 \\  
All &  \nodata &  FSat & 0.999 \\      
All &  \nodata &  FlagMask & 4 \\       
All &  \nodata &  ApCor & 1  \\         
All &  \nodata &  Force1 & 1 \\        
All &  \nodata &  Align & 2  \\         
All &  \nodata &  aligntol & 4  \\          
All &  \nodata &  alignstep & 2 \\           
All &  \nodata &  Rotate & 1  \\        
All &  \nodata  & RCentroid & 1  \\     
All &  \nodata  & PosStep & 0.1 \\      
All &  \nodata &  dPosMax & 2.5 \\     
All &  \nodata &  RCombine & 1.415 \\   
All &  \nodata &  SigPSF & 3.0  \\      
All &  \nodata &  PSFres & 1  \\        
All &  \nodata &  psfstars &  \nodata \\       
All &  \nodata &  psfoff & 0.0 \\      
All  & \nodata &  DiagPlotType & PNG\\
WFC &  1,2 &  ACSpsfType & 1 \\
IR &  1 &  WFC3IRpsfType & 1 \\
UVIS &  1,2 &  WFC3UVISpsfType & 1 \\
\enddata
\label{parameters}
\tablenotetext{a}{Parameter descriptions available at
http://americano.dolphinsim.com/dolphot/dolphot.pdf}
\end{deluxetable}

\clearpage

\begin{deluxetable}{ccc}
\tablecaption{Crowding and square of sharpness values used for measurement culling and flagging\tablenotemark{a}}
\tablehead{
\colhead{Camera} &
\colhead{Sharpness Squared} &
\colhead{Crowding}
}
\startdata
UVIS & 0.15 & 1.3\\ 
ACS & 0.2 & 2.25\\
IR & 0.15 & 2.25\\
\enddata
\tablenotetext{a}{Maximum allowed values for the crowding and the square of the sharpness value applied to the combined measurement for each star in each band independently.  Measurements with lower values are given a GST value of 1, and those with higher values are given a GST flag of 0. Measurements that did not pass the sharpness criterion in any band were completely rejected.}
\label{gstcuts}
\end{deluxetable}

\begin{deluxetable}{lccccccc}
\tablecaption{Number of stars with reliable measurements in each band in each brick}
\tablehead{
\colhead{Brick} &
\colhead{Total} &
\colhead{F275W} &
\colhead{F336W} &
\colhead{F475W} &
\colhead{F814W} &
\colhead{F110W} &
\colhead{F160W}
}
\startdata
1  &  7739068  &  18310  &  268608  &  6587303  &  7276966  &  5605734  &  4547719 \\
2  &  5389720  &  32745  &  150598  &  4064105  &  4635696  &  3753250  &  2740961 \\
3  &  7420074  &  25280  &  126152  &  6296160  &  6969507  &  5327618  &  4161662 \\
4  &  5699925  &  38868  &  165470  &  4430452  &  4966957  &  3949133  &  2947135 \\
5  &  7017006  &  27125  &  121143  &  5871771  &  6513479  &  5070041  &  4000621 \\
6  &  5737431  &  35452  &  154325  &  4413826  &  4969404  &  3928435  &  2898493 \\
7  &  6548176  &  31481  &  110925  &  5307224  &  5917345  &  4654753  &  3696269 \\
8  &  5619644  &  33584  &  151109  &  4256729  &  4817534  &  3841721  &  2787050 \\
9  &  6107219  &  38838  &  144052  &  4589795  &  5361598  &  4255312  &  3282843 \\
10  &  5286304  &  37461  &  114350  &  3826009  &  4445694  &  3666566  &  2604794 \\
11  &  2944424  &  16185  &  41962  &  2257404  &  2560174  &  2007530  &  1490975 \\
12  &  5037810  &  32830  &  130383  &  3715558  &  4286834  &  3433973  &  2421219 \\
13  &  5701760  &  37060  &  100222  &  4178398  &  4886377  &  3907437  &  2844294 \\
14  &  5381346  &  37474  &  139786  &  3945878  &  4583746  &  3649396  &  2565388 \\
15  &  5506459  &  47311  &  201660  &  3579821  &  4638748  &  3876490  &  2831302 \\
16  &  4978208  &  36893  &  135707  &  3453638  &  4152284  &  3371484  &  2324596 \\
17  &  4928251  &  35865  &  144164  &  3299131  &  4072788  &  3462695  &  2398277 \\
18  &  3988686  &  31269  &  87305  &  2588023  &  3199622  &  2691382  &  1724250 \\
19  &  4030587  &  27500  &  76068  &  2638976  &  3259447  &  2740509  &  1762750 \\
20  &  3116348  &  38093  &  70767  &  1843219  &  2390934  &  2064556  &  1273885 \\
21  &  3207788  &  31108  &  90925  &  1897030  &  2535071  &  2228104  &  1414521 \\
22  &  2718684  &  25772  &  55894  &  1495505  &  2002241  &  1739825  &  1063914 \\
23  &  2756854  &  22383  &  53262  &  1613141  &  2134913  &  1780956  &  1114743 \\
\enddata
\label{stars}
\end{deluxetable}

\clearpage

\begin{turnpage}

\begin{deluxetable}{llcccccccccccccccccc}
\tablecaption{Example of the Photometry Catalog (full catalog
  available online)}
\tablehead{
\colhead{RA(J2000)} &
\colhead{DEC(J2000)} &
\colhead{F275W} &
\colhead{S/N} &
\colhead{GST} &
\colhead{F336W} &
\colhead{S/N} &
\colhead{GST} &
\colhead{F475W} &
\colhead{S/N} &
\colhead{GST} &
\colhead{F814W} &
\colhead{S/N} &
\colhead{GST} &
\colhead{F110W} &
\colhead{S/N} &
\colhead{GST} &
\colhead{F160W} &
\colhead{S/N} &
\colhead{GST} 
}
\startdata
10.580256333  &  41.2441757179  &  26.68  &  1.0  &  F  &  99.999  &  -0  &  F  &  24.603  &  42.3  &  T  &  21.85  &  125.7  &  T  &  20.981  &  117.5  &  T  &  20.027  &  124.2  &  T \\
10.5802572159  &  41.2457103763  &  23.848  &  8.2  &  T  &  27.801  &  0.8  &  F  &  25.392  &  19.4  &  T  &  23.603  &  35.7  &  T  &  22.538  &  33.2  &  T  &  21.777  &  31.8  &  F \\
10.58025753  &  41.245922544  &  99.999  &  -0.2  &  F  &  99.999  &  -0.1  &  F  &  25.573  &  20.7  &  T  &  23.428  &  40.8  &  T  &  22.814  &  28.7  &  T  &  22.279  &  22.2  &  T \\
10.5802579935  &  41.2460588596  &  99.999  &  -0.4  &  F  &  26.706  &  2.2  &  F  &  25.415  &  24.7  &  T  &  23.329  &  45.6  &  T  &  22.106  &  49.5  &  T  &  21.329  &  48.7  &  T \\
10.5802580564  &  41.2434928097  &  26.553  &  1.1  &  F  &  99.999  &  -1.7  &  F  &  99.999  &  -1.4  &  F  &  21.861  &  123.8  &  T  &  20.065  &  235.2  &  T  &  18.969  &  300.9  &  T \\
10.5802582443  &  41.2474459936  &  99.999  &  -1.3  &  F  &  27.188  &  1.4  &  F  &  24.931  &  34.0  &  T  &  21.997  &  108.1  &  T  &  21.097  &  112.3  &  T  &  20.149  &  133.1  &  T \\
10.5802583084  &  41.2479862568  &  99.999  &  -0.3  &  F  &  26.528  &  2.5  &  F  &  25.488  &  21.9  &  T  &  22.83  &  63.8  &  T  &  22.022  &  54.5  &  T  &  21.264  &  51.9  &  T \\
10.5802586378  &  41.2493851795  &  99.999  &  -0.3  &  F  &  99.999  &  -1.2  &  F  &  26.932  &  6.7  &  T  &  24.43  &  17.2  &  T  &  23.229  &  19.1  &  T  &  23.048  &  10.3  &  F \\
10.5802594502  &  41.2488740433  &  99.999  &  -1.6  &  F  &  99.999  &  -1.6  &  F  &  25.714  &  18.5  &  T  &  23.245  &  48.7  &  T  &  22.496  &  40.2  &  T  &  21.367  &  46.5  &  T \\
10.5802601961  &  41.2440702789  &  99.999  &  -0.5  &  F  &  27.688  &  0.9  &  F  &  26.722  &  7.4  &  T  &  25.029  &  10.0  &  T  &  99.999  &  -1.5  &  F  &  99.999  &  -0.7  &  F \\
10.580261801  &  41.2473107079  &  99.999  &  -0.3  &  F  &  27.534  &  1.0  &  F  &  25.032  &  30.4  &  T  &  22.741  &  59.8  &  T  &  21.947  &  53.0  &  T  &  21.201  &  49.9  &  T \\
...  & ...   &  ...   &  ...   &  ...  & ...  & ...  & ...   & ...  &...   &...   & ...  &  ... & ...  & ...  & ...  & ...  & ...  & ...  & ...  \\
\enddata
\label{catalog}
\end{deluxetable}

\end{turnpage}

\clearpage

\begin{deluxetable}{ccccccccc}
\tablecaption{Standard deviation of magnitude offsets and median sharpness for different magnitude bins in different filters.}
\tablehead{
\colhead{Mag} &
\colhead{F475W Offset} &
\colhead{F475W Sharp} &
\colhead{F814W Offset} &
\colhead{F814W Sharp} &
\colhead{F110W Offset} &
\colhead{F110W Sharp} &
\colhead{F160W Offset} &
\colhead{F160W Sharp} 
}
\startdata
22.5 & \nodata & \nodata & 0.016 & 0.010 & 0.013 & 0.006 & 0.014 & 0.006\\
23.5 & 0.019 & 0.018 & 0.017 & 0.009 & 0.013 & 0.006 & 0.017 & 0.006\\
24.5 & 0.022 & 0.017 & 0.018 & 0.007 & 0.023 & 0.006 & 0.049 & 0.009\\
25.5 & 0.027 & 0.014 & 0.021 & 0.004 & 0.048 & 0.007 & 0.051 & 0.016\\
26.5 & 0.031 & 0.010 & 0.044 & 0.003 & 0.041 & 0.012 & \nodata & \nodata \\
27.5 & 0.043 & 0.004 & 0.077 & 0.007 & \nodata & \nodata & \nodata & \nodata\\
28.5 & 0.113 & 0.012 & \nodata & \nodata & \nodata & \nodata & \nodata & \nodata\\
\enddata
\label{systematics}
\end{deluxetable}

\clearpage

\begin{turnpage}

\begin{deluxetable}{llcccccccccccc}
\tablecaption{Example of the Artificial Star Catalog (full catalog
  available online)}
\tablehead{
\colhead{RA(J2000)} &
\colhead{DEC(J2000)} &
\colhead{F275W$_{\rm IN}$} &
\colhead{F275W$_{\rm OUT}$} &
\colhead{F336W$_{\rm IN}$} &
\colhead{F336W$_{\rm OUT}$} &
\colhead{F475W$_{\rm IN}$} &
\colhead{F475W$_{\rm OUT}$} &
\colhead{F814W$_{\rm IN}$} &
\colhead{F814W$_{\rm OUT}$} &
\colhead{F110W$_{\rm IN}$} &
\colhead{F110W$_{\rm OUT}$} &
\colhead{F160W$_{\rm IN}$} &
\colhead{F160W$_{\rm OUT}$} 
}
\startdata
10.86064898  &  41.46919167  &  25.084  &  99.999  &  23.212  &  23.205  &  22.026  &  22.016  &  20.207  &  20.189  &  19.586  &  19.573  &  19.013  &  18.993 \\
10.86064901  &  41.44335114  &  26.924  &  99.999  &  26.251  &  25.919  &  25.976  &  25.77  &  24.465  &  24.192  &  23.986  &  24.018  &  23.658  &  99.999 \\
10.86065098  &  41.46782277  &  28.915  &  99.999  &  26.728  &  99.999  &  22.813  &  22.839  &  19.456  &  19.475  &  18.288  &  18.296  &  17.202  &  17.214 \\
10.86065183  &  41.45522179  &  31.883  &  99.999  &  29.876  &  99.999  &  26.091  &  26.198  &  22.439  &  22.445  &  21.091  &  21.11  &  19.9  &  19.913 \\
10.86065268  &  41.43842322  &  22.525  &  22.616  &  22.362  &  22.359  &  22.729  &  22.704  &  22.114  &  22.021  &  21.965  &  21.865  &  21.827  &  21.585 \\
10.860653  &  41.43563822  &  28.765  &  99.999  &  26.253  &  99.999  &  22.481  &  22.427  &  19.303  &  19.308  &  18.169  &  18.175  &  17.086  &  17.099 \\
10.86065355  &  41.44038206  &  31.363  &  99.999  &  31.311  &  99.999  &  26.655  &  26.603  &  22.013  &  22.006  &  20.306  &  20.306  &  18.973  &  18.988 \\
10.86065372  &  41.43920687  &  26.731  &  99.999  &  26.185  &  99.999  &  25.609  &  25.802  &  24.889  &  25.148  &  24.625  &  99.999  &  24.457  &  99.999 \\
10.86065452  &  41.43584972  &  28.891  &  99.999  &  26.218  &  99.999  &  24.649  &  24.703  &  22.634  &  22.649  &  21.929  &  22.029  &  21.152  &  21.145 \\
10.86065506  &  41.44638759  &  30.342  &  99.999  &  28.582  &  99.999  &  24.538  &  24.628  &  20.886  &  20.891  &  19.591  &  19.584  &  18.445  &  18.454 \\
10.86065509  &  41.45877434  &  30.293  &  99.999  &  27.207  &  99.999  &  25.15  &  24.777  &  22.683  &  22.459  &  21.79  &  21.713  &  20.892  &  20.835 \\
... & ... & ... & ... & ... & ... & ... & ... & ... & ... & ... & ... & ... & ... \\
\enddata
\label{fake-catalog}
\end{deluxetable}

\end{turnpage}

\clearpage

\begin{deluxetable}{ccccccc}
\tablecaption{RMS, Bias, and DOLPHOT-reported uncertainties from our artificial star tests as a function of stellar density, filter, and magnitude.  Full table available in electronic form only.}
\tablehead{
\colhead{Density} &
\colhead{Filter} &
\colhead{Magnitude} &
\colhead{Bias} &
\colhead{RMS} &
\colhead{DOLPHOT} &
\colhead{Ratio} 
}
\startdata
 0.04 & F275W &  16.0 & 0.005 & 0.015 & 0.002 &   7.4\\
 0.04 & F275W &  16.5 & 0.005 & 0.012 & 0.003 &   4.0\\
 0.04 & F275W &  17.0 & 0.006 & 0.012 & 0.004 &   3.1\\
 0.04 & F275W &  17.5 & 0.007 & 0.011 & 0.005 &   2.3\\
 0.04 & F275W &  18.0 & 0.009 & 0.012 & 0.006 &   2.0\\
 0.04 & F275W &  18.5 & 0.011 & 0.016 & 0.008 &   1.9\\
 0.04 & F275W &  19.0 & 0.013 & 0.018 & 0.010 &   1.8\\
 0.04 & F275W &  19.5 & 0.015 & 0.023 & 0.013 &   1.7\\
 0.04 & F275W &  20.0 & 0.018 & 0.029 & 0.017 &   1.7\\
 0.04 & F275W &  20.5 & 0.022 & 0.037 & 0.022 &   1.7\\
 ... &   ... &  ... &  ... &  ... & ... & ...\\
\enddata
\tablenotetext{a}{Density of stars with 18.5$<$F160W$<$19.5 per arcsec$^2$.}
\label{ast_results}
\end{deluxetable}

\begin{deluxetable}{ccccccccc}
\tablecaption{50\% completeness limits for each band at a variety of stellar densities.}
\tablehead{
\colhead{Brick} &
\colhead{Field} &
\colhead{Stellar Density\tablenotemark{a}} &
\colhead{F275W$_{\rm lim}$} &
\colhead{F336W$_{\rm lim}$} &
\colhead{F475W$_{\rm lim}$} &
\colhead{F814W$_{\rm lim}$} &
\colhead{F110W$_{\rm lim}$} &
\colhead{F160W$_{\rm lim}$} 
}
\startdata
21 & 15 & 0.041 & 25.06 & 26.04 & 27.97 & 26.85 & 26.11 & 25.16\\
9 & 2 & 0.185 & 24.96 & 25.95 & 27.60 & 26.03 & 24.99 & 23.99\\
5 & 10 & 0.516 & 24.83 & 25.90 & 27.34 & 25.17 & 23.80 & 22.88\\
3 & 5 & 1.549 & 24.81 & 25.87 & 26.25 & 23.89 & 22.72 & 21.70\\
1 & 5 & 4.304 & 24.64 & 25.34 & 25.41 & 23.05 & 21.89 & 20.68\\
1 & 10 & 7.532 & 24.41 & 25.08 & 24.72 & 22.39 & 21.07 & 19.99 \\
\enddata
\tablenotetext{a}{Density of stars with 18.5$<$F160W$<$19.5 per arcsec$^2$.}
\label{comp_table}
\end{deluxetable}

\begin{deluxetable}{ccc}
\tablecaption{Fields with elevated IR background}
\tablehead{
\colhead{Brick} &
\colhead{Field} &
\colhead{Filter} 
}
\startdata
2 & 7 & F110W\\
4 & 2 & F110W\\
4 & 13 & F110W\\
7 & 16 & F110W\\
7 & 17 & F110W\\
8 & 7 & F110W\\
8 & 9 & F110W\\
10 & 17 & F110W\\
13 & 7 & F110W\\
16 & 2 & F160W\\
16 & 15 & F110W\\
17 & 4 & F110W\\
17 & 18 & F110W\\
18 & 3 & F110W\\
18 & 7 & F110W\\
18 & 13 & F110W\\
18 & 14 & F110W\\
18 & 15 & F110W\\
21 & 6 & F110W\\
22 & 2 & F160W\\
22 & 3 & F160W\\
22 & 7 & F160W\\
22 & 8 & F110W\\
22 & 13 & F160W\\
22 & 14 & F110W\\
23 & 11 & F110W\\
23 & 13 & F110W\\
23 & 14 & F110W\\
23 & 14 & F110W\\
23 & 15 & F110W\\
23 & 15 & F160W\\

\enddata
\label{highir}
\end{deluxetable}

\clearpage

\clearpage

\clearpage

\clearpage

\clearpage

\clearpage

\clearpage

\end{document}